\documentclass[amsmath,superscriptaddress,nofootinbib,
amssymb,aps,prd,floatfix,twocolumn]{revtex4}

\usepackage{graphicx}
\usepackage{mathrsfs}
\usepackage[bookmarks=false]{hyperref}
\usepackage[utf8]{inputenc}
\usepackage{dcolumn}
\usepackage{bm}
\usepackage{soul}

\begin{document}

\title{From quasinormal modes of rotating black strings\\
to hydrodynamics of a moving CFT plasma}

\author{Luis A. H. Mamani}%
 \email{luis.mamani@ufabc.edu.br}
\affiliation {Centro de Ci\^encias Naturais e Humanas, Universidade
Federal do ABC, Avenida dos Estados 5001, 09210-580 Santo Andr\'e, SP,
Brazil.}
\affiliation{Laborat\'orio de Astrof\'{\i}sica Te\'orica e Observacional, 
Departamento de Ci\^encias Exatas e Tecnol\'ogicas,
Universidade Estadual de Santa Cruz, Rodovia Jorge Amado, km 16, 45650-000 Ilh\'eus,
BA, Brazil.}

\author{Jaqueline Morgan}
\email{jaqueline.morgan@caxias.ifrs.edu.br}
\affiliation{Instituto Federal de Educação, Ciência e Tecnologia do Rio Grande do Sul, Rua
Avelino Antônio de Souza 1730, 95043-700 Caxias do Sul, RS, Brazil.}

\author{Alex S. Miranda}
\email{asmiranda@uesc.br}
\affiliation{Laborat\'orio de Astrof\'{\i}sica Te\'orica e Observacional, 
Departamento de Ci\^encias Exatas e Tecnol\'ogicas,
Universidade Estadual de Santa Cruz, Rodovia Jorge Amado, km 16, 45650-000 Ilh\'eus,
BA, Brazil.}

\author {Vilson T. Zanchin}
\email{zanchin@ufabc.edu.br}
\affiliation {Centro de Ci\^encias Naturais e Humanas, Universidade
Federal do ABC, Avenida dos Estados 5001, 09210-580 Santo Andr\'e, SP,
Brazil.}

\begin{abstract}
A certain identification of points in a planar Schwarzschild-anti de Sitter
(AdS) black hole generates a four-dimensional static black string. In turn,
a rotating black string can be obtained from a static one by means of 
a local boost along the compact direction. On the basis of the gauge/gravity duality,
these black strings are dual to rotating thermal states of a strongly interacting
conformal field theory (CFT) that lives on a cylinder. In this work, we obtain the
complete quasinormal mode (QNM) spectrum of the gravitational perturbations of
rotating black strings. Analytic solutions for the dispersion relations are found
in the hydrodynamic limit, characterized by fluctuations with wavenumber and frequency
much smaller than the Hawking temperature of the string (or the temperature of the CFT in the
dual description). We obtain these dispersion relations both by studying the gravitational
perturbations of rotating black strings and by investigating relativistic wave vectors in
a moving fluid living on the boundary of the AdS spacetime. Relativistic effects like the Doppler
shift of the frequencies, wavelength contraction, and dilation of the thermalization time are shown
explicitly in such a regime. 
The numerical solutions for the fundamental QNMs show a crossover (a transition) from a
hydrodynamic-like behavior to a linear relativistic scaling for large wavenumbers.
Additionally, we find a new family of QNMs which are purely damped in the zero wavenumber
limit and that does not follow as a continuation of QNMs of the static black string,
but that appears to be closely related to the algebraically special perturbation modes.

\end{abstract}

\maketitle

\section{Introduction}
\label{introd}

Since its advent in the late $1990$'s, the celebrated anti-de Sitter/conformal 
field theory (AdS/CFT) correspondence 
\cite{Maldacena:1997re,Witten:1998qj,Gubser:1998bc,Aharony:1999ti} has been 
extended and applied to different areas of physics. Such developments have lead 
to what is now known as the AdS/QCD 
\cite{Polchinski:2001tt,Sakai:2004cn,BoschiFilho:2006pt,Erdmenger:2007cm}, the 
AdS/condensed matter \cite{Herzog:2009xv,Hartnoll:2009sz,McGreevy:2009xe}, and 
the fluid/gravity \cite{Rangamani:2009xk,Hubeny:2010ry,Kovtun:2012rj} 
correspondences. Over the last two decades, the AdS/CFT duality has allowed the 
study of properties of strongly coupled systems in a $n$-dimensional flat 
spacetime by mapping them to a weakly coupled gravitational theory in an 
asymptotically AdS$_{n+1}$ spacetime. In applications to particle physics, 
top-down and bottom-up models were used to study, among other things, the mass 
spectrum, the correlation functions, and the deep inelastic scattering of 
glueballs, vector and scalar mesons  
\cite{Karch:2006pv,BallonBayona:2008zi,Miranda:2009uw,Colangelo:2009ra, 
BallonBayona:2010ae,Colangelo:2012jy,Mamani:2013ssa,Capossoli:2015sfa,
Ballon-Bayona:2017sxa}. Some phenomena in condensed matter, such as the 
high-temperature superconductivity 
\cite{Gubser:2008px,Hartnoll:2008vx,Hartnoll:2008kx}, the classical and quantum 
Hall effects 
\cite{Hartnoll:2007ai,Blake:2014yla,Lindgren:2015lia,Bergman:2010gm,
Bayntun:2010nx,Jokela:2011eb} and the (non-)Fermi liquid behavior of certain 
materials \cite{Lee:2008xf,Bergman:2011rf,Davison:2013uha}, were also object of 
study in the literature. In relation to plasma physics, the fluid/gravity 
correspondence establishes a one-to-one map between solutions of the 
relativistic Navier-Stokes equation and asymptotically AdS black hole solutions 
of Einstein equations 
\cite{Bhattacharyya:2007vs,Bhattacharyya:2008jc,Bhattacharyya:2008xc,
Bredberg:2010ky,Bredberg:2011jq}. Among the important results obtained so far, 
one may cite the universality of the ratio between the shear viscosity and 
entropy density of a holographic CFT plasma \cite{Policastro:2002se, 
Policastro:2002tn,Kovtun:2003wp,Herzog:2002fn,Herzog:2003ke,Kovtun:2004de,
Son:2007vk}.

\begin{figure}[!ht]
\includegraphics[width=3.5cm]{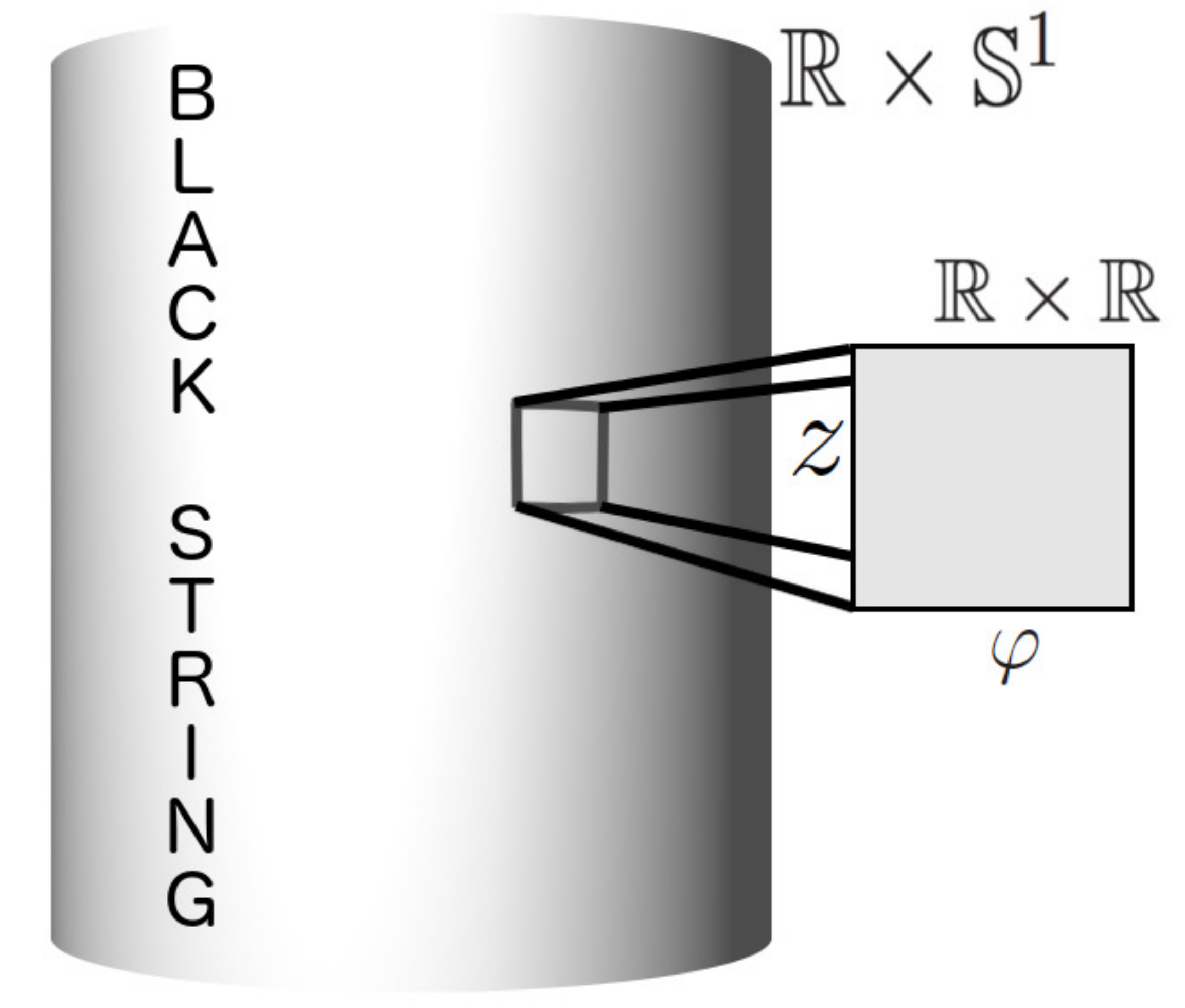}
\caption{The global topology of the black string $\mathbb{R}\times\mathbb{S}^1$, 
and the visualization of the local flat geometry.}
\label{BlackStringBG}
\end{figure}

In the gravity side of the correspondence, the Einstein equations with a 
negative cosmological constant admit four-dimensional black hole solutions 
\cite{Lemos:1994xp,Lemos:1995cm} with cylindrical horizon topology (see 
Fig.~\ref{BlackStringBG}). These objects, known also as black strings, can be 
put to rotate through a boost in the compact direction, which is an improper 
coordinate transformation as discussed by Stachel \cite{Stachel:1981fg}. 
Although locally equivalent, static and rotating black strings are globally 
different solutions of the Einstein equations. Just as static black strings are 
dual to static thermal states of a strongly coupled CFT on the boundary of the 
AdS$_4$ spacetime, the rotating black strings correspond to rotating thermal 
states of this CFT. Such a theory is defined on a three-dimensional Minkowski 
spacetime with one compact dimension.

According to the AdS/CFT dictionary, the classical field perturbations in the 
AdS bulk represent out-of-equilibrium linear excitations in the CFT. In 
particular, we can obtain finite-temperature correlation functions by evaluating 
the action of the scalar, electromagnetic or gravitational perturbations in the 
AdS spacetime, following the Son-Starinets prescription or its generalizations 
\cite{Son:2002sd,Herzog:2002pc,Skenderis:2008dh,Skenderis:2008dg}. By explicitly 
calculating the momentum-space retarded Green's functions in the dual field 
theory, it has been noticed that the poles of these functions are precisely the 
frequencies of the quasinormal modes (QNMs) of field perturbations in the AdS 
spacetime \cite{Cardoso:2001hn,Birmingham:2001pj,Birmingham:2002ph}. The 
standard boundary conditions used for asymptotically flat black hole spacetimes 
are just incoming waves at the horizon and outgoing waves at the spatial 
infinity, which lead to a complex spectrum for the QNMs 
\cite{Kokkotas:1999bd,Nollert:1999ji}. In the case of asymptotically AdS 
backgrounds, the presence of a negative cosmological constant changes the 
spacetime asymptotic structure and the outgoing wave condition at spatial 
infinity is, in general, replaced by a Dirichlet boundary condition 
\cite{Horowitz:1999jd,Cardoso:2001bb,Nunez:2003eq,Kovtun:2005ev}.

The study of QNMs of AdS black holes is a two-decade old topic, and so there are 
a significant number of papers in the literature investigating QNMs in 
asymptotically AdS spacetimes. See Refs. \cite{Berti:2009kk,Konoplya:2011qq} for 
reviews on the subject and 
\cite{Cardoso:2001vs,Starinets:2002br,Miranda:2005qx,Miranda:2007bv,
Miranda:2008vb,Morgan:2009vg,Morgan:2009pn,Morgan:2013dv} for references on the 
study of QNMs of black branes and black strings. In spite of this long history, 
the gravitational QNMs of rotating black strings studied in the present work 
have not been investigated yet. The frequencies of these QNMs correspond to the 
poles of the stress-energy tensor correlators in the dual CFT. We see that it is 
possible to separate the gravitational perturbations of the rotating black 
strings into two sectors, which can be called as transverse and longitudinal 
perturbations. The main goal of this work is to explore the complete QNM 
spectrum associated to these perturbations and investigate its connections with 
the AdS/CFT correspondence. 

The structure of the paper is as follows. Section \ref{secBack} is a review of 
the main properties of the rotating black string background, such as the Hawking 
temperature and the relation with the spacetime of a static black string. In 
Section \ref{sec-equations} the fundamental differential equations for the 
transverse and longitudinal gravitational perturbations are presented, including 
a study of the symmetries of these equations. Section \ref{sec-qnmresults} is 
dedicated to obtain the solutions of the fundamental perturbation equations in 
the hydrodynamic limit. In such an approximation, analytical solutions are found 
and used to build the dispersion relations for both sectors of the gravitational 
perturbations. Numerical solutions are also shown for a comparison to the 
analytical expressions. Several results for non-hydrodynamic QNMs are presented 
and discussed in section \ref{sec-numericalQNM}. Since we cannot find a general 
exact solution for the differential equations, we use the Horowitz-Hubeny method 
\cite{Horowitz:1999jd}. Section \ref{sec-purelyDamped} is devoted to investigate 
the relation between a class of highly damped QNMs and the algebraically special 
frequencies of the rotating black strings.  Section \ref{secfinal} contains the 
final comments and conclusion.

\section{The background spacetime}\label{secBack}

The gravitational background considered here is the spacetime of a
rotating AdS$_4$ black string, whose metric may be written
in the form  \cite{Lemos:1994xp}
\begin{equation}\label{background1}
\begin{split}
ds^2=\frac{\alpha ^2 \gamma ^2\,r^2_{h}}{u^2}\Bigg[&-(dt-a d\varphi)^2f
      + \left(\frac{d\varphi}{\alpha }-a \alpha dt\right)^2\\ 
  & + \frac{dz^2}{\gamma^2}\Bigg]
   +\frac{du^2}{\alpha ^2 u^2 f},
 \end{split}
\end{equation}
where $\alpha$ is a parameter related to the AdS radius $R$ by 
$\alpha = 1/R=\sqrt{-3/\Lambda}$, with $\Lambda$ being the negative 
cosmological constant, $r_h$ is a constant with units of length,
and we defined
\begin{gather}
\gamma= \left(1-a^2\alpha^2 \right)^{-1/2}, \label{gamma}\\
    f\equiv f(u)=1-u^3 \label{f(u)}, 
\end{gather}
with $a$ being the rotation parameter.
The ranges of the coordinates are $-\infty< t< +\infty$, $0\leq u < \infty$, 
$0\leq\varphi < 2\pi$, and $-\infty < z< +\infty$, so that metric 
\eqref{background1} represents a rotating black 
hole with cylindrical topology.   
In the present coordinates, the AdS boundary, where 
the dual CFT$_3$ lives, is located at $u= 0$.

The spacetime described by the metric \eqref{background1} presents an event 
horizon located at $u=u_h=1$, which is the real positive zero of the equation 
$f(u)=0$. To establish a relation between the constant $r_h$ and the 
circumferential radius of the event horizon, we consider the length $l$ of the 
circular curve of constant $t$, $u$, and $z$, and obtain
\begin{equation}
l=2\pi\frac{\gamma\, r_h}{u}\left[1-a^2\alpha^2 f\right]^{1/2}.    
\end{equation}
In the limit of $u\rightarrow u_h$, it follows $l=2\pi \gamma r_h$,
and hence $\gamma r_h$ can be 
identified with the circumferential radius of the cylindrical 
surface of the event horizon.

For a rotating black string, the mass $M$ and the angular
momentum $J$ per unit length along the string are well 
defined quantities.
In terms of the parameters $r_h$ and $a$, these quantities
are given by \cite{Lemos:1994xp}
\begin{equation}
M=r_h^3\alpha^3\gamma^2\left(\frac{2+a^2\alpha^2}{8}\right),
\qquad J=\frac{3r_h^3\alpha^3a\gamma^2}{8}.
\end{equation}
Inverting the last relations, it results in 
\begin{equation}\label{eqhori}
\begin{split}
r_{h}^3&=
2\frac{\sqrt{9M^2-8J^2\alpha^2}-M}{\alpha^3},\\
\qquad 
a&=\frac{3M-\sqrt{9M^2-8\alpha^2J^2}}{2\alpha^2J}.
\end{split}
\end{equation}
As can be seen from the above equations, an event horizon exists
if and only if $0\leq J^2\alpha^2\leq M^2$, or, 
equivalently, if and only if $0\leq a^2\alpha^2\leq1$. 

The Hawking temperature of a rotating black string can be written
as \cite{Lemos:1994xp}
\begin{equation}\label{TemperatureBS}
T=\frac{\mathcal{T}}{\gamma},
\end{equation}
where $\mathcal{T}=3\alpha^{2}r_h/4\pi$ is the temperature of a static black 
string with horizon radius $r_h$. From the dual field-theory perspective, 
$\mathcal{T}$ is the local rest-frame temperature of the CFT plasma 
\cite{Bhattacharyya:2007vs,Caldarelli:2008ze}. As emphasized by Cardoso {\it{et 
al.}} \cite{Cardoso:2013pza}, Eq. \eqref{TemperatureBS} ``gives the redshift 
factor relating measurements done in the laboratory and comoving frames".

It is worth mentioning also that the extremely rotating solution is obtained 
when $a\alpha =1$, or, equivalently, $J\alpha= M$. In the extremal case, the 
radius $r_h$ vanishes, which means that the singularity at $u\rightarrow \infty$ 
becomes lightlike. Moreover, the Hawking temperature of the black string 
vanishes as expected for extremal black holes.
 
It is important to determine the angular 
velocity of matter particles and photons around the rotating black strings. In 
particular, we are interested in obtaining the angular velocity at the event 
horizon, the static limit surface and the AdS boundary. We start with the 
general expression for the angular velocity of lightlike particles in a circular 
orbit (see, e.g., Ref.~\cite{Semerak:1993}),
\begin{equation}\label{omega}
\Omega_{\pm}=\omega_\varphi\pm \sqrt{\omega_\varphi^2-\frac{g_{tt}}
{g_{\varphi\varphi}}},
 \end{equation}
where
 \begin{equation}\label{angularVe}
\omega_\varphi=-\frac{g_{t\varphi}}{g_{\varphi\varphi}}
=\frac{a\alpha^2\left(1-f\right)}{1-a^2\alpha^2f},
 \end{equation}
and is interpreted as the angular velocity of a locally non-rotating 
observer~\cite{Misner:01}. The metric coefficients in Eqs.~\eqref{omega} and 
\eqref{angularVe} are
\begin{equation}
\begin{split}
g_{tt}&=\alpha^2\gamma^2r_h^2\left(a^2\alpha^2-f\right)/u^2,\\    
g_{\varphi\varphi}&=\gamma^2r_h^2\left(1-a^2\alpha^2f\right)/u^2,\\   
g_{t\varphi}&=-a\alpha^2\gamma^2r_h^2\left(1-f\right)/u^2.
\end{split}
\end{equation}
As it is well known, any time-like particle
in a circular orbit is constrained to travel with angular velocity 
between $\Omega_{+}$ and $\Omega_{-}$.

Substituting the metric (\ref{background1}) into Eqs.
(\ref{omega}) and (\ref{angularVe}), we get
\begin{equation}\label{angularHorizon}
\begin{split}
R\,\Omega_{\pm}\Big|_{u=u_h}&=a\alpha, \quad\quad
R\, \omega_\varphi\Big|_{u=u_h}=a\alpha.
\end{split}
\end{equation}
This is similar to the Kerr and Kerr-Newman black-hole cases, for which the 
angular velocities $\Omega_\pm$ and $\omega_\varphi$ of equatorial circular 
orbits coalesce at the horizon. From this we conclude that the angular velocity 
of the rotating black string is $a\alpha$. The expressions for the angular 
velocities at the event horizon of rotating charged black strings are found in 
Refs. \cite{Lemos:1995cm,Awad:2002cz}. Taking the zero charge limit of the 
expressions presented in those papers, and considering the different notations, 
we get the results (\ref{angularHorizon}). 

The stationary limit surface of a rotating black string, defined by the 
condition $g_{tt}=0$, is located at $u=u_s=(1-a^2\,\alpha^2)^{1/3}$. It 
coincides with the horizon in the nonrotating black string case, when 
$a\alpha=0$, and is located outside the horizon for $a\alpha\neq 0$. This means 
that there is an ergosphere in the rotating black string. On this surface, it 
results
\begin{equation}
R\,\Omega_{+}=R\,\omega_\varphi=\frac{2a\, \alpha}{1+a^2\,\alpha^2},\quad\quad
R\,\Omega_{-}=0.
\end{equation}
Any particle or observer inside the surface delimited by $u_s$ must rotate along 
the direction of the black string rotation, an effect associated with the 
inertial frame dragging~\cite{Misner:01}. In the extremely rotating case ($a\alpha=1$),
$R\,\Omega_{+}$ reaches the speed of light.
It is worth noticing that the rotating black string has only one event horizon and one 
stationary limit surface \cite{Lemos:1994xp}, differently from the Kerr black 
hole, which has event and Cauchy horizons and an ergosphere with two 
surfaces, the outer surface being located outside the event horizon
\cite{Visser:2007fj}.

At the AdS boundary ($u=u_B=0$) we get
\begin{equation}\label{angularBound}
\begin{split}
R\, \Omega_{\pm}\big|_{u=u_B}=\pm 1, \quad \quad
R\, \omega_\varphi\big|_{u=u_B}=0.
\end{split}
\end{equation}
These results show a difference in respect to the asymptotically flat spacetime 
where the angular velocity vanishes at the spatial infinity as $r\, 
\Omega_{\pm}=\pm 1$, where $r$ is the radial Boyer-Lindquist coordinate. In 
turn, the result $\omega_\varphi(u_B)=0$ means that locally stationary observers 
at the AdS boundary is really at rest.

\section{Fundamental equations for the gravitational perturbations}
\label{sec-equations}

In this section, we present the fundamental differential equations that govern 
the gravitational perturbations of the background spacetime \eqref{background1}. 
The starting point is the wave equations for the gravitoelectromagnetic 
perturbations of rotating charged black strings obtained in Ref. 
\cite{Miranda:2014vaa}. In the particular case we are interested in here, i.e., 
for zero electric charge and without source terms, these general differential 
equations can be written as 
\begin{equation}\label{eqRWZ}
\left[f\partial_{u}\left(f\partial_{u}\right)-\gamma^2  \left(\mathfrak{w} 
-a \alpha^{2} \mathfrak{m}\right)^2-V_{\scriptscriptstyle{T,L}}
\right]\Phi_{\scriptscriptstyle{T,L}}=0.
\end{equation}
Here, $\Phi_{\scriptscriptstyle{T}}(u)$ and $\Phi_{\scriptscriptstyle{L}}(u)$ 
stand for the Regge-Wheeler-Zerilli (RWZ) master variables associated, 
respectively, to the transverse and longitudinal sectors of the gravitational 
perturbations,
while $V_{\scriptscriptstyle{T}}(u)$ and 
$V_{\scriptscriptstyle{L}}(u)$ are the corresponding effective potentials, given by
\begin{align}
V_{\scriptscriptstyle{T}}(u)=&f\left(\mathfrak{p}\,^2 -3u\right),
\label{potential_T}\\ 
V_{\scriptscriptstyle{L}}(u)=&\dfrac{f}{\mathfrak{p}\,^2 +3u}
\left[\mathfrak{p}^4 + \dfrac{9\,\big(2+\mathfrak{p}\,^2 
u^2+u^3\big)}{\mathfrak{p}\,^2+3\, u}\right],
\label{potential_L}
\end{align}
where
\begin{equation}
\mathfrak{p}\,^2 = \mathfrak{q}^{2} +\gamma^{2} \alpha^{2} 
(\mathfrak{m}-a \mathfrak{w})^{2}.
\label{freq_wavenumber}
\end{equation}
The frequency $\mathfrak{w}$ and the wavenumbers $\mathfrak{m}$ and 
$\mathfrak{q}$ are normalized by the temperature $\mathcal{T}$
according to the relations
\begin{equation} \label{eq:norm}
\mathfrak{w}=\frac{3\,\omega}{4\pi \mathcal{T}},
\qquad \mathfrak{m}=\frac{3\,m}{4\pi \mathcal{T}},
\qquad \mathfrak{q}=\frac{3\,q}{4\pi \mathcal{T}}, 
\end{equation}
where $\omega$ is the frequency, $m$ is the wavenumber along the rotation 
direction $\varphi$, and $q$ is the wavenumber along the direction $z$.

In the limit of $a\rightarrow 0$, the transverse and 
longitudinal perturbations, labeled by $(-)$ and $(+)$ in Ref. 
\cite{Miranda:2014vaa}, correspond respectively to the odd (axial) and even 
(polar) perturbations under the parity transformation $\varphi\rightarrow 
-\varphi$. As shown along this work, the $(T)$ sector gives rise to 
shear modes in the hydrodynamic regime, while the $(L)$ sector gives rise to 
sound wave modes.

By comparing Eqs. \eqref{eqRWZ}-\eqref{freq_wavenumber} to the corresponding 
differential equations obtained in the static black string case 
\cite{Cardoso:2001vs,Miranda:2005qx}, we find that the RWZ variables governing 
the perturbations of a rotating black string satisfy fundamental equations of 
the same form as the equations for the perturbations of a static black string, 
provided we consider the change of the radial coordinate $r$ by $u=r_{h}/r$ and 
establish the following relation between the frequencies and wavenumbers,
\begin{equation}\label{boost2}
\left\{
\begin{aligned}
&\bar{\mathfrak{w}}=\gamma(\mathfrak{w}-a\alpha^{2}\mathfrak{m}),\\
&\bar{\mathfrak{m}}=\gamma(\mathfrak{m}-a\mathfrak{w}),\\
&\bar{\mathfrak{q}}=\mathfrak{q},\\
\end{aligned}
\right.
\end{equation}
where the barred and unbarred quantities refer, respectively, to the static and 
rotating cases. Such a connection between the perturbation equations of a 
rotating and a static black string are expected in advance, since the metrics of 
the corresponding background spacetimes are related by an `illegitimate' linear 
transformation, which mixes the time with the angle $\varphi$ 
\cite{Lemos:1994xp,Lemos:1995cm}. The same result was obtained in the study of 
electromagnetic perturbations of rotating black strings \cite{Morgan:2013dv}. 
Hence, in principle, the equations of motion for the gravitational perturbations 
of the rotating black string, Eq.~\eqref{eqRWZ}, could be obtained directly by 
replacing relations (\ref{boost2}) into the fundamental differential equations 
for the perturbations of the static black string.

In addition to the fundamental equations, the perturbation problem in an anti-de 
Sitter spacetime requires the imposition of boundary conditions at the horizon 
and at the AdS boundary. Whereas the natural boundary condition at $u=u_h$ is 
that of an ingoing wave only, since classically the horizon acts like a one-way 
membrane, the boundary condition at the spatial infinity can be Dirichlet, 
Neumann or Robin according to which the field perturbations, their derivatives 
or a combination of both are required to vanish at the anti-de Sitter boundary.

Another important issue in the determination of the quasinormal spectrum of 
black holes is the choice of appropriate gauge-invariant quantities to describe 
the perturbations of the black hole. As an example, it is known that the 
electromagnetic and gravitational QNM spectra of the even (polar) sector, 
obtained by using the RWZ and Kovtun-Starinets (KS) \cite{Kovtun:2005ev} master 
variables, are different if one uses the same boundary condition at the spatial 
infinity (see, for instance, 
Refs.~\cite{Michalogiorgakis:2006jc,Miranda:2008vb}). As first argued in 
Ref.~\cite{Michalogiorgakis:2006jc} and elaborated in details in 
Ref.~\cite{Dias:2013sdc} for a global Schwarzschild-$\mbox{AdS}_4$ black hole, 
the criterion of non-deformation of the boundary metric requires the imposition 
of Robin-type conditions on the RWZ master fields. These boundary conditions are 
translated into Dirichlet conditions for the KS variables \cite{Morgan:2009pn}.

Therefore, to describe gravitational perturbations of AdS backgrounds, the KS 
variables have at least two advantages over the RWZ variables. The first one is 
the correspondence between the QNM frequencies and the poles of the 
stress-energy tensor correlators in the dual field theory \cite{Kovtun:2005ev}, 
which is easier to be established by using the KS than by using the RWZ master 
functions. The second one is that numerically it is easier to deal with 
Dirichlet than with Robin boundary conditions.

In this work, we use the KS gauge-invariant quantities, and so it is important 
to express the fundamental equations of the gravitational perturbations in terms 
of such variables. Our task is then to rewrite the differential equations 
(\ref{eqRWZ}) in terms of the KS gauge invariant variables. To do that we need 
the relations between the RWZ and KS variables as it was done in 
Ref.~\cite{Morgan:2009pn} for a $d$-dimensional black brane, which is a 
topologically trivial version (for $d=4$) of the static black string. Since the 
form of such relations are quite different depending on the perturbation sector, 
we consider the transverse and longitudinal sectors separately.

We start with the transverse sector. In this case, the relation between the RWZ 
and KS variables can be written as
\begin{equation}\label{vectconnect}
Z_{\scriptscriptstyle{T}}(u)=P_{\scriptscriptstyle{T}}(u)\,
\partial_{u}\Phi_{\scriptscriptstyle{T}}(u)
+ Q_{\scriptscriptstyle{T}}(u)\,\Phi_{\scriptscriptstyle{T}}(u),
\end{equation}
where the coefficients are polynomials in $u$ given by
\begin{equation}
\begin{split}
P_{\scriptscriptstyle{T}}(u)=u f, \qquad\quad
Q_{\scriptscriptstyle{T}}(u)=-f.
\end{split}
\end{equation}

Notice that relation (\ref{vectconnect}) depends only on the holographic coordinate 
$u$. Hence, the differential equation for the KS
variable $Z_{T}$ may straightforwardly be obtained by replacing relation (\ref{vectconnect}) into 
the corresponding equation for the RWZ variable $\Phi_{\scriptscriptstyle{T}}$, 
Eq.~\eqref{eqRWZ}.
After some algebraic manipulations, it yields
\begin{equation}\label{fund-eq-vector}
\begin{split}
\partial_{u}^{2}Z_{\scriptscriptstyle{T}}&+
\left[\frac{\gamma^2  \left(\mathfrak{w} 
-a \alpha^{2} \mathfrak{m}\right)^2\;\partial_{u}\ln f}
{\gamma^2  \left(\mathfrak{w} 
-a \alpha^{2} \mathfrak{m}\right)^2-f\,\mathfrak{p}^2}-\frac{2}{u}\right]
\partial_{u}Z_{\scriptscriptstyle{T}}\\
&+\left[\frac{\gamma^2  \left(\mathfrak{w} 
-a \alpha^{2} \mathfrak{m}\right)^2-f\,
\mathfrak{p}^2}{f^2}\right]Z_{\scriptscriptstyle{T}}=0.
\end{split}
\end{equation}

Now we consider the longitudinal sector, whose analysis is not 
straightforward as in the transverse sector, because the relation between the 
RWZ and KS master functions depends also on the frequencies and wavenumbers
\cite{Morgan:2009pn}. In this situation, we can use the relationship connecting 
the frequencies and wavenumbers of static and rotating black strings, cf. Eqs.~\eqref{boost2},
and the relation between the RWZ and KS variables found in the static case. 
Working out such relations, it follows

\begin{equation}\label{scaconnect}
\begin{split}
Z_{\scriptscriptstyle{L}}(u)&=P_{\scriptscriptstyle{L}}(u)\,
\partial_{u}\Phi_{\scriptscriptstyle{L}}(u)
+Q_{\scriptscriptstyle{L}}(u)\,\Phi_{\scriptscriptstyle{L}}(u),
\end{split}
\end{equation}
where the coefficients are given by 
\begin{equation}
\begin{split}
P_{\scriptscriptstyle{L}}(u)&=
-\frac{f^{2}(u)\mathfrak{p}^2}{\mathfrak{p}^2+3 u},\\
Q_{\scriptscriptstyle{L}}(u)&=
\frac{-\mathfrak{p}^2}{4\left(\mathfrak{p}^2+3 u\right)^2}\Bigg\{2\,u\,
\mathfrak{p}^2 \left[2 \gamma^2  \left(\mathfrak{w} 
-a \alpha^{2} \mathfrak{m}\right)^2-9\, u\right]\\
& +3\,\left[4+8\, u^2 \,\gamma^2  \left(\mathfrak{w} 
-a \alpha^{2} \mathfrak{m}\right)^2+u^3\left(u^3-14\right)\right]\\
&+\left(u^3-4\right)u\,\mathfrak{p}^4
+\frac{36\, u^3 \,\gamma^2  \left(\mathfrak{w} 
-a \alpha^{2} \mathfrak{m}\right)^2}{\mathfrak{p}^2}\Bigg\}.
\end{split}
\end{equation}
Hence, manipulating Eqs.~(\ref{eqRWZ}), \eqref{potential_L}, 
and (\ref{scaconnect}) we obtain the fundamental differential 
equation for the KS master variable of the longitudinal sector, 
\begin{equation}\label{fund-eq-scalar}
\begin{split}
\partial_{u}^{2}Z_{\scriptscriptstyle{L}}&+
\frac{\mathfrak{p}^{2}\,Y_{1}+\gamma^2  \left(\mathfrak{w} 
-a \alpha^{2} \mathfrak{m}\right)^2\,Y_2} {u\,X\,f}\,
\partial_{u}Z_{\scriptscriptstyle{L}}\\
&+\frac{\mathfrak{p}^2\,Y_3+ 
\mathfrak{p}^4\,Y_4+4\,\gamma^4  \left(\mathfrak{w} 
-a \alpha^{2} \mathfrak{m}\right)^4}{X\,f^{2}}\,Z_{\scriptscriptstyle{L}}=0,
\end{split}
\end{equation}
where we have introduced the coefficients\footnote{There are some typos in the 
coefficients presented in Eqs.~(3.22) and~(4.7) of Ref. \cite{Morgan:2009pn}. 
Here we write the correct expressions of the coefficients.}
\begin{equation}
\begin{split}
&X = 4\,\gamma^2  \left(\mathfrak{w} 
-a \alpha^{2} \mathfrak{m}\right)^2\!-
\!\left(f+3\right)\,\mathfrak{p}^2,\\
&Y_1 =8f^2+3\left(f+3\right)u^{3}, \\
&Y_2 =4\,\left(f-3\right),\\
&Y_3 =-\left(\partial_{u}f\right)^2f-\left(5f+3\right)\,\gamma^2 
\left(\mathfrak{w} -a \alpha^{2} \mathfrak{m}\right)^2,\\
&Y_4 =\left(f+3\right)f.
\end{split}
\end{equation} 

For numerical purposes it is important to know the singular points and 
symmetries of the differential equations (\ref{fund-eq-vector}) and 
(\ref{fund-eq-scalar}). Let us start with the transverse-sector differential 
equation (\ref{fund-eq-vector}). The singularities of this equation are
\begin{equation}
u_{0}=0,\quad u_{1}=1,\quad
u_3 =\left[\frac{\mathfrak m^2\alpha^2+\mathfrak q^2
-\mathfrak w ^2}{\mathfrak{p}^2}\right]^{1/3}, 
\end{equation}
and all of them are regular singular points. The same happens with the 
longitudinal-sector differential equation (\ref{fund-eq-scalar}), whose regular 
singular points are
\begin{equation}
u_{0}=0, \quad u_{1}=1,\quad
u_{3}=\sqrt[3]{4}\left[\frac{\mathfrak m^2\alpha^2
+\mathfrak q^2
-\mathfrak w ^2}{\mathfrak{p}^2}\right]^{1/3}.
\end{equation}
The regular nature of the singularities allows us to expand the solutions near 
these regular singular points. That is the core of the power series method that 
we use in the numerical procedure of this work.

Finally, let us look at the symmetries of the equations
(\ref{fund-eq-vector}) and (\ref{fund-eq-scalar}). 
Both differential equations are invariant under the following (symmetry)
transformations, 
\begin{equation}\label{EqSym}
\begin{aligned}
&\{\mathfrak{m},a\}\to \{-\mathfrak{m},-a\};&\quad 
&\{\mathfrak{m},\mathfrak{w}\}\to \{-\mathfrak{m},-\mathfrak{w}^{*}\};\\
&\{\mathfrak{w},a\}\to \{-\mathfrak{w}^{*},-a\};&\quad
&\{\mathfrak{q}\}\to \{-\mathfrak{q}\};
\end{aligned}
\end{equation}
where $\mathfrak{w}^*$ is the complex conjugate of the frequency $\mathfrak{w}$.
These symmetry transformations are going to be used when we present the 
numerical results in the next sections. We split the complex frequency as 
$\mathfrak{w}=\mathfrak{w}_{R}-i\mathfrak{w}_{I}$, so that the imaginary part 
$\mathfrak{w}_{I}$ is positive for decreasing perturbations. With this convention,
negative $\mathfrak{w}_{I}$ would imply instability against
linear perturbations, what does not
happen for the rotating AdS black strings in four-dimensional spacetimes.
Additionally, we choose to fix the 
signals of $a$ and $\mathfrak{w}_R$ as being positive, such that (for 
$\mathfrak{q}=0$) the signal of the wavenumber component $\mathfrak{m}$ 
indicates the propagation direction of the wave on the cylinder surfaces of 
constant $u$, along or contrary to the string rotation. Hence, we present graphs 
for numerical results in the first and second quadrants of the planes 
$(\mathfrak{m},\mathfrak{w}_R)$ and $(\mathfrak{m},\mathfrak{w}_I)$. In the case 
of $\mathfrak{m}=0$, a similar procedure is followed to present the numerical 
results in terms of the wavenumber component $\mathfrak{q}$. The symmetric 
solutions may be obtained using the transformations of Eq.~\eqref{EqSym}.

\section{Hydrodynamic quasinormal modes}
\label{sec-qnmresults}

\begin{figure*}[ht]
\begin{tabular}{*{2}{>{\centering\arraybackslash}p{.45\textwidth}}}
\includegraphics[width=7.5cm]{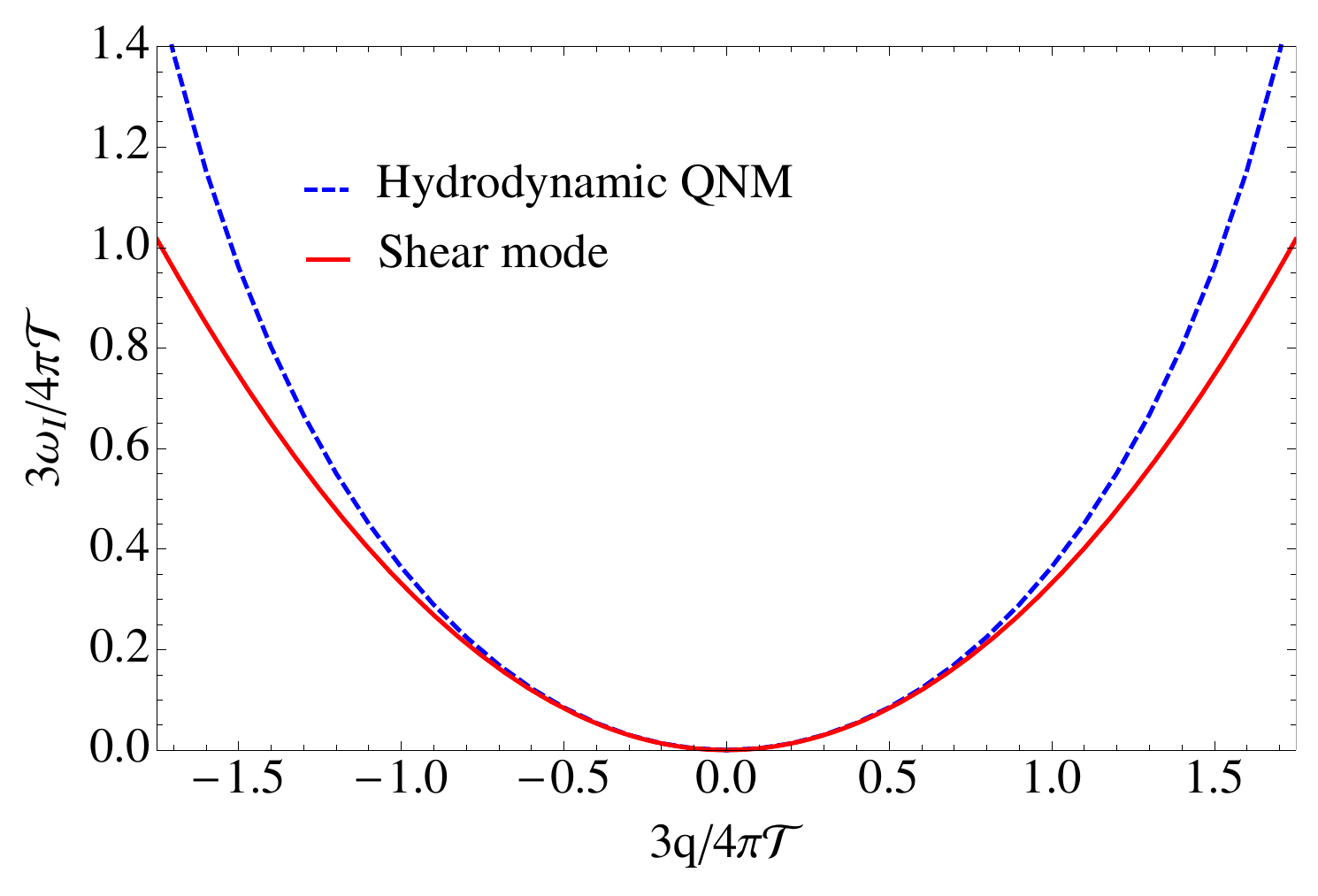}&
\includegraphics[width=7.5cm]{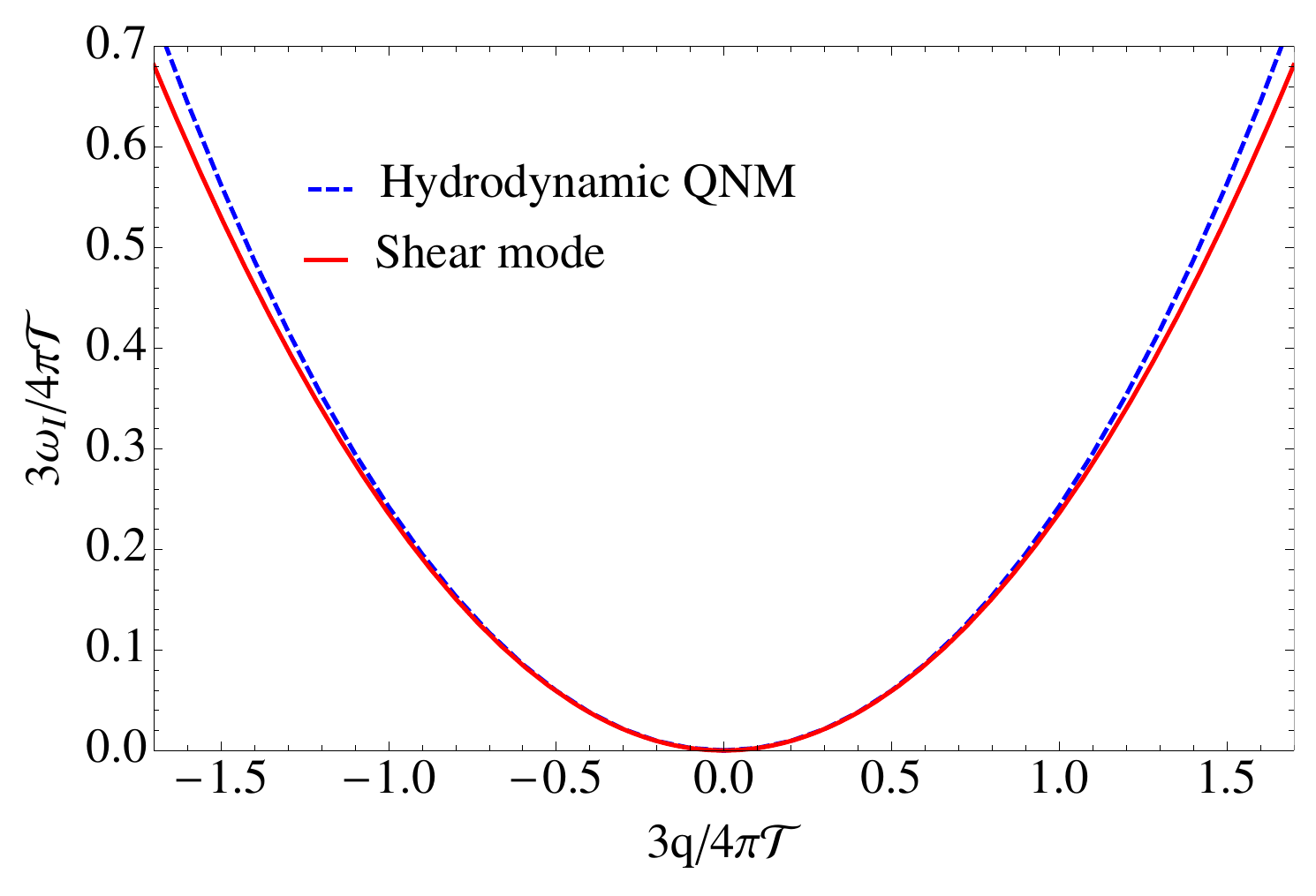}
\\
\end{tabular}
\caption{Dispersion relations of the hydrodynamic QNMs (solid lines), cf. Eq.~\eqref{movvec}, and shear 
modes (dashed lines) of the transverse sector, with $\mathfrak{m}=0$, for
$a\alpha=0.1$ (left panels) and $a\alpha=1/\sqrt{2}$ (right panel). These 
solutions are purely damped.} \label{DampingAxial}
\end{figure*}
The hydrodynamic approximation is characterized by a regime where the 
frequencies and wavenumbers are much smaller than the local temperature 
$\mathcal{T}$, i.e., $\mathfrak{w}\ll 1$,  $\mathfrak q \ll 1$ and $\mathfrak m 
\ll 1$. In such a regime, it is possible to express the dispersion relations
$\mathfrak{w}(\mathfrak{m},\mathfrak{q})$ as power series in the
wavenumbers $\mathfrak q$ and $\mathfrak m$. In general, the 
hydrodynamic frequencies are the lowest frequencies in the spectrum of QNMs.
The main characteristic of these modes is that the frequency vanishes as the 
wavenumbers go to zero, i.e., $\mathfrak{w} \rightarrow0$ as ($\mathfrak{m}, 
\mathfrak{q})\rightarrow 0$.

In this section, we develop a complete analysis of the hydrodynamic quasinormal modes
for the transverse sector (also called shear channel), and for the longitudinal 
sector (also called sound channel). Firstly, we present the analytical results 
for both sectors, then we solve numerically the differential equations 
\eqref{fund-eq-vector} and \eqref{fund-eq-scalar} with the appropriate boundary conditions and, at the end, we 
compare the numerical and analytical results.

\begin{figure*}[ht!]
\begin{tabular}{*{2}{>{\centering\arraybackslash}p{.45\textwidth}}}
\includegraphics[width=7.5cm,angle=0]{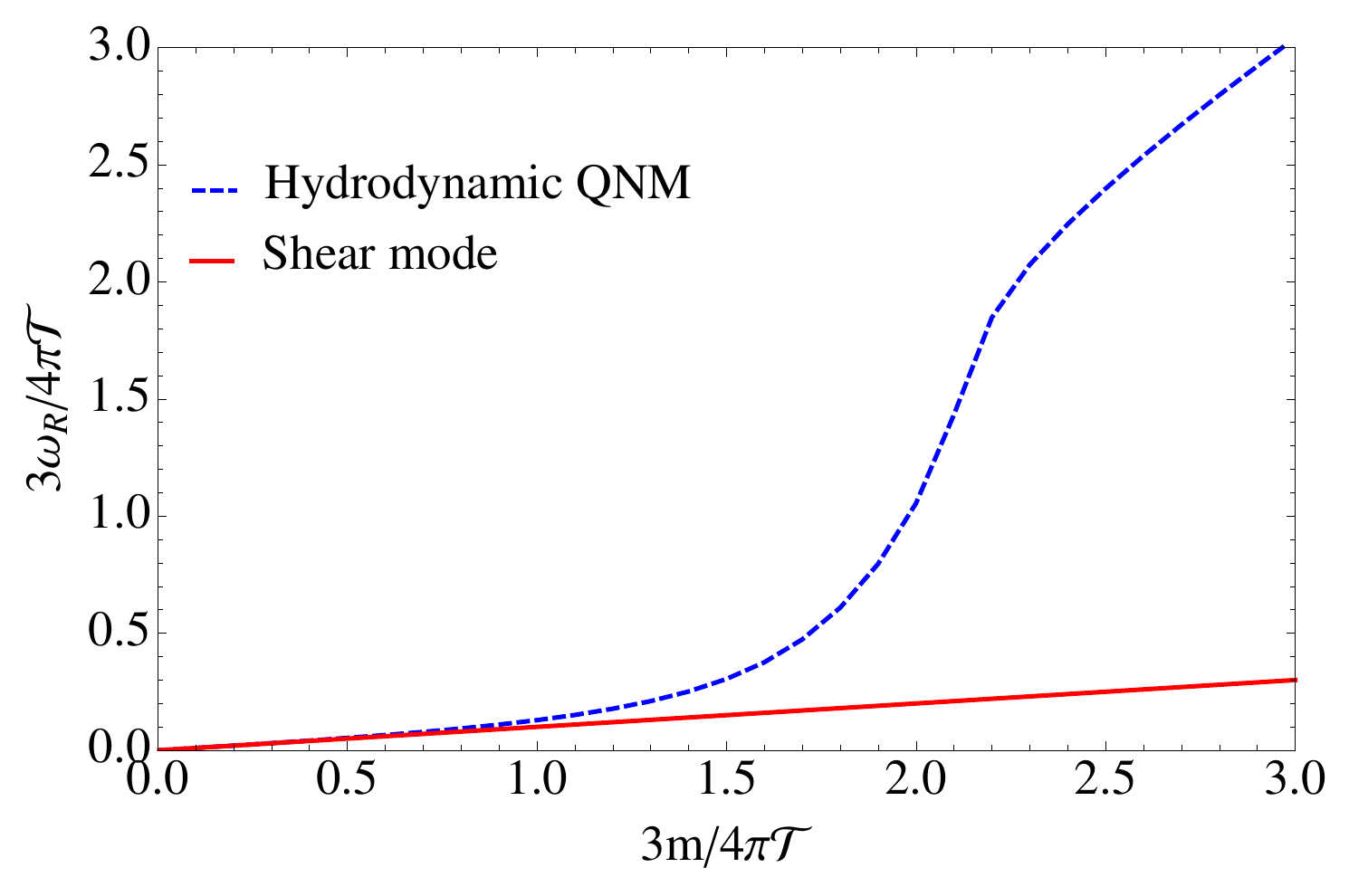} &
\includegraphics[width=7.5cm,angle=0]{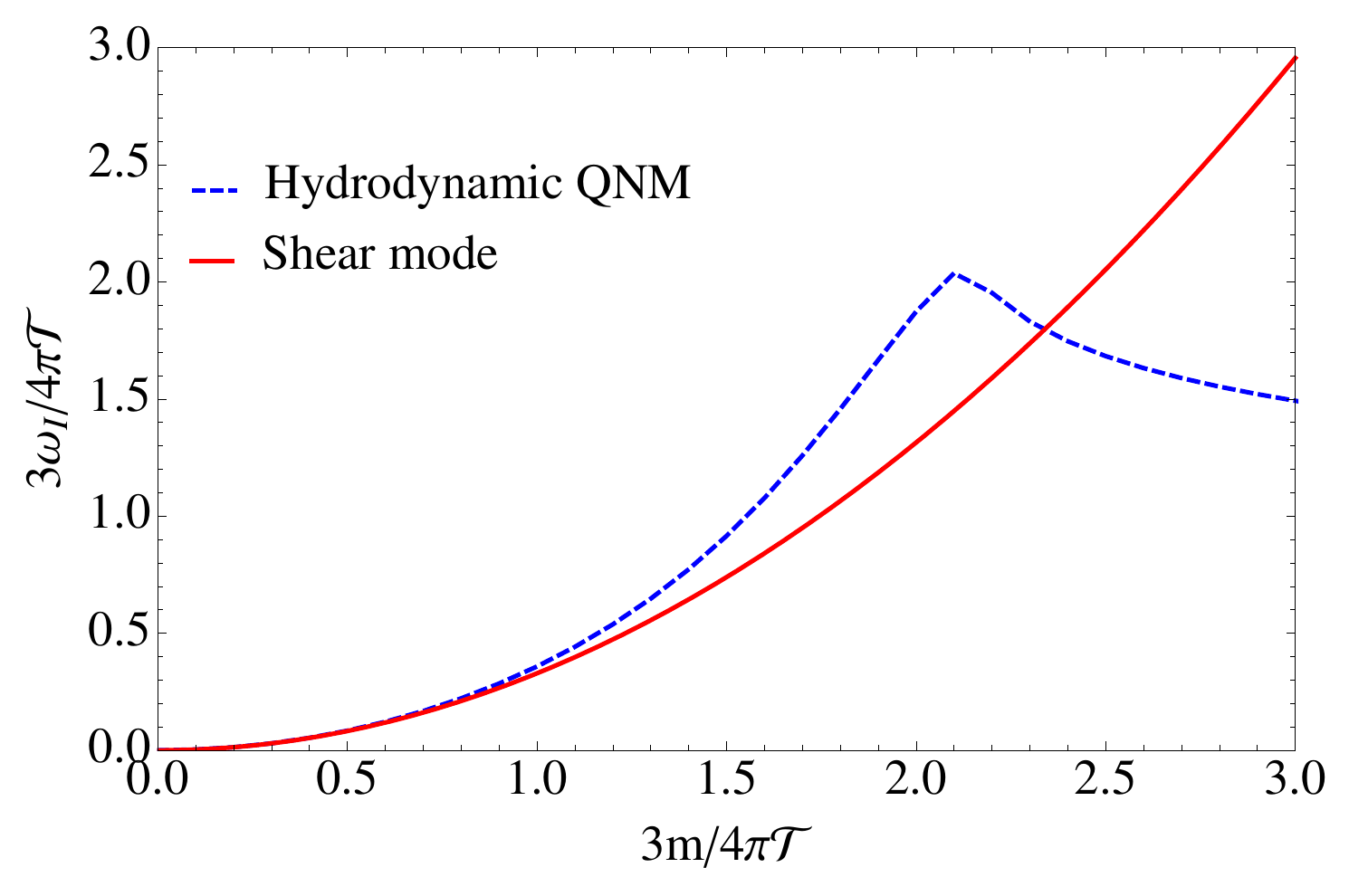}\\
\includegraphics[width=7.5cm,angle=0]{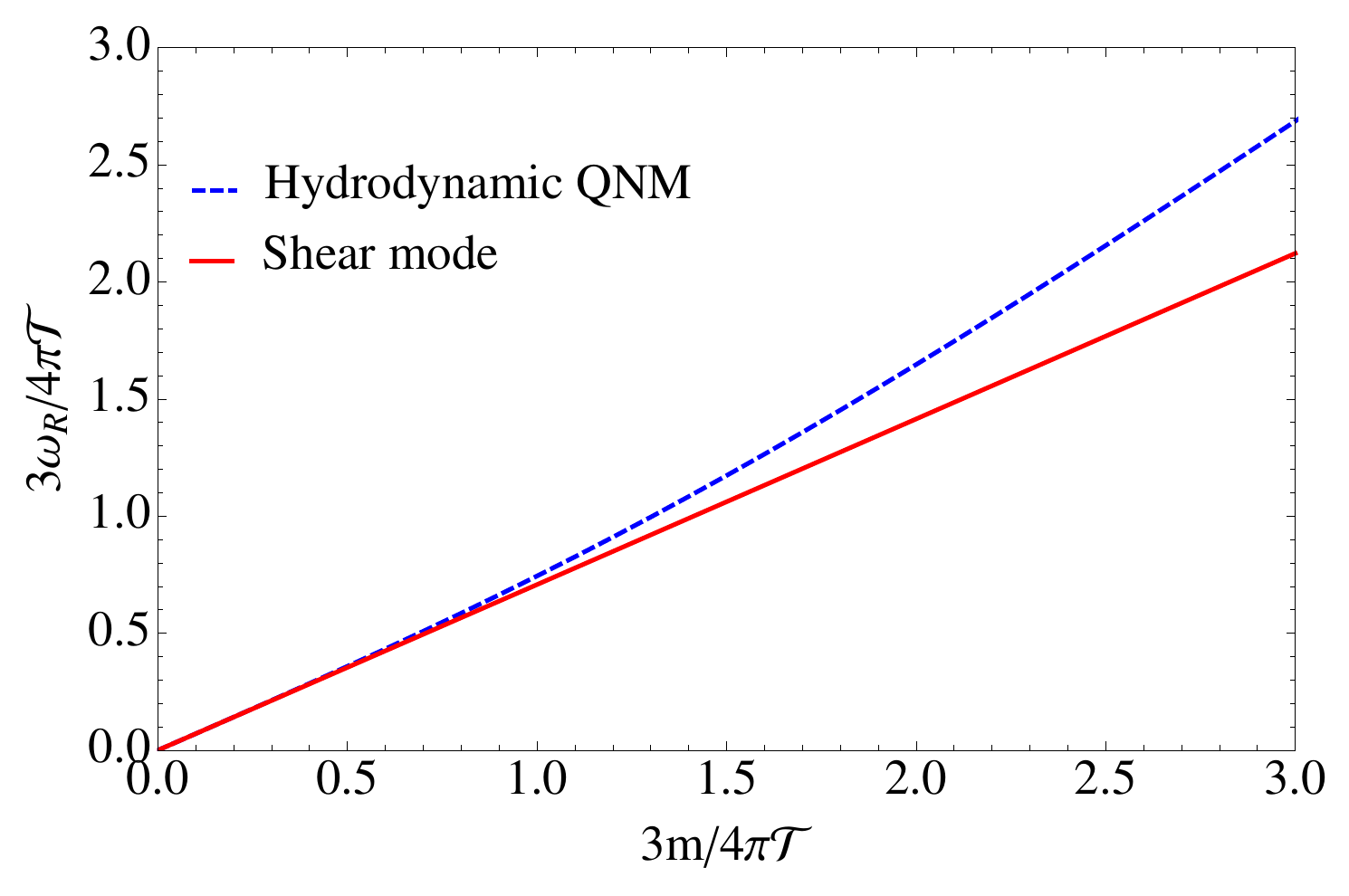} &
\includegraphics[width=7.5cm,angle=0]{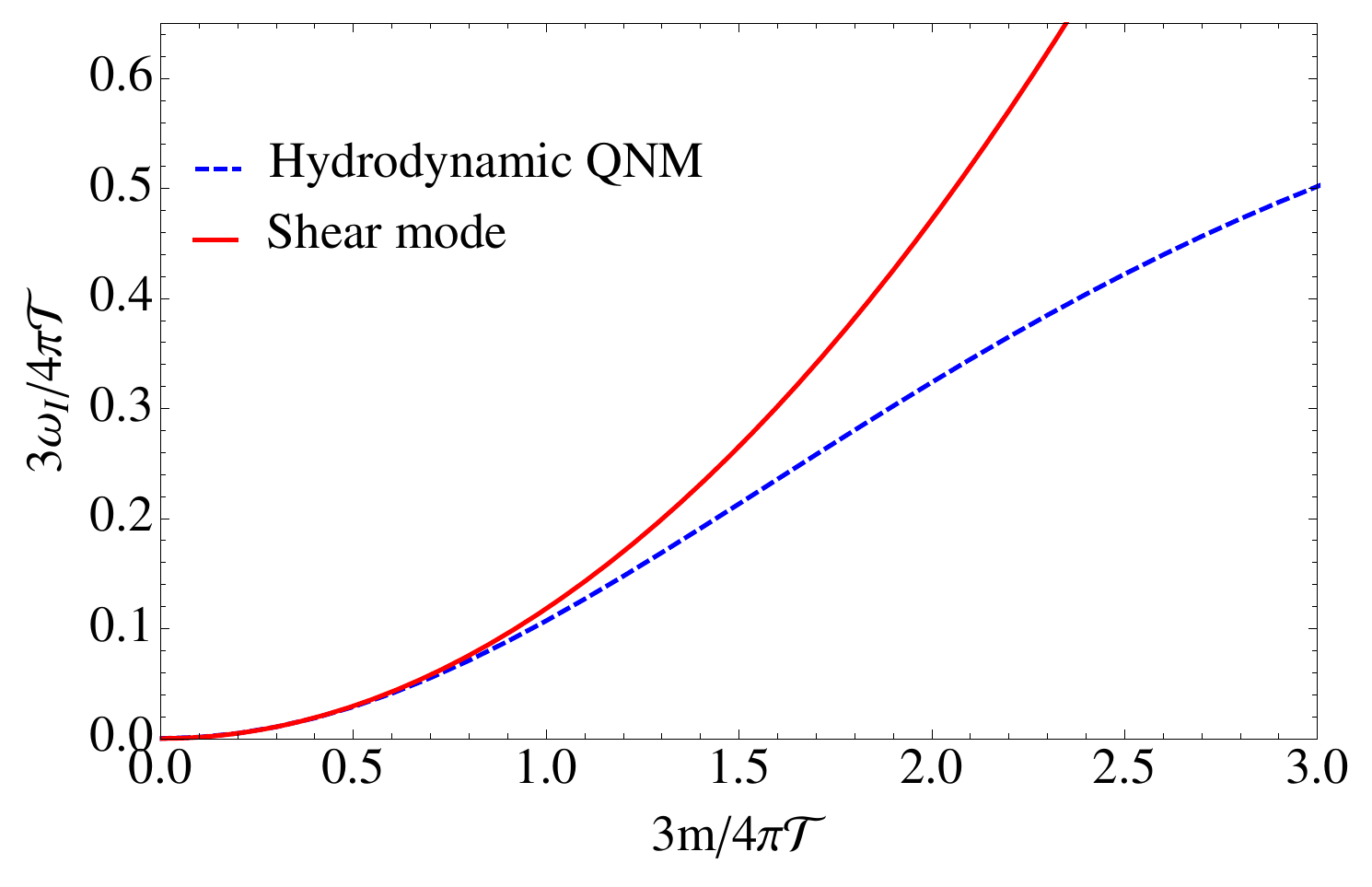} \\
\end{tabular}
\caption{Dispersion relations of the hydrodynamic QNMs (solid lines), cf. Eq.~\eqref{movvec}, and shear 
mode (dashed lines) of the transverse sector, with $\mathfrak{q}=0$, for
$a\alpha=0.1$ (top panels) and $a\alpha=1/\sqrt{2}$ (bottom panels).} 
\label{DampingAxial2}
\end{figure*}

\subsection{Transverse sector - analytical results}
\label{sec-analiticShear}

Here we start by writing the dispersion relation for the transverse sector (or 
shear channel) of the static black string, which is characterized by being a 
purely damped mode \cite{Herzog:2002fn,Miranda:2005qx},
\begin{equation}\label{staticvet}
\mathfrak{\bar{w}}= -\frac{i}{3}\left(\mathfrak{\bar q}^{2}+\alpha^{2}  
\mathfrak{\bar m}^2\right)+\cdots,
\end{equation}
where the ellipses denotes higher powers of $\mathfrak{\bar q}$ and 
$\mathfrak{\bar m}$. In the specific case of the hydrodynamic QNMs, the fact 
that $\mathfrak{\bar{w}}\rightarrow 0$ as $(\bar{\mathfrak{m}}, 
\bar{\mathfrak{q}})\rightarrow 0$ allows us to obtain the dispersion relations 
of the rotating black strings by substituting the relations (\ref{boost2}) 
directly into Eq.~\eqref{staticvet} and then solving the resulting equations for 
$\mathfrak{w}$. To do that, we introduce an additional dimensionless parameter 
$\lambda$, that scales the wavenumbers as $\mathfrak m \rightarrow \mathfrak m 
\lambda$ and $\mathfrak q \rightarrow \mathfrak q \lambda$, so that the 
frequency can be expanded in terms of this parameter as
\begin{equation}\label{freqexp}
\mathfrak w=c_0+c_1(\mathfrak m,\mathfrak q) \lambda+
c_2(\mathfrak m,\mathfrak q)\lambda^2
+c_3(\mathfrak m,\mathfrak q)\lambda^3+\cdots\,.
\end{equation}
We then solve for the coefficients $c_1(\mathfrak m,\mathfrak q), c_2(\mathfrak 
m,\mathfrak q),\dots,$ order by order in Eq.~\eqref{staticvet} and set the 
coefficient $c_0=0$ because the frequency of a hydrodynamic mode goes to zero in 
the limit ($\mathfrak m, \mathfrak q) \rightarrow 0$. After re-introducing the 
original wavenumbers (or, equivalently, by setting $\lambda=1$), we obtain the 
following dispersion relation,
\begin{equation}\label{movvec}
\mathfrak{w}=a\alpha^{2}\mathfrak{m} -\frac{i}{3\gamma}
\left(\frac{\alpha^{2}\mathfrak{m}^{2}}{\gamma^{2}}+\mathfrak{q}^{2}
\right)+\cdots\,.
\end{equation}
Rewriting Eq.~\eqref{movvec} in terms of the non-normalized frequency and
wavenumbers, it follows that
\begin{equation}\label{movvec1}
\omega=a \alpha^{2}{m}- \frac{i}{4\pi\gamma {\mathcal{T}}} 
\left(\frac{\alpha^{2}{m}^{2}} {\gamma^{2}}+{q}^{2} \right)+\cdots\,.
\end{equation}

The result \eqref{movvec1} is the dispersion relation of a diffusion problem, 
where the first term on the right hand side is the convective term (with 
velocity $a\alpha$). The coefficient of the quadratic term is the shear-diffusion 
coefficient, given by $D=1/(4\pi\gamma\,\mathcal{T})$, while the perturbation 
damping time is defined as $\tau=1/\omega_I$. 

It is worth comparing the above results with those obtained for the static black 
string \cite{Herzog:2002fn,Miranda:2005qx}. Our convention here is that black 
strings rotate counterclockwise (we called this as the positive direction). The 
wavenumber parallel to rotation is $\mathfrak m$ and the perpendicular component 
is $\mathfrak q$.
After comparing the dispersion relation \eqref{staticvet} with the
non-normalized version of \eqref{movvec1}, we see there is a Lorentz
contraction of the wavelength $2\pi/(\alpha\, m)$, i.e., $m=\gamma\,\bar{m}$. 
There is also a dilation of the
damping time, i.e., $\tau = \gamma\, \bar\tau$. These are 
expected  effects for relativistic systems described in different frames.
At last, let us mention that Eq.~\eqref{movvec} reduces to the static result, 
Eq.~\eqref{staticvet}, in the limit of zero angular velocity ($a=0$).

\subsection{Transverse sector - numerical results}
\label{sec-numericalShear}

In this subsection, we solve numerically the differential equation 
\eqref{fund-eq-vector} and compare the resulting quasinormal
frequencies to the analytical solution, Eq. \eqref{movvec},
obtained in Sec.~\ref{sec-analiticShear} for the hydrodynamic regime
($\mathfrak{w},\mathfrak{m},\mathfrak{q}\ll 1$).

The numerical solutions are found by means of
the power series method developed in Ref.~\cite{Horowitz:1999jd}. 
We notice that it is difficult to visualize and interpret the numerical results
by plotting directly the relations $\mathfrak w_{R}(\mathfrak m,\mathfrak q)$ and  
$\mathfrak w_{I}(\mathfrak m,\mathfrak q)$, since the results are three-dimensional 
graphics. For that reason, we initially split the analysis in two cases, as follows.

Firstly, we look for results with $\mathfrak{m}=0$, i.e., for perturbations 
propagating along the direction $z$. The corresponding numerical results are 
displayed in Fig.~\ref{DampingAxial}. In this case the frequencies are pure 
imaginary numbers. These results are expected because by setting 
$\mathfrak{m}=0$ in Eq.~\eqref{movvec} the real part of the frequency vanishes. 
In Fig.~\ref{DampingAxial} the dashed lines represent  the numerical results, 
while the solid lines represent the analytical solution given by 
Eq.~\eqref{movvec}. As expected, we observe a good agreement between the 
numerical and analytical results in the regime of small wavenumbers and 
frequencies. It is also seen that the hydrodynamic approximation is no longer
valid for large wavenumbers.

The effects of the rotation on the damping time can be noticed in 
Fig.~\ref{DampingAxial}. In fact, it is perceived that, given a 
specific wavenumber value, the imaginary part of the frequency for  
$a\alpha=0.1$ (left panel) is larger than the imaginary part of the 
frequency for $a\alpha=1/\sqrt{2}\approx 0.71$ (right panel). 
This means that perturbations on a slowly rotating black string are
damped faster than the perturbations on a fast rotating black string. 
In view of the symmetry of the 
differential equations \eqref{eqRWZ} under the change $\mathfrak{q}\rightarrow 
-\mathfrak{q}$, commented at the end of Sec.~\ref{sec-equations}, the curves for 
negative $\mathfrak{q}$ are mirror-symmetrical to the curves for positive 
$\mathfrak{q}$.

Secondly, we look for results with $\mathfrak{q}=0$, i.e., for perturbations 
propagating just along the $\varphi$ direction. The results obtained in this 
case are displayed in Fig.~\ref{DampingAxial2}. The real part of the frequency 
is interpreted as the convective term, cf. Eq.~\eqref{movvec}. This term is 
positive when the perturbation wave vector has the same direction as the 
rotation velocity ($\mathfrak{m}>0$), and is negative when the perturbation wave 
vector is in the opposite direction ($\mathfrak{m}<0$). Both cases represent 
propagation in the direction of increasing $\varphi$, the same direction of the 
rotation velocity,  as expected for a convection process. Following our choice 
of non-negative $\mathfrak{w}_R$, the graphs in Fig.~\ref{DampingAxial2} contain 
only the quasinormal frequencies for $\mathfrak{m}\geq 0$.

Furthermore, by comparing quantitatively the results of the imaginary parts for 
different values of the rotation parameter, e.g., $a\alpha=0.1$ and 
$a\alpha=1/\sqrt{2}$ (top and bottom panels in Fig.~\ref{DampingAxial2}), we 
observe that $\mathfrak{w}_{I}^{\scriptscriptstyle{a\alpha=0.1}} 
>\mathfrak{w}_{I}^{\scriptscriptstyle{a\alpha=0.71}}$ (for a specific value of 
the wavenumber, e.g., $\mathfrak{m}=1$). This means that the damping time for 
perturbations propagating on a black string with smaller rotation parameter are 
damped faster than the perturbations propagating on a black string with larger 
rotation parameter, i.e., $\tau^{\scriptscriptstyle{a\alpha=0.1}} 
<\tau^{\scriptscriptstyle{a\alpha=0.71}}$.
Such a result is in agreement with the expected relativistic effect of time dilation.
This kind of behavior has already been observed in a previous study of electromagnetic
perturbations of rotating black strings \cite{Morgan:2013dv}.

\subsection{Longitudinal sector - analytical results}
\label{sec-analiticSound}

The longitudinal sector (or sound channel) of the gravitational perturbations is 
governed by the differential equation \eqref{fund-eq-scalar}. It is possible to 
solve this differential equation by using a perturbative expansion of 
$Z_{\scriptscriptstyle{L}}$ in powers of $\mathfrak{w}$, $\mathfrak{m}$ and 
$\mathfrak{q}$, as it was originally done in Refs.~\cite{Policastro:2002se, 
Policastro:2002tn}. However, here we use the relations 
between the frequencies and wavenumbers of the static and rotating black 
strings, given by Eq.~\eqref{boost2}, and the dispersion relation for the static 
black string, as obtained in Refs.~\cite{Herzog:2003ke,Miranda:2008vb}, 
namely,
\begin{equation}\label{rotscalarest}
 \mathfrak{\bar w}=\pm\frac{1}{\sqrt{2}} \sqrt{\mathfrak{\bar 
q}^{2}+\alpha^2 \mathfrak{\bar m}^2}-\frac{i}{6}\left(\mathfrak{\bar 
q}^{2}+\alpha^2 \mathfrak{\bar m}^2\right)+\cdots\,.
\end{equation}
First we note that the general dispersion relation for the longitudinal momentum
fluctuations of a fluid is given by the sound wave mode
%
%
\begin{equation} \label{sound-dr}
\bar\omega=\pm\bar{c}\,\sqrt{\bar{q}^2+\alpha^{2}\bar m^{2}}-i\,{\bar{\Gamma}}
\left(\bar{q}^2+\alpha^{2}\bar m^{2}\right)+\cdots,
\end{equation}
where $\bar{c}$ is the (phase) speed of sound and $\bar\Gamma$ is the 
sound wave damping constant. By comparing Eqs.~\eqref{rotscalarest} and 
\eqref{sound-dr}, the phase speed and the damping constant are identified, 
respectively, by ${\bar c}=1/\sqrt{2}$ and $\bar{\Gamma}=1/(8\pi\,{\mathcal{T}})$. 
As in the transverse sector, the damping time is given by the inverse of the 
imaginary part, i.e., $\bar \tau=1/{\bar \omega}_{I}$.

Therefore, by replacing relations \eqref{boost2} into Eq.~\eqref{rotscalarest} 
and following the same procedure as in the case of the transverse sector, 
specifically considering the expansion \eqref{freqexp}, we obtain
\begin{equation}\label{rotscalar}
\begin{split}
\mathfrak{w}&=\frac{a\alpha\;\gamma^{2}}{1+\gamma^{2}} \alpha\mathfrak{m}
\pm\frac{1}{1+\gamma^{2} } \sqrt{2\,\alpha^{2}\mathfrak{m}^{2}
+\left(1+\gamma^{2}\right)\mathfrak{q}^{2}}\\
& - \frac{2i\;\gamma^{3}} 
{3\left(1+\gamma^{2}\right)^{2} }\;\mathfrak{q}^{2}
-\frac{2i\;\gamma^{3}}{\left(1+\gamma^{2}\right)^{3}} 
\left(\frac{2}{3}+a^{2}\alpha^{2}\right)\alpha^{2}\mathfrak{m}^{2}\\
&\pm
\frac{i\,a\alpha\,\gamma^{3}\left(6+a^{2}\alpha^{2}\right)} {3\left(1+\gamma^{2}\right)^{3}}
\alpha\mathfrak{m} \sqrt{2\;\alpha^{2}
\mathfrak{m}^{2} +\left(1+\gamma^{2}\right)\;\mathfrak{q}^{2}}\\ 
&\mp\frac{ia^{3}\alpha^{3}\gamma^{3}}{\left(1+\gamma^{2}\right)^{2}}
\frac{\alpha\mathfrak m\, \mathfrak q^{2}}{\sqrt{2\,\alpha^{2}\mathfrak{m}^{2}
+\left(1+\gamma^{2}\right)\mathfrak{q}^{2}}}+\cdots\, .
\end{split}
\end{equation}
 
In order to interpret this result, we rewrite it in a more compact form. To do 
that, we introduce a new parameter $c$, defined by
\begin{equation}\label{soundvelo}
\begin{split}
c=&\frac{(1-\bar c\,^{2})a\alpha\cos \theta }{1-a^{2}\alpha^{2}\bar c\,^{2}}\pm
\frac{\bar c\,\sqrt{1-a^{2}\alpha^{2}}}{1-a^{2}\alpha^{2}\bar c\,^{2}}\\
&\times  \sqrt{1-a^{2}\alpha^{2}\bar c\,^{2}
-\left(1-\bar c\,^{2}\right)a^{2}\alpha^{2}\cos^{2} \theta }\,,
\end{split}
\end{equation}
where $\cos \theta = \mathfrak{m}\alpha/(\mathfrak{m}^{2}\alpha^{2}+ 
\mathfrak{q}^{2})^{1/2}$. In the dual CFT description, the modulus of the parameter
$c$ can be interpreted as the ``speed of sound'' in the cylinder rest frame $K$
(the rotating black-string frame in the bulk). Differently from the value of the speed of 
sound $\bar{c}=1/\sqrt{2}$ in the fluid rest frame $\bar K$ (the static black-string 
frame) that is positive, by definition, $c$ may be positive, zero or negative, 
depending on the propagation direction, because $\mathfrak{m}$ may assume all 
values in the real line (see, for instance, Fig.~\ref{Sound2}).

By writing Eq.~\eqref{rotscalar} in terms of the parameter $c$ and reintroducing the original 
quantities $\omega$, $q$, and $m$ from Eq.~\eqref{eq:norm}, it gives (see also
Appendix~\ref{sec-RelativisticWaveVectors} for more details)
\begin{equation}\label{rotscalarSimple}
\begin{split}
\omega=c\,&\sqrt{m^2\alpha^2+q^2} -\dfrac{i}{8\pi{\mathcal{T}} 
\gamma\left(1-{a\alpha}{\bar c}  \cos{\bar\theta}\right)}\\
& \times\left[\gamma^2\left(1-\frac{a\alpha\,c\,} 
{\cos\theta}\right)^2\,m^2\alpha^{2}+q^2\right],
\end{split}
\end{equation}
where $\cos\bar\theta =\bar m\alpha/(\bar m^{2}\alpha^{2}+\bar q^{2})^{1/2}$.

Differently from the transverse sector, here it is not straightforward to
identify the 
relativistic effects as the Lorentz contraction and the time dilation. By 
comparing, term by term,  Eqs.~\eqref{rotscalarest} 
and~\eqref{rotscalarSimple}, we can identify the generalized effects of time 
dilation and wavelength contraction in a situation where the wave is moving in 
both reference frames,
\begin{equation}\label{Eq:Transformations}
\begin{gathered}
\tau=\gamma\left(1-{a\,\alpha}{\bar c}\, \cos{\bar\theta}\right)\bar\tau,\\
\bar{m}\alpha=\gamma\left(1-\frac{a\alpha\,c\,} 
{\cos\theta}\right)\,m\alpha\,.\\
\end{gathered}
\end{equation}

To simplify the analysis we consider some particular cases. By setting 
$\bar{\theta}=0$, $a\alpha=\bar{c}$, and by choosing the minus sign solution from 
Eq.~\eqref{soundvelo}, we obtain the usual time dilation between the damping 
times, $\bar{\tau}=\gamma\tau$. Notice that, in this case, the frame $K$ is comoving 
with the plane wave. Besides, the relation between the wavenumber components 
parallel to the rotation ($\bar{\theta}=0\Rightarrow\theta=0$) reduces to $\bar 
m=\gamma\, m$ or, in terms of the wavelengths, 
$(2\pi/\bar{m}\alpha)=\gamma^{-1}(2\pi/m\alpha)$ (see also Appendix 
\ref{sec-RelativisticWaveVectors}).

\subsection{Longitudinal sector - numerical results}
\label{sec-numericalSound}

\begin{figure*}[ht!]
\begin{tabular}{*{2}{>{\centering\arraybackslash}p{.45\textwidth}}}
\includegraphics[width=7.5cm,angle=0]{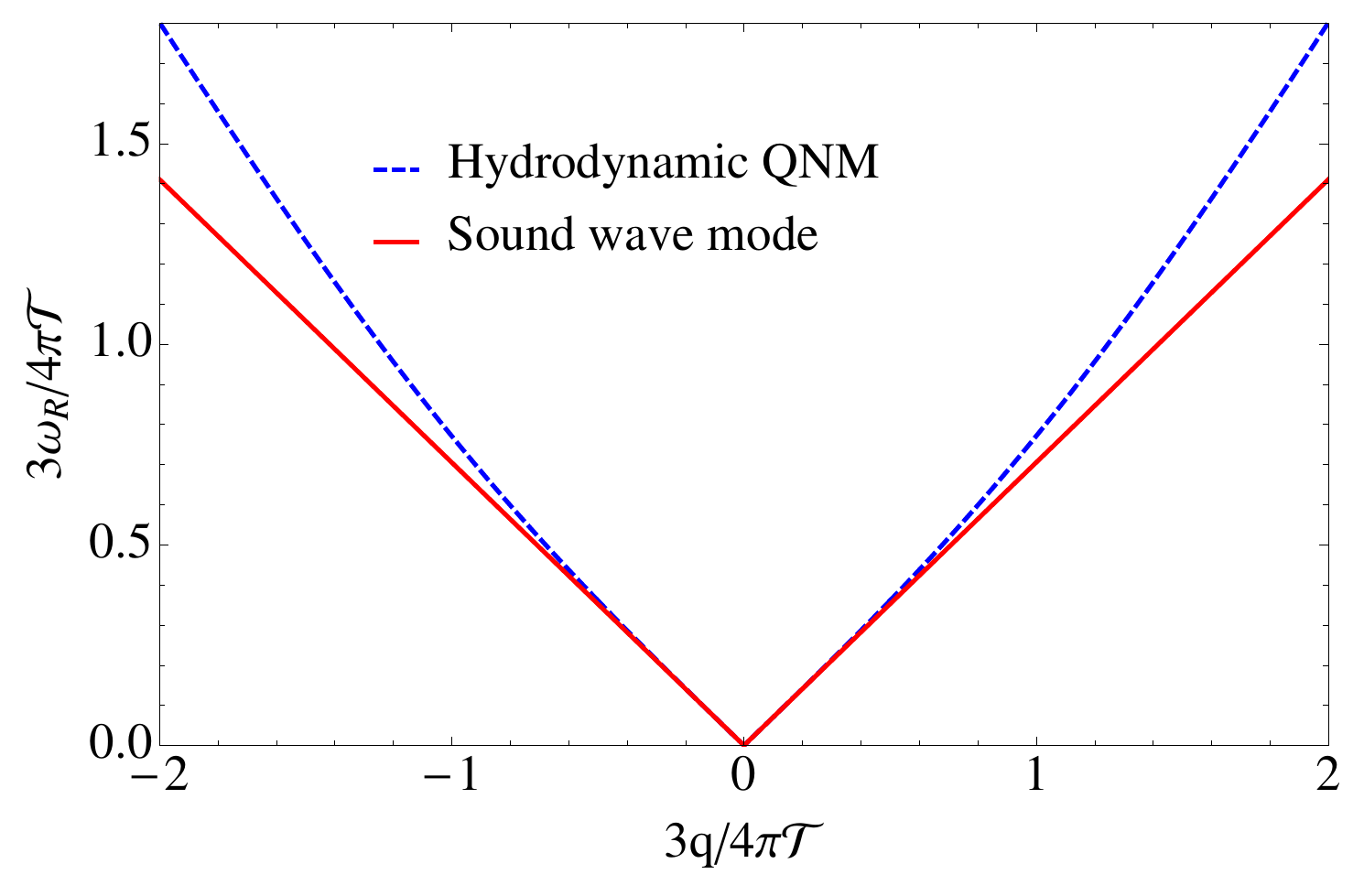} &
\includegraphics[width=7.5cm,angle=0]{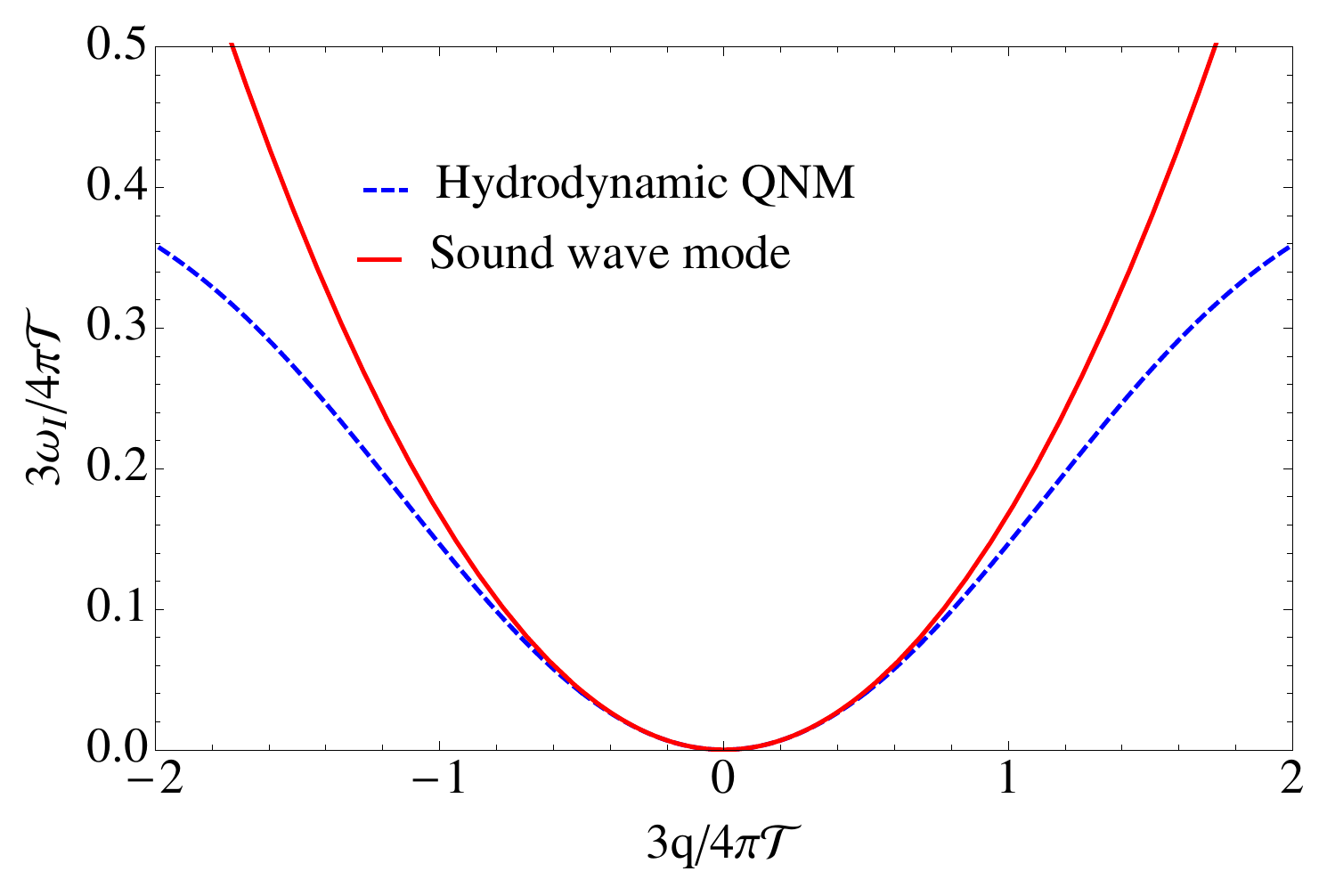}\\
\includegraphics[width=7.5cm,angle=0]{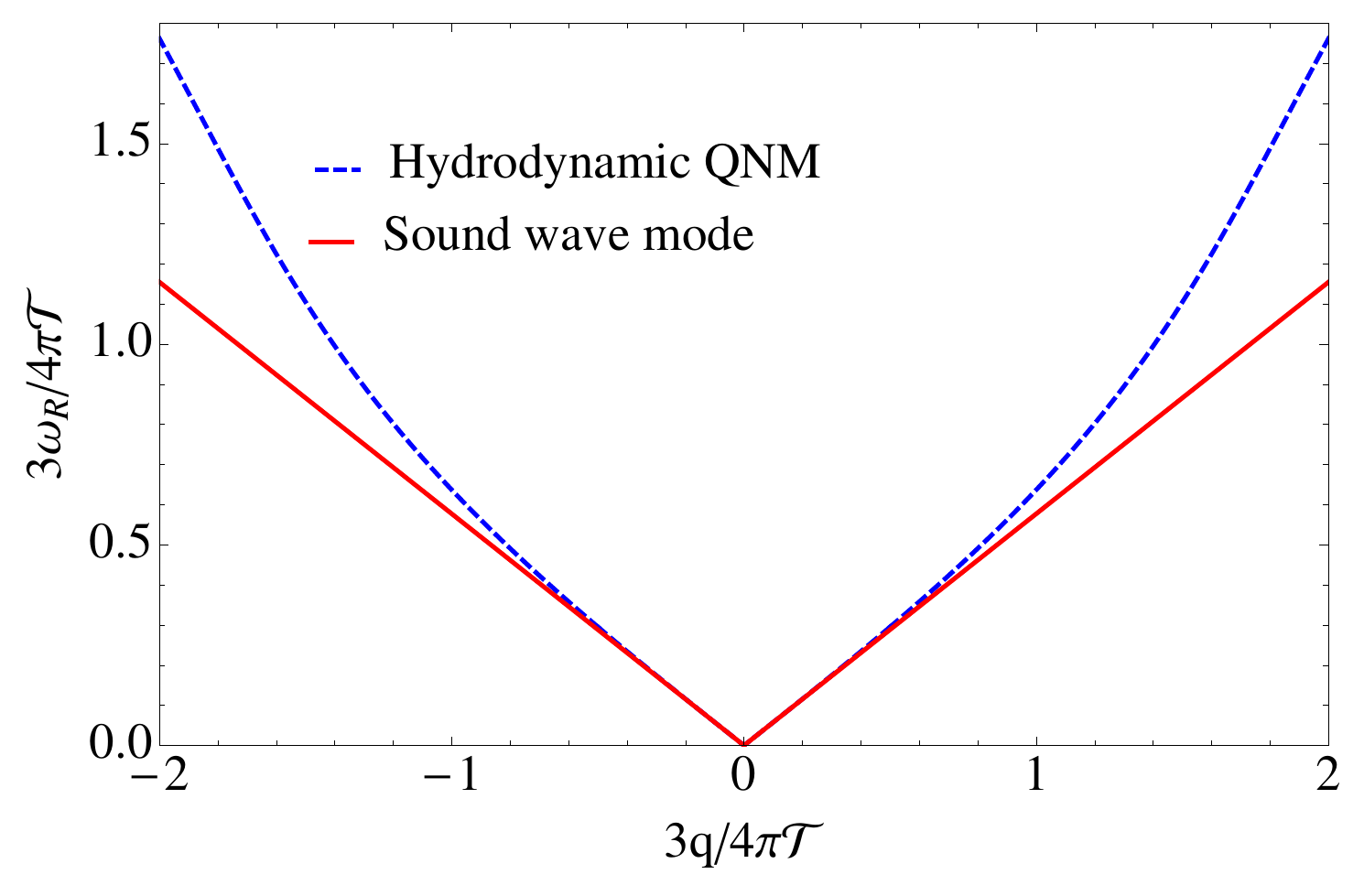} &
\includegraphics[width=7.5cm,angle=0]{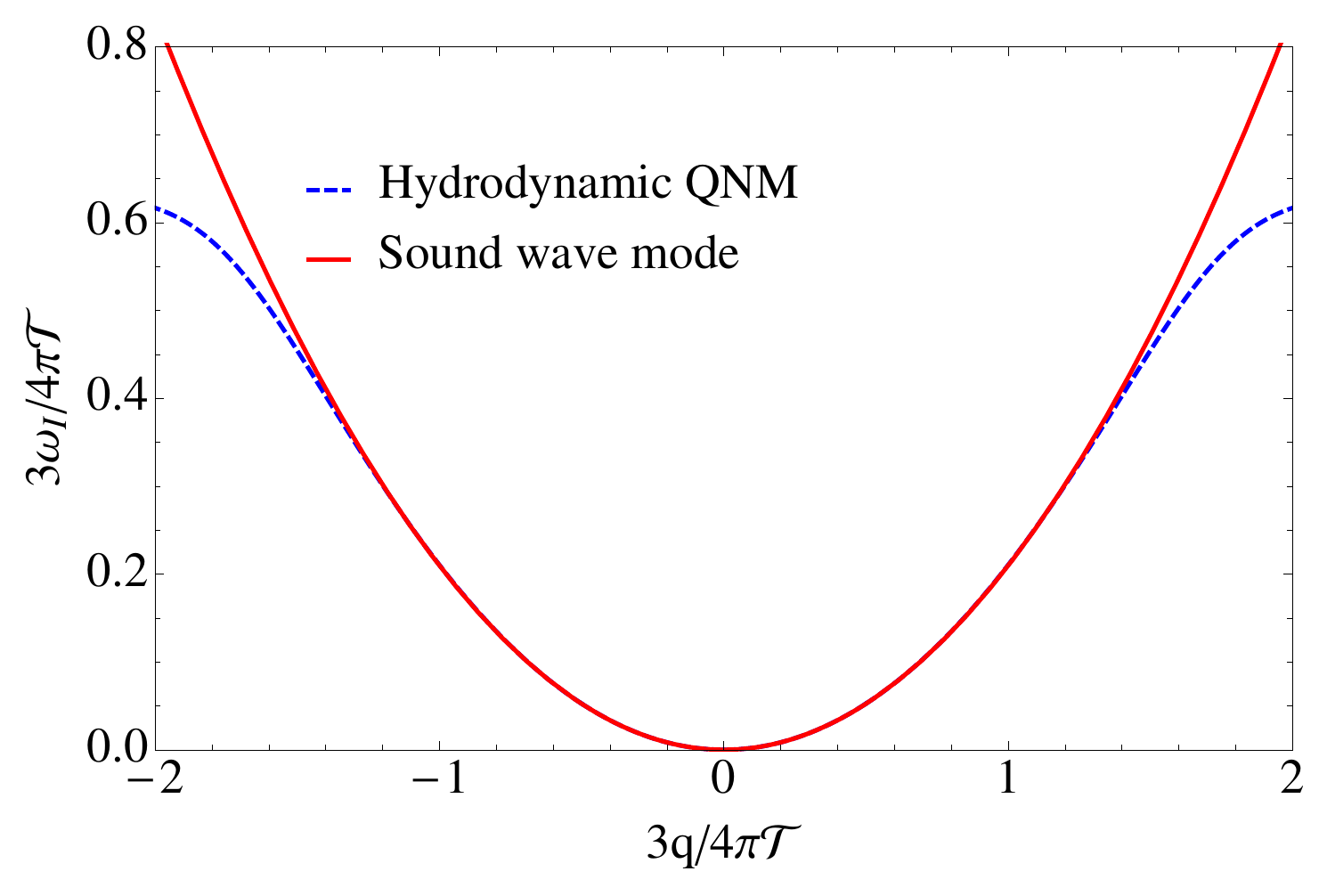}\\
\includegraphics[width=7.5cm,angle=0]{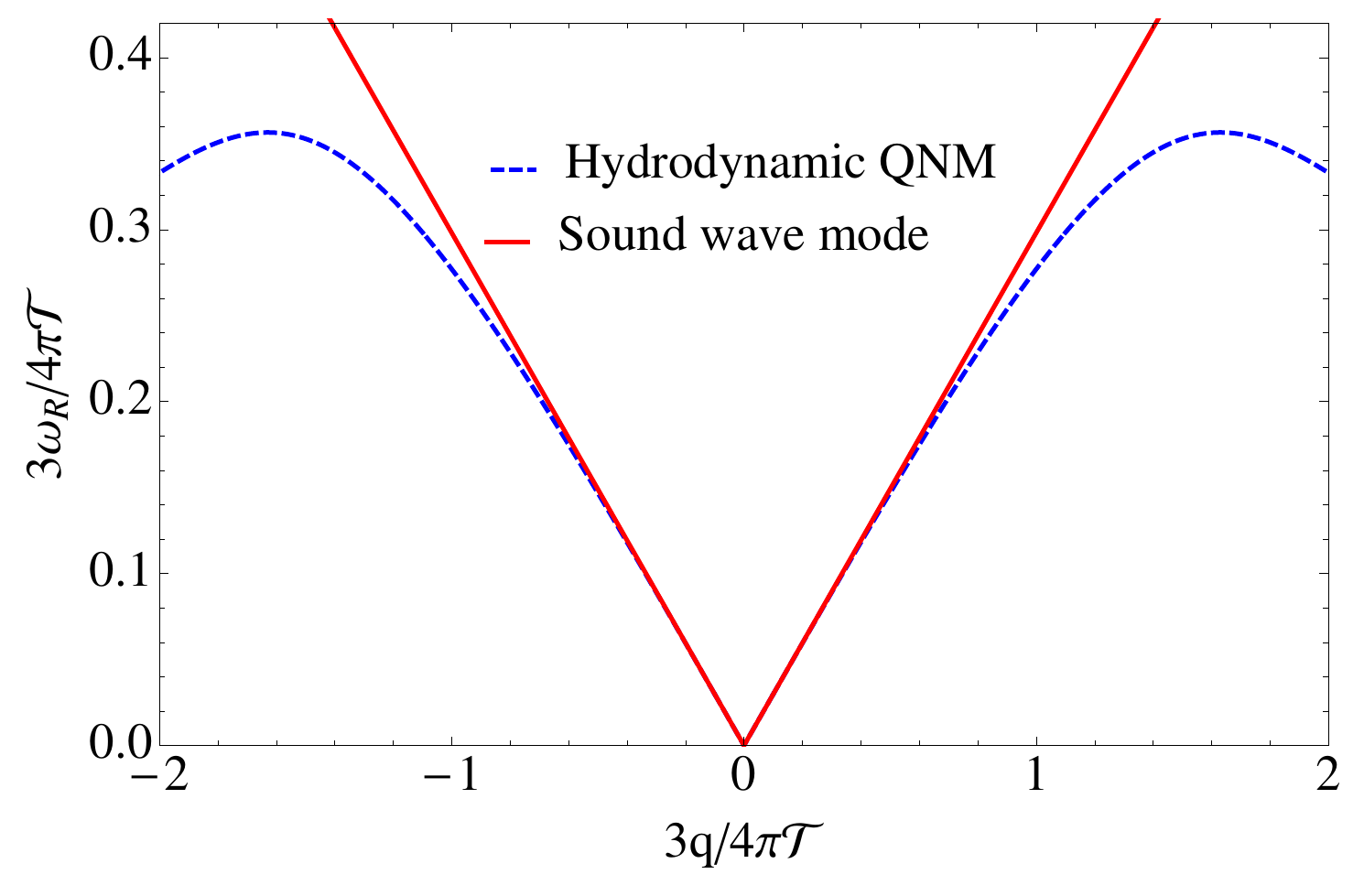}& 
\includegraphics[width=7.5cm,angle=0]{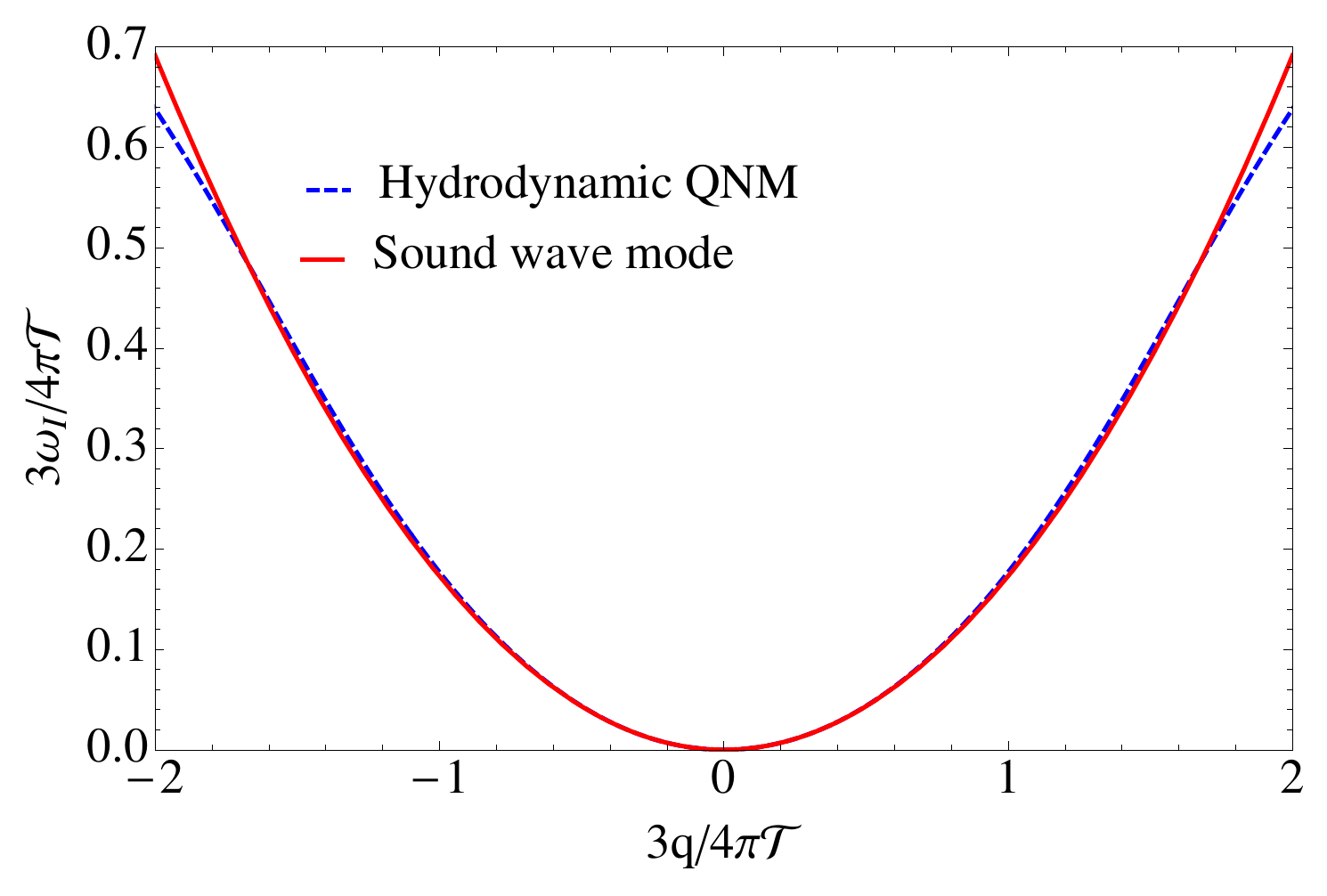}
\\
\end{tabular}
\caption{Hydrodynamic QNMs (solid lines), cf. Eq.~\eqref{rotscalar}, 
and sound mode (dashed lines) of the 
longitudinal sector with $\mathfrak{m}=0$ for $a\alpha=0.1$ (top panels), 
$a\alpha=1/\sqrt{2}$ (middle panels), and $a\alpha=0.95$ (bottom 
panels).}\label{Sound1}
\end{figure*}

\begin{figure*}[ht!]
\begin{tabular}{*{2}{>{\centering\arraybackslash}p{.45\textwidth}}}
\includegraphics[width=7.5cm,angle=0]{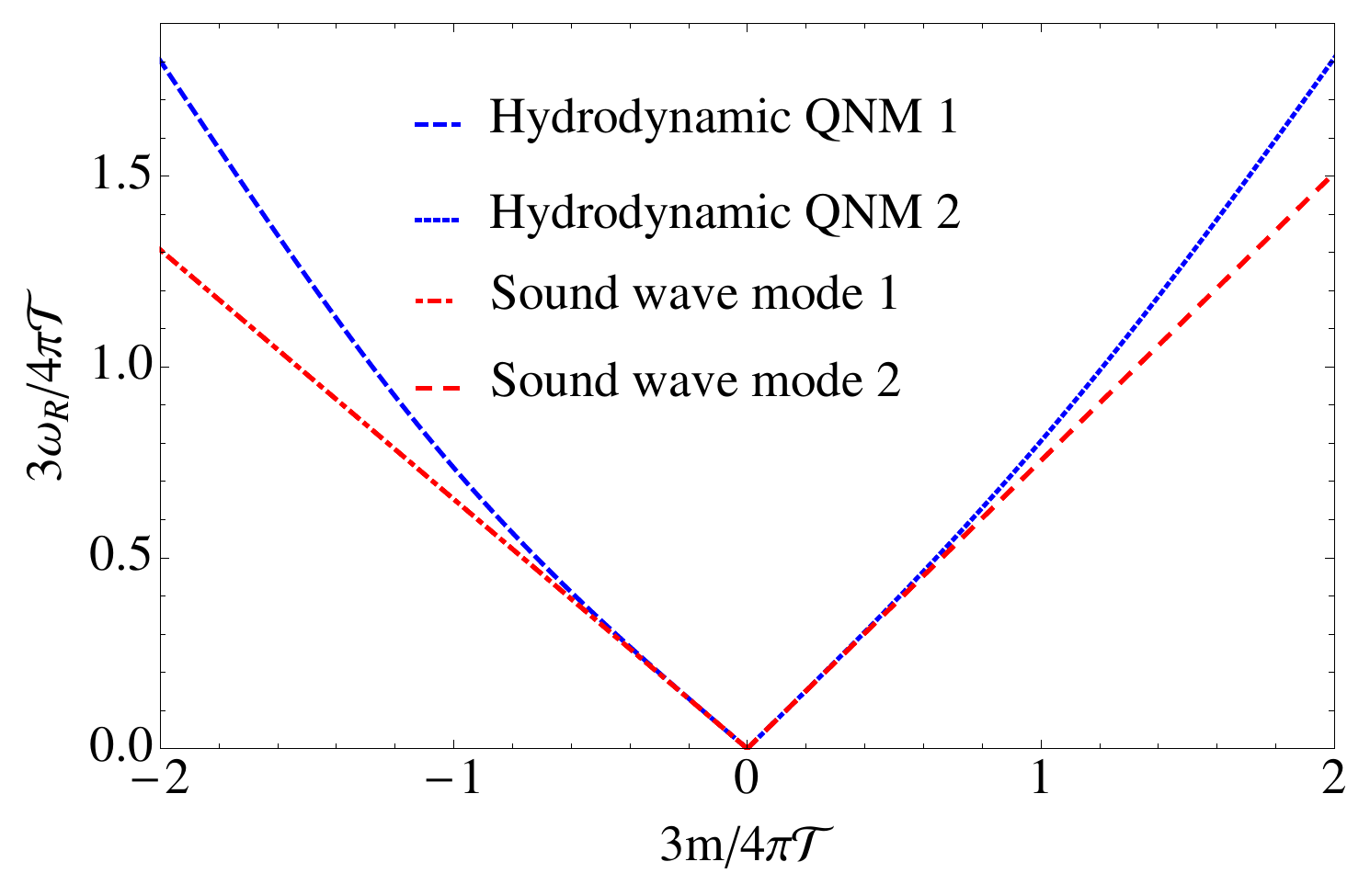}&
\includegraphics[width=7.5cm,angle=0]{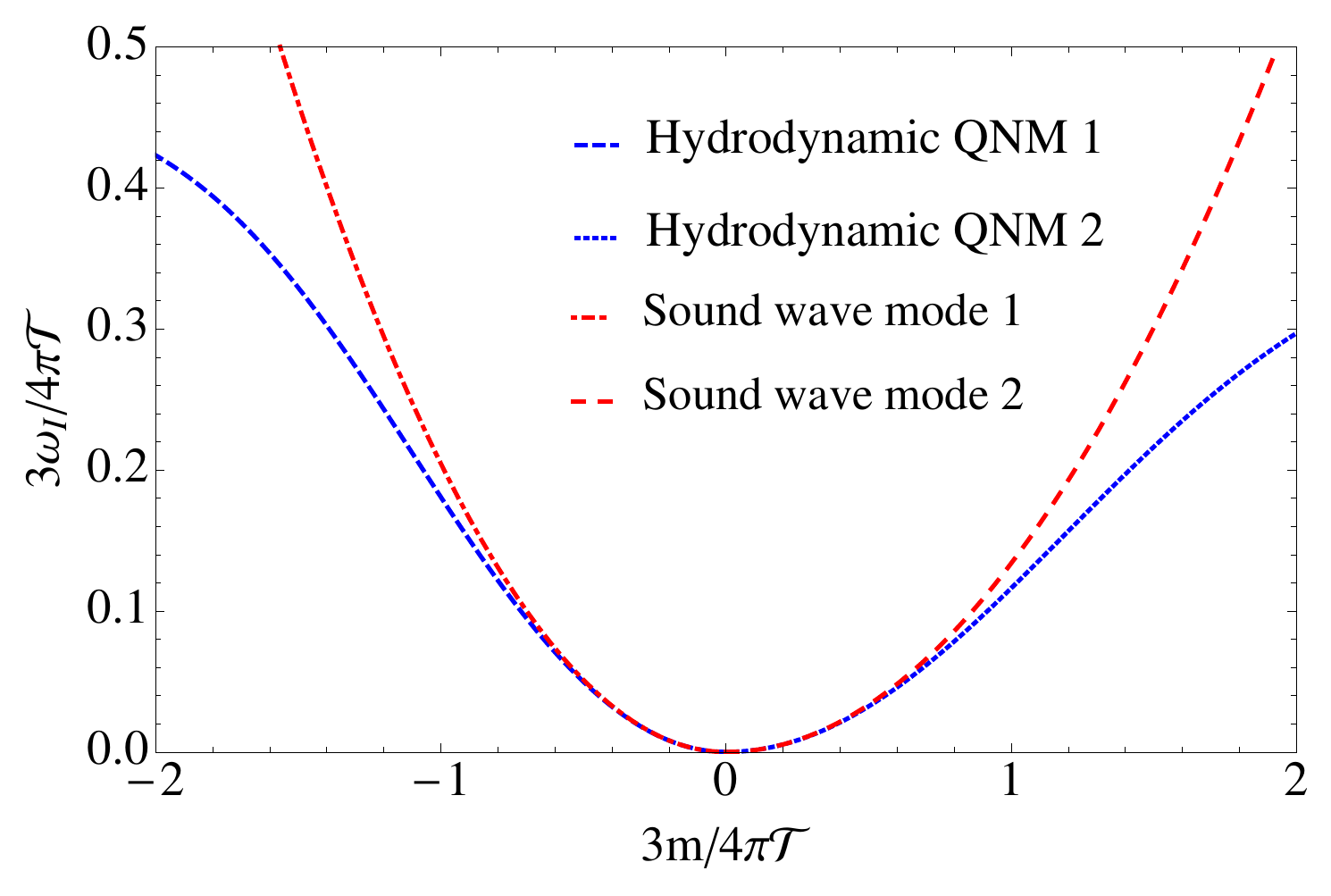}\\
\includegraphics[width=7.5cm,angle=0]{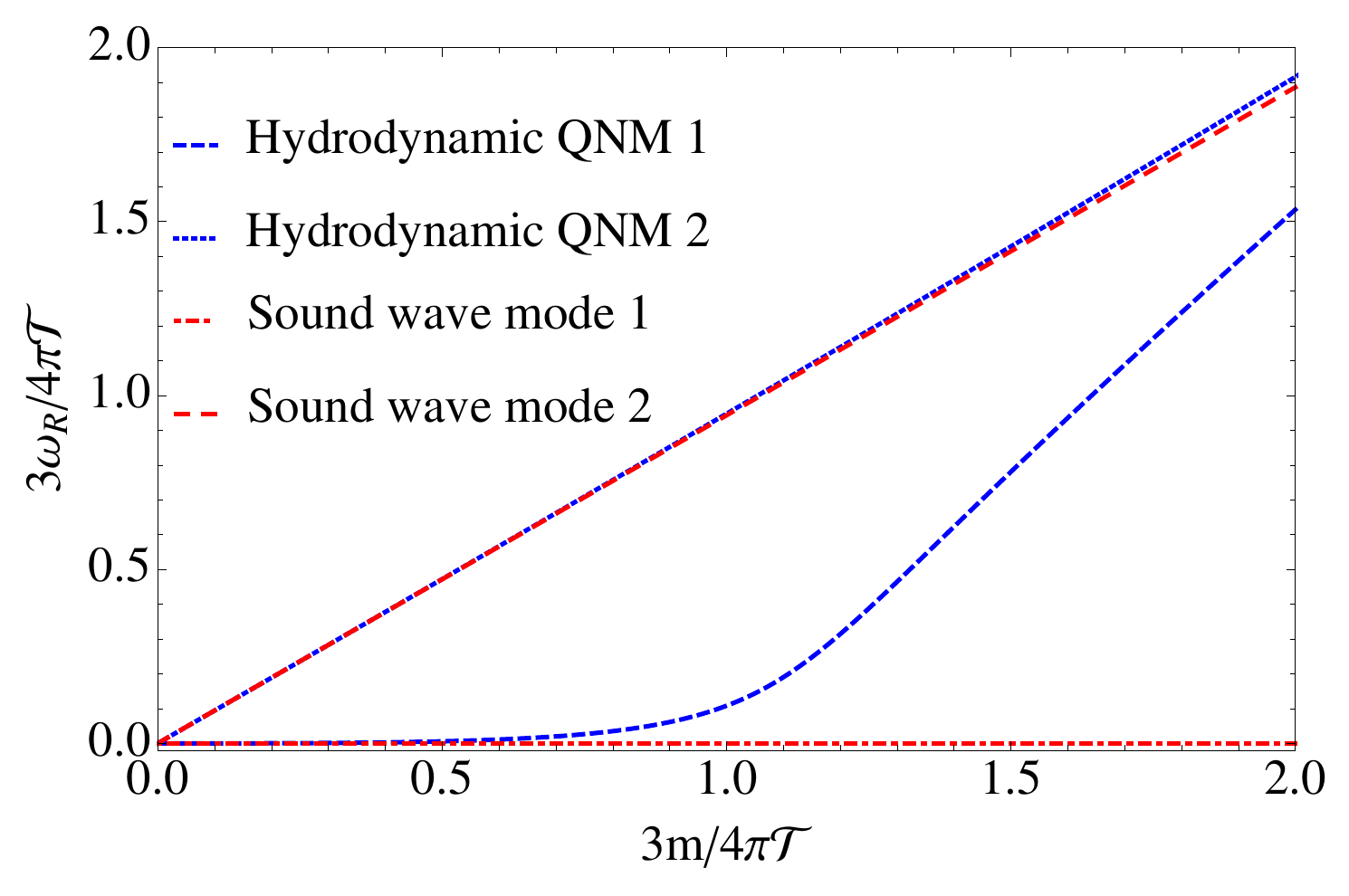}&
\includegraphics[width=7.5cm,angle=0]{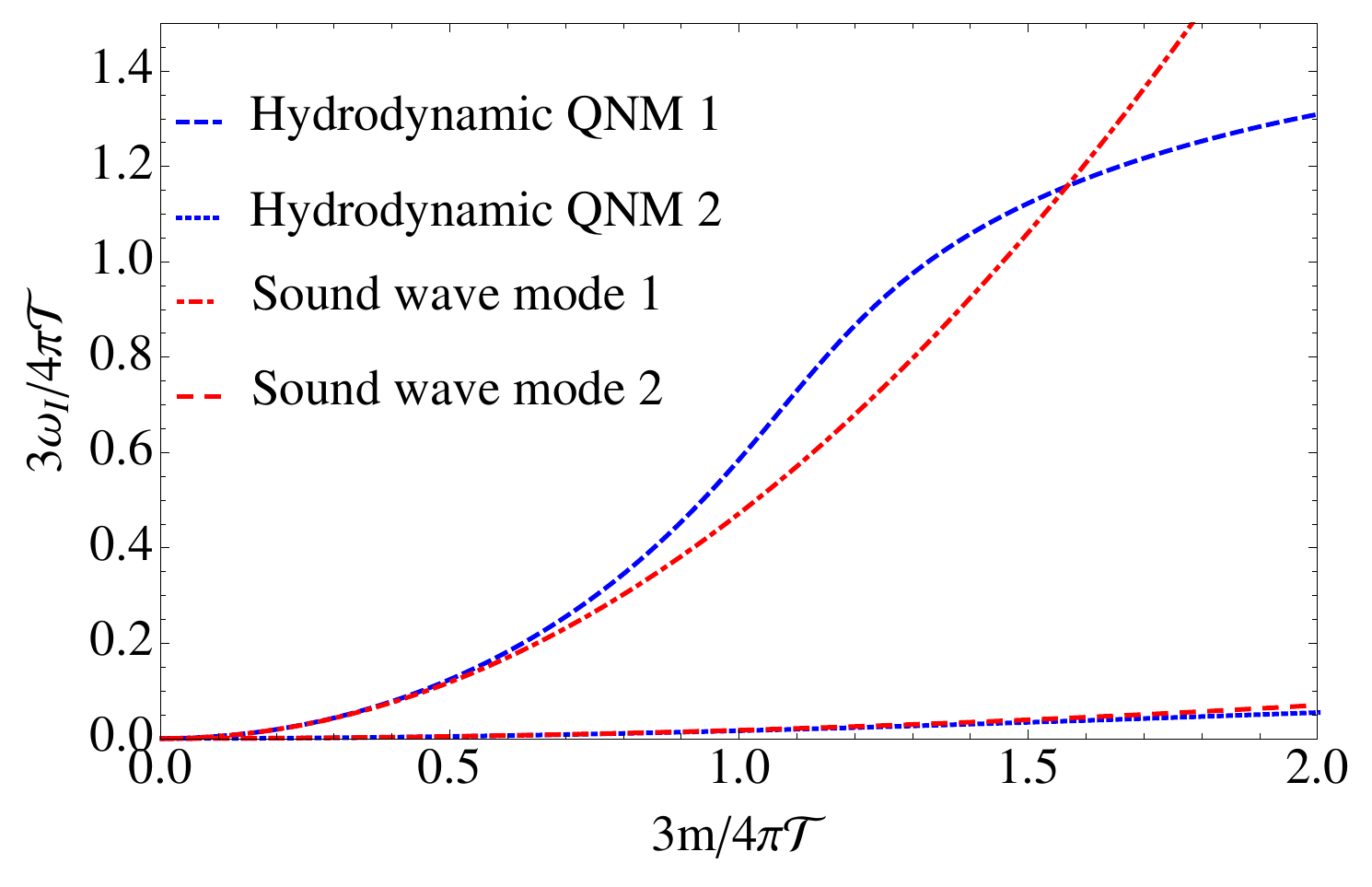}\\
\includegraphics[width=7.5cm,angle=0]{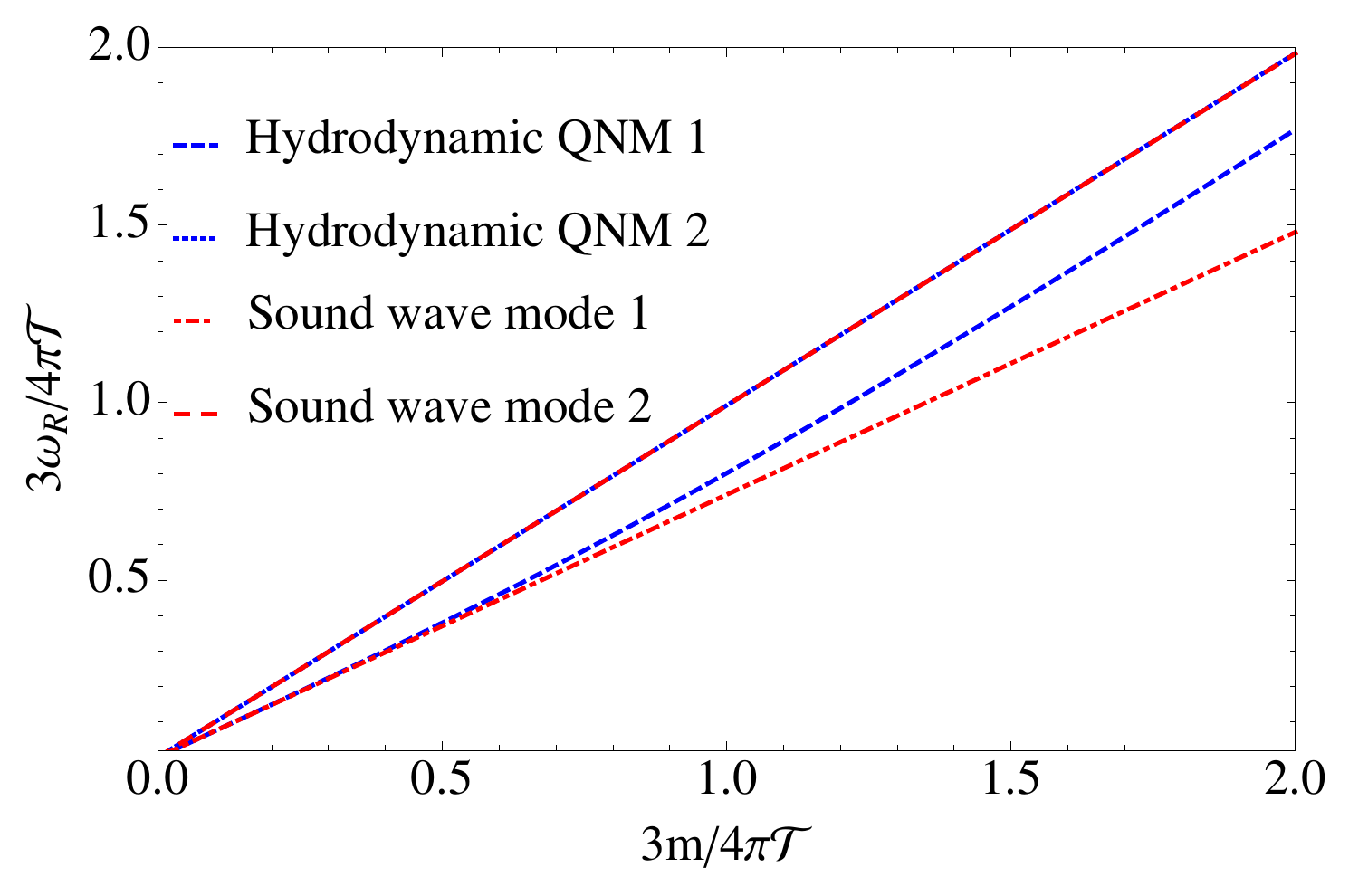}&
\includegraphics[width=7.5cm,angle=0]{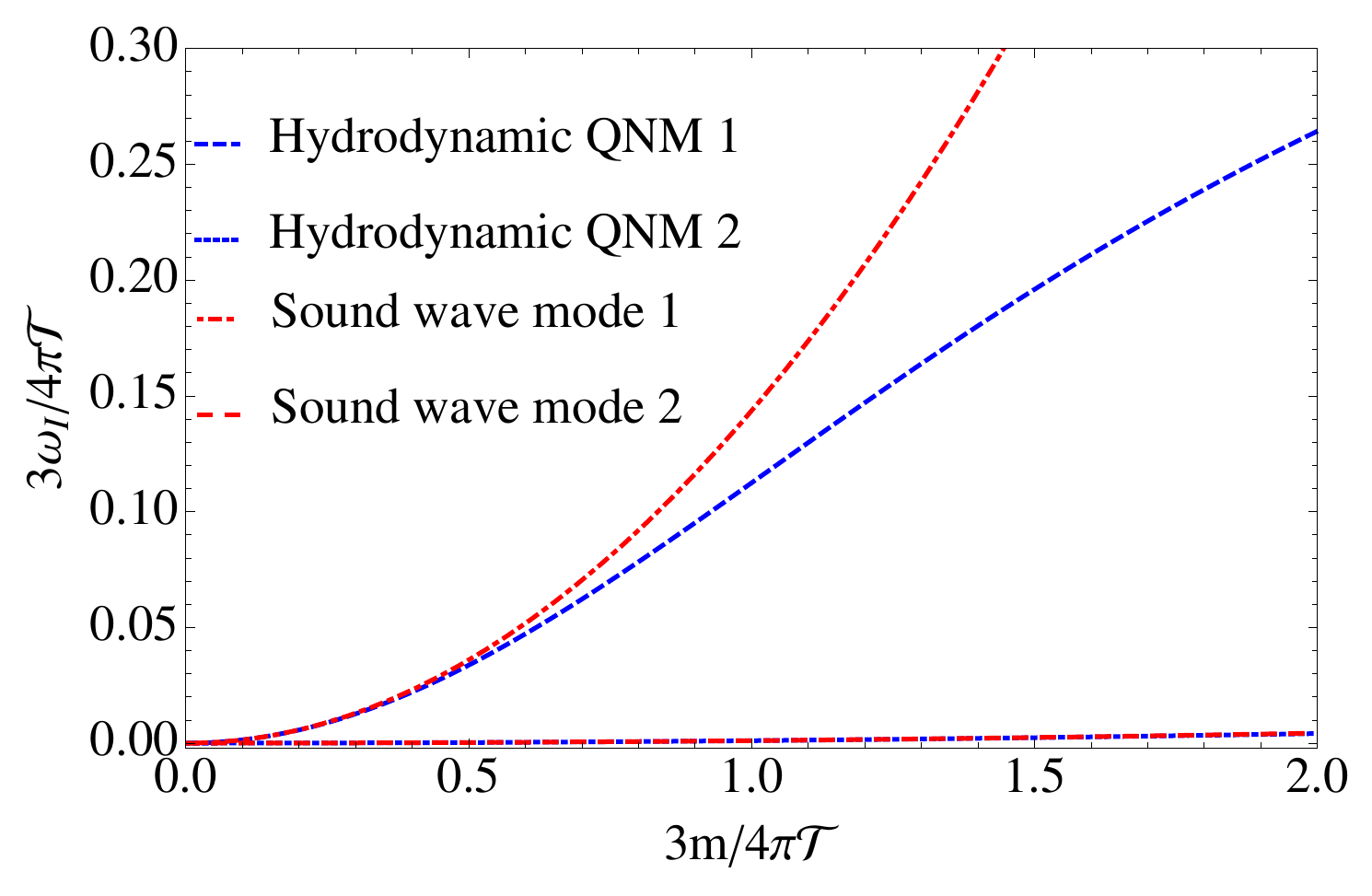}
\\
\end{tabular}
\caption{Hydrodynamic QNMs (solid lines), cf. Eq.~\eqref{rotscalar}, 
and sound mode (dashed lines) of the 
longitudinal sector with $\mathfrak{q}=0$ for $a\alpha=0.1$ (top panels), 
$a\alpha=1/\sqrt{2}$ (middle panels), and $a\alpha=0.95$ (bottom 
panels).}\label{Sound2}
\end{figure*}

In this subsection we show the numerical solutions of the differential equation 
\eqref{fund-eq-scalar} for the lowest lying mode and compare the results to 
the analytical solutions~\eqref{rotscalar}. The results are displayed in 
Figs.~\ref{Sound1} and \ref{Sound2}, where the dashed lines represent the 
numerical results while solid lines represent the analytical solutions, 
respectively. As it was done for the transverse sector, to simplify the analysis 
and trying to get a clear physical interpretation of the results, the analysis 
is split into two cases. 


Firstly, we consider perturbations propagating perpendicularly to the rotation 
direction ($\mathfrak{m}=0$). The results for this case 
are displayed in Fig.~\ref{Sound1}, where we observe that the real part of the 
frequency is proportional to $\mathfrak{q}$ in the hydrodynamic limit.
From Eq.~\eqref{rotscalar} we realize that the proportionality factor depends on
the value of the rotation parameter and, to observe the effect of such a parameter,
we plot three different cases: $a\alpha=0.1$ (top panels),
$a\alpha=1/\sqrt{2}\approx 0.71$ (middle 
panels), and $a\alpha=0.95$ (bottom panels). By comparing the real 
parts of the frequencies in this figure for a fixed value of the wavenumber, 
e.g., for $\mathfrak{q}=1$, it is verified that when the 
rotation increases the real part of the frequency decreases, i.e., 
$\mathfrak{w}_R^{\scriptscriptstyle{a\alpha=0.1}} 
>\mathfrak{w}_R^{\scriptscriptstyle{a\alpha=0.71}} 
>\mathfrak{w}_R^{\scriptscriptstyle{a\alpha=0.95}}$. This behavior is understood 
by noticing that, in the hydrodynamic approximation, given by 
Eq.~\eqref{rotscalar}, the leading term of the real part of the frequency is 
$\mathfrak{w}_R =\mathfrak{q}/\sqrt{1+\gamma^2}$, which decreases when $a 
\alpha$ increases.

On the other hand, the imaginary part of the frequency is proportional to 
$q^2$, and the corresponding coefficient,
\begin{equation}
\Gamma=\frac{\gamma^3}{2\pi\mathcal{T}(1+\gamma^2)^2}\,,
\end{equation}
is associated to the damping of sound waves in the dual field theory. It is observed that the 
imaginary part of the frequency increases with the rotation parameter, as can be 
seen in the top and middle panels of Fig.~\ref{Sound1}, since
$\mathfrak{w}_I^{\scriptscriptstyle{a\alpha=0.1}} 
<\mathfrak{w}_I^{\scriptscriptstyle{a\alpha=0.71}}$. This behavior remains until 
the rotation parameter reaches the critical value 
$\left(a\alpha\right)_c=\sqrt{2/3}$,
where the sound wave damping coefficient attains a maximum value,
decreasing from then on
($\mathfrak{w}_I^{\scriptscriptstyle{a\alpha=0.71}} 
>\mathfrak{w}_I^{\scriptscriptstyle{a\alpha=0.95}}$). This means that 
perturbations with the same wavenumbers decay faster for higher rotation 
parameters if $a\alpha < (a\alpha)_c$, but decay more slowly if $a\alpha > 
(a\alpha)_c$.  It is interesting to note that in the 
extremely rotating case, for $a\alpha= 1$, the damping coefficient vanishes,
eliminating the imaginary part of the frequency, such that the QNMs become 
normal modes.

Secondly, we consider the case $\mathfrak{q}=0$, which turns out to be a more 
interesting situation, cf. Fig.~\ref{Sound2}. In this case, the hydrodynamic
approximation to the real part of the frequency is linear in $\mathfrak{m}$,
with the proportionality coefficient being the transformation of the velocity $\bar c$
to the moving frame, see Appendix \ref{sec-RelativisticWaveVectors} for more details. 
Figure~\ref{Sound2} displays three cases: $a\alpha=0.1$ (top panels), 
$a\alpha=1/\sqrt{2}$ (middle panels), and $a\alpha=0.95$ (bottom panels). 
The graphs in all cases show good agreement between the analytical 
and the numerical results in the hydrodynamic limit. The deviation of the numerical
results from the sound wave mode is also observed. Such deviation is easily understood by
noticing that the analytical solutions are no longer valid beyond the hydrodynamic regime.

According to our conventions, assuming positive $\mathfrak{w}_R$, 
the perturbations with positive $\mathfrak m$ propagate in the same direction as
the rotation, while the perturbations with a 
negative $\mathfrak m$ value propagate in the opposite direction. 
In the former case the real part of the frequency, obtained from
the positive solution of Eq.~\eqref{rotscalar}, reduces to 
$\left[(\bar c+a\alpha)/(1+a\alpha\bar c)\right]\mathfrak m\alpha$,
while in the opposite direction it becomes
$\left[(\bar c-a\alpha)/(1-a\alpha\bar c)\right]\mathfrak m\alpha$.
With these  simplifications on hand we analyze in the following some 
interesting situations.

An interesting situation appears when the value of the rotation 
parameter equals the value of the speed of sound in the rest frame of the fluid,
i.e., $a\alpha=\bar{c}=1/\sqrt{2}$. The results for this situation are
presented in the middle panels of Fig.~\ref{Sound2}. In this case, 
there are two hydrodynamic quasinormal modes for positive values of
$\mathfrak{m}$, indicated respectively by QNM 1 and QNM 2, 
and none for negative values of $\mathfrak{m}$ with $\mathfrak{w}_{R}>0$. 
The QNM 1 can be understood from Eq.~\eqref{rotscalar}. 
For negative $\mathfrak{m}$ and $a> 1/\sqrt{2}$ such equation 
furnishes frequencies with negative real part, representing in fact 
waves (positive frequencies) travelling along $+\varphi$ (with positive 
$\mathfrak{m}$, and hence these modes, like QNM 1, are plotted in the first 
quadrant of Fig.~\ref{Sound2}.
Furthermore, by using the transformations \eqref{Eq:Transformations}, it is seen
that the QNM 1 and QNM 2 correspond, respectively, to waves propagating in the 
$-\bar{\varphi}$ and $+\bar{\varphi}$ direction in the frame $\bar{K}$.
For small wavenumber values, the real part of the QNM 1 frequency is 
essentially zero (within our numerical precision), while the real part
of the QNM 2 frequency is $\mathfrak  w_{R}\approx2/\sqrt{2}$.
Also, the QNM 1 frequency has relatively large 
imaginary part (in comparison to QNM 2), implying a small damping time 
($1/\omega_{I}$). Contrarily, the damping time of the hydrodynamic QNM 2 is
large (in comparison to QNM 1), which means that the dissipative effects can be 
neglected in this case. Once the QNM 1 frequency has real part
close to zero, it looks like a diffusion problem in the hydrodynamic regime.
Such a result is expected in advance, since in this situation the frame $K$ is
comoving with the sound wave.

Bottom panels of Fig.~\ref{Sound2} show the results for the case $a\alpha=0.95$. 
We observe again the existence of two hydrodynamic QNMs for a given $\mathfrak{m}> 0$,
one of them being associated to a wave propagating along the $-\bar{\varphi}$
direction of the frame $\bar{K}$.  This effect can be interpreted in analogy to the case of an
observer moving away from a source of sound waves that is at rest in relation to the medium. 
In the present case, as the velocity of the observer $K$, i.e., the rotation
velocity of the black string, is larger than the speed of sound in the medium ($1/\sqrt{2}$),
the observer detects all the wave fronts moving away from him in the opposite direction.
Additionally, the imaginary part of the QNM 2 is very small, indicating that this perturbation mode
is weakly damped. 

\begin{figure*}[ht!]
\begin{tabular}{*{2}{>{\centering\arraybackslash}p{.45\textwidth}}}
\includegraphics[width=7.9cm,angle=0]{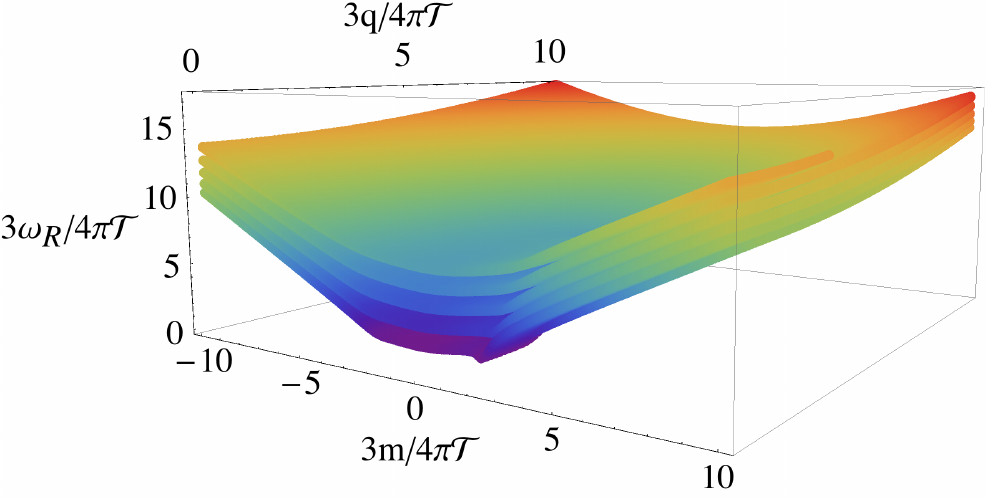}& 
\includegraphics[width=7.9cm,angle=0]{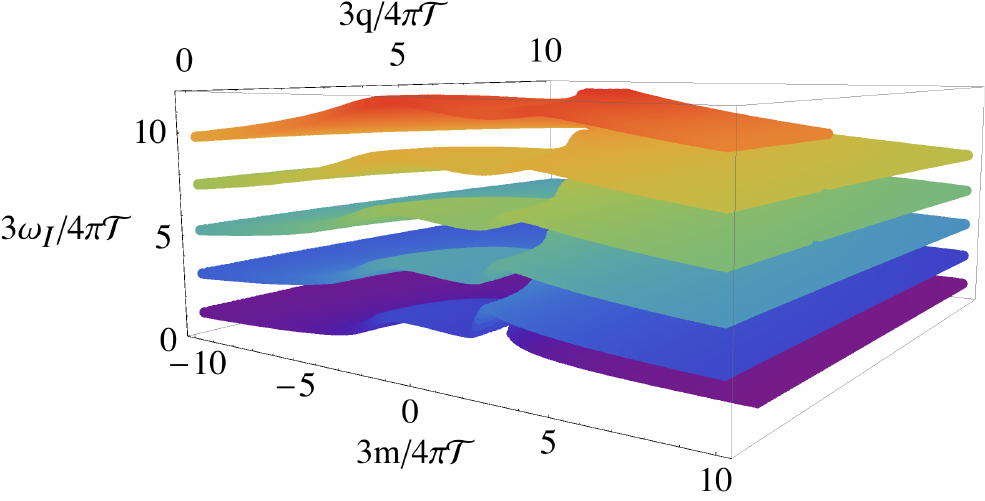}\\
 \end{tabular}
\caption{The first five non-hydrodynamic modes of the transverse sector obtained 
by setting $a\alpha=0.1$. The left (right) panel shows the real (imaginary) part 
of the frequency.}
\label{modosaxial}
\end{figure*}

\begin{figure*}[ht!]
\begin{tabular}{*{2}{>{\centering\arraybackslash}p{.45\textwidth}}}
\includegraphics[width=7.5cm,angle=0]{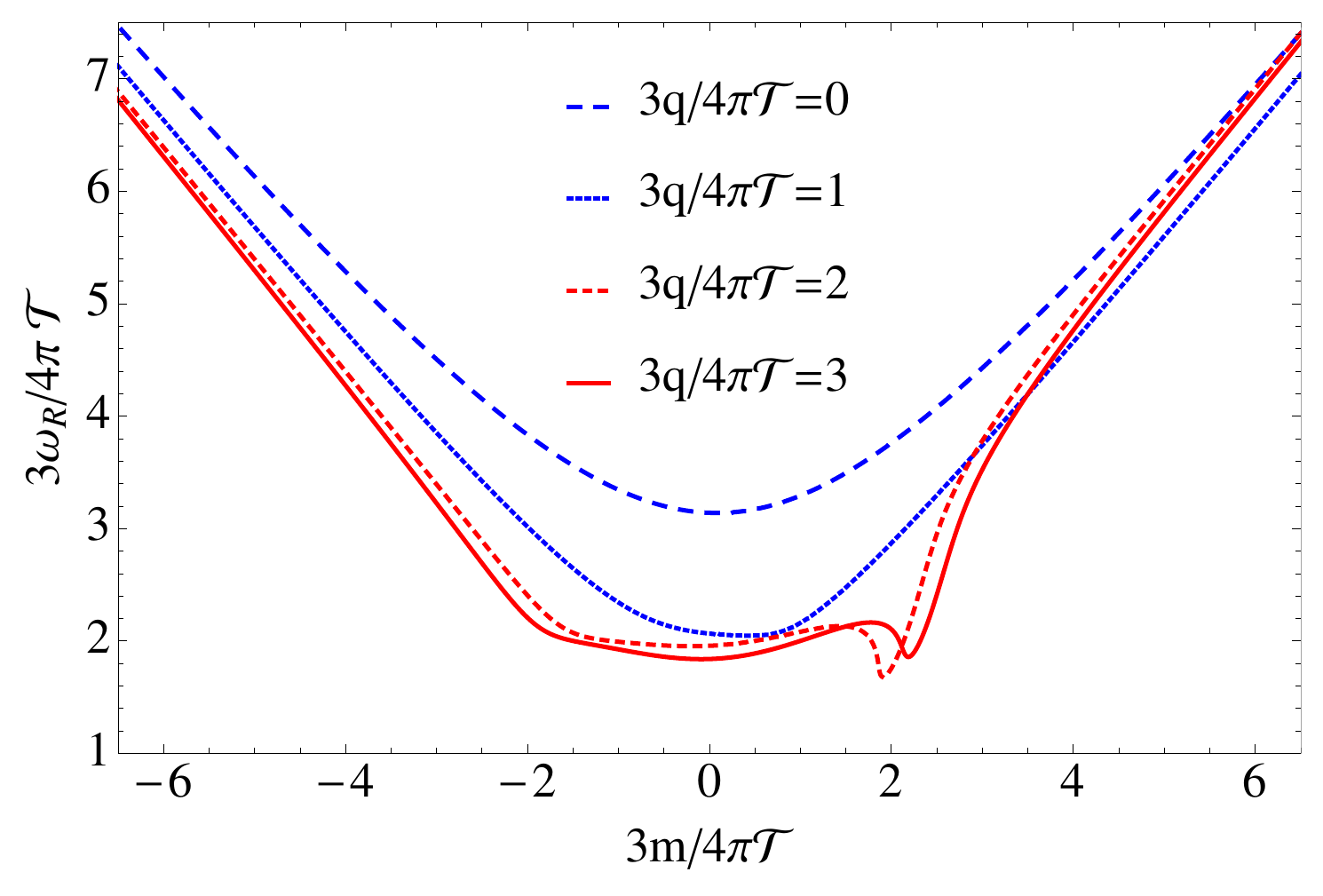}&
\includegraphics[width=7.5cm,angle=0]{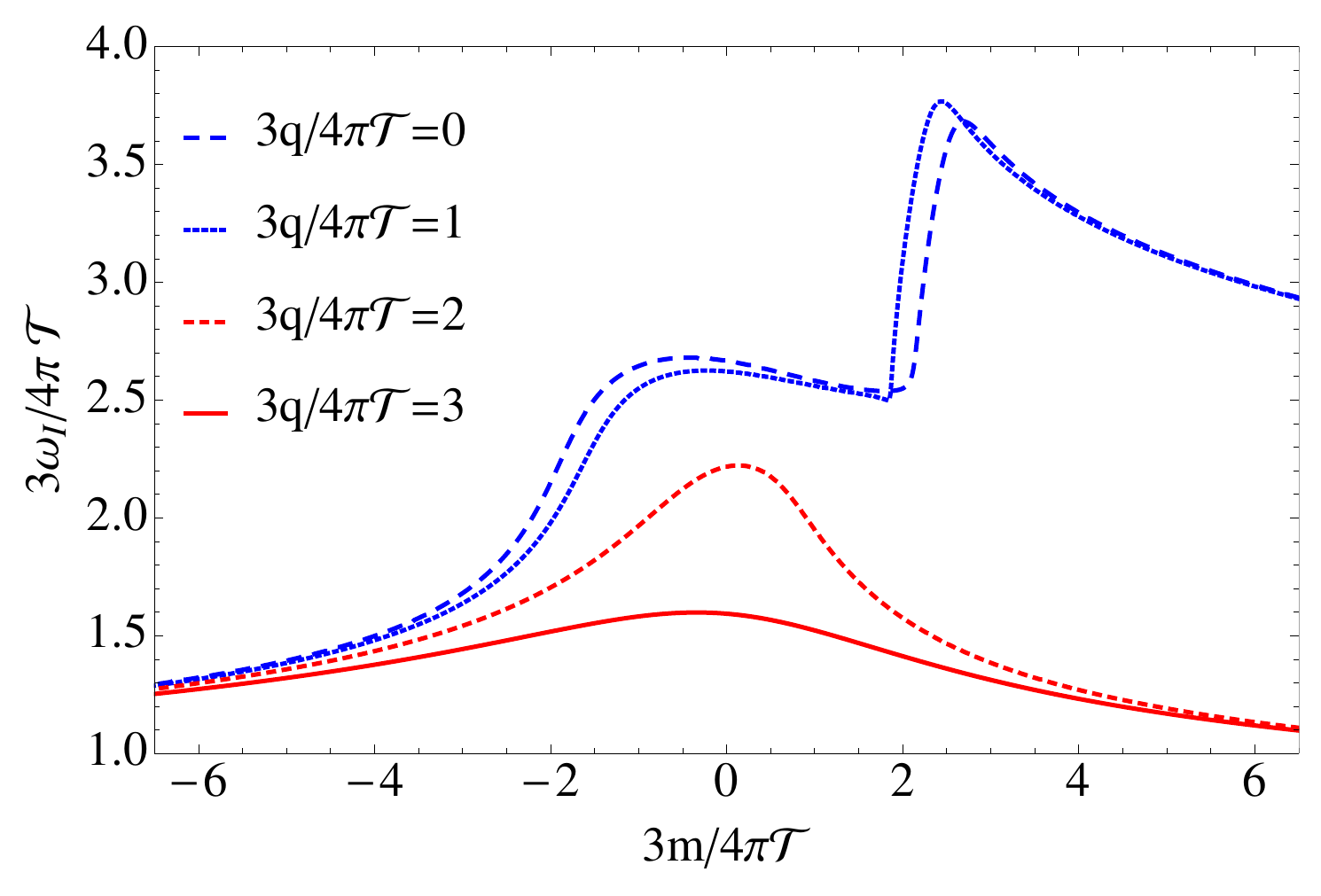}\\
\end{tabular}
\caption{The real (left panel) and imaginary 
(right panel) parts of the 
frequency of the first non-hydrodynamic QNM for 
different values of 
$\mathfrak{q}=3q/4\pi{\cal T}$. This are two dimensional 
cuts of Fig.~\ref{modosaxial} at the given values of 
$\mathfrak{q}$.}
\label{modo1q0a}
\end{figure*}

\section{Non-hydrodynamic quasinormal modes}
\label{sec-numericalQNM}

In this section, we investigate the non-hydrodynamic QNMs of the transverse and 
longitudinal sectors for a few different values of the rotation parameter, 
namely, for $a\alpha=0.1,\;0.2,\;1/\sqrt{2}$ and $0.8$. The frequencies of the 
non-hydrodynamic (or ordinary) modes are obtained by solving numerically the 
differential equations (\ref{fund-eq-vector}) and (\ref{fund-eq-scalar}) for 
specific values of the wavenumbers $\mathfrak{m}$ and $\mathfrak{q}$. The main 
difference between the non-hydrodynamic and the hydrodynamic QNMs is the 
behavior in the small wavenumber limit, with the hydrodynamic frequencies going 
to zero, while the non-hydrodynamic frequencies tend to nonzero values. The 
ration between the imaginary parts of successive frequencies at zero wavenumbers 
is used to rank in increasing order the different non-hydrodynamic QNMs. Since 
their damping time $\tau= 1/\omega_{I}$ is relatively short for small wavenumbers,
the  non-hydrodynamic modes are the first to disappear in this regime. 
Moreover, numerical studies on the time evolution of the perturbations show that 
such modes dominate the initial response of the system, i.e., 
they are the most relevant modes at early times after the perturbation takes 
place (see Ref. \cite{Morgan:2009pn} as an example).
Several works in this direction were published in the last few years due to
its interpretation and relevance in the dual field theory, and mainly to get
more information about the quark-gluon plasma before the hydrodynamic regime 
\cite{Noronha:2011fi, Heller:2014wfa, Janik:2014zla}.

\subsection{Transverse sector}

The non-hydrodynamic QNM frequencies of the transverse sector are found by solving 
numerically the differential equation \eqref{fund-eq-vector}. The results 
displayed here are for wavenumbers in the intervals $\mathfrak{m}\in [-10,10]$ 
and $\mathfrak{q}\in [0,10]$. In Fig.~\ref{modosaxial} we show the frequencies 
of the first five transverse non-hydrodynamic QNMs obtained for $a\alpha=0.1$. 
The first difference to note in comparison to the static case studied in 
Ref.~\cite{Miranda:2008vb} is that the frequencies are now asymmetric under the 
change $\mathfrak{m}\to -\mathfrak{m}$. This result for the real parts of the 
frequencies may be understood in terms of the difference in the Doppler shift 
produced by the rotation as the waves propagate in the $+\varphi$ or $-\varphi$ 
directions.

\begin{figure*}[ht!]
\begin{tabular}{*{2}{>{\centering\arraybackslash}p{.45\textwidth}}} 
\includegraphics[width=7.5cm,angle=0]{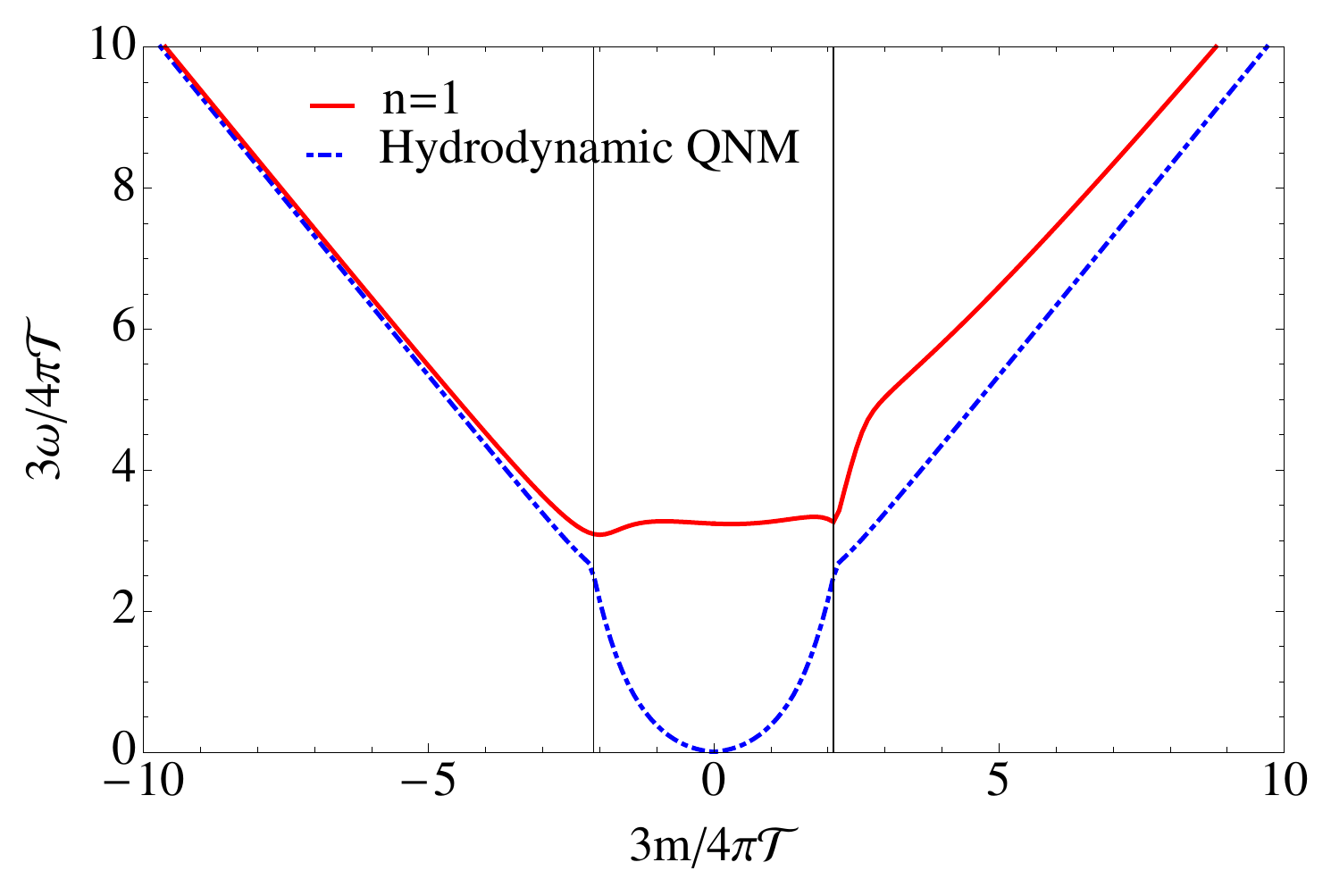}&
\includegraphics[width=7.5cm,angle=0]{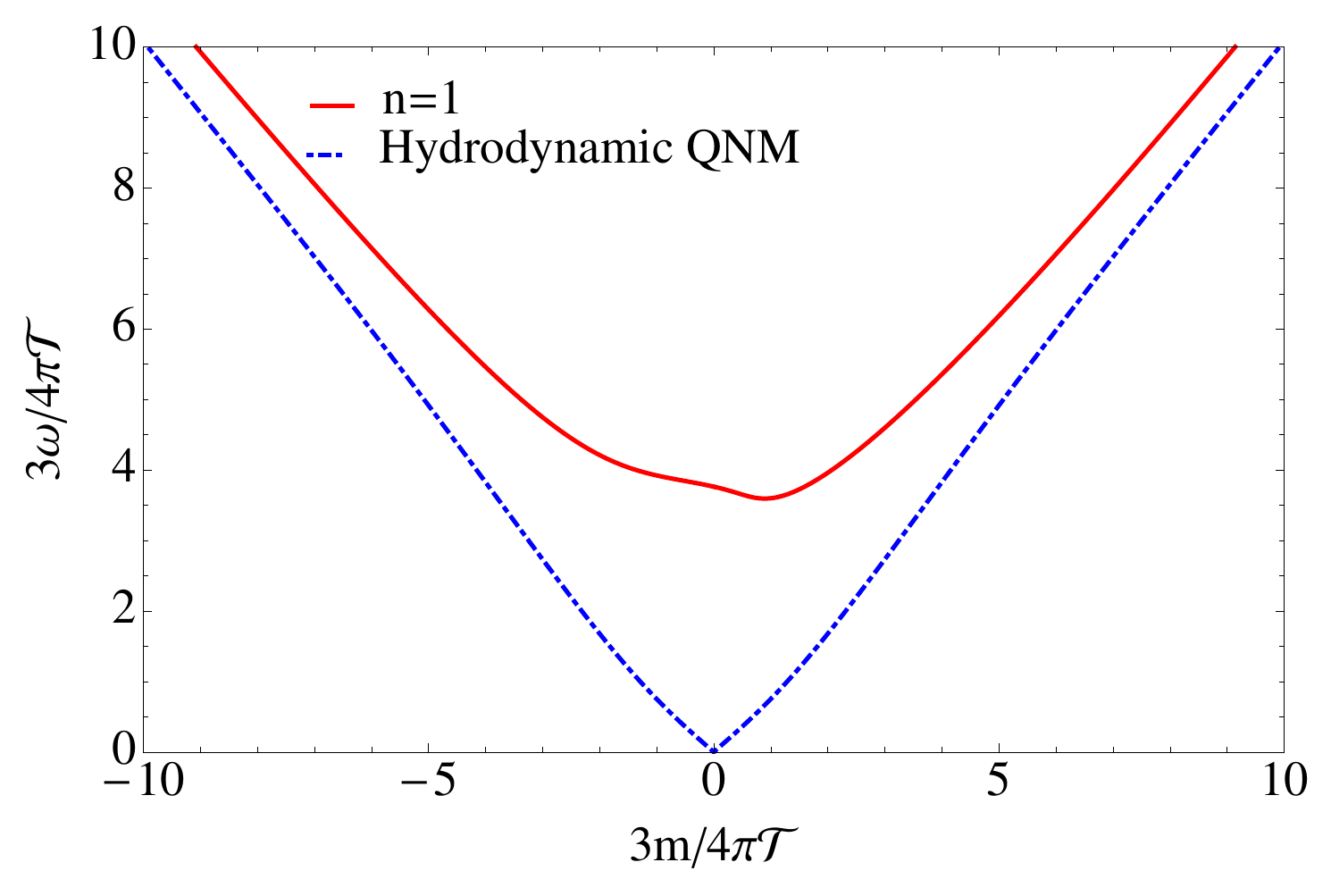}\\
\includegraphics[width=7.5cm,angle=0]{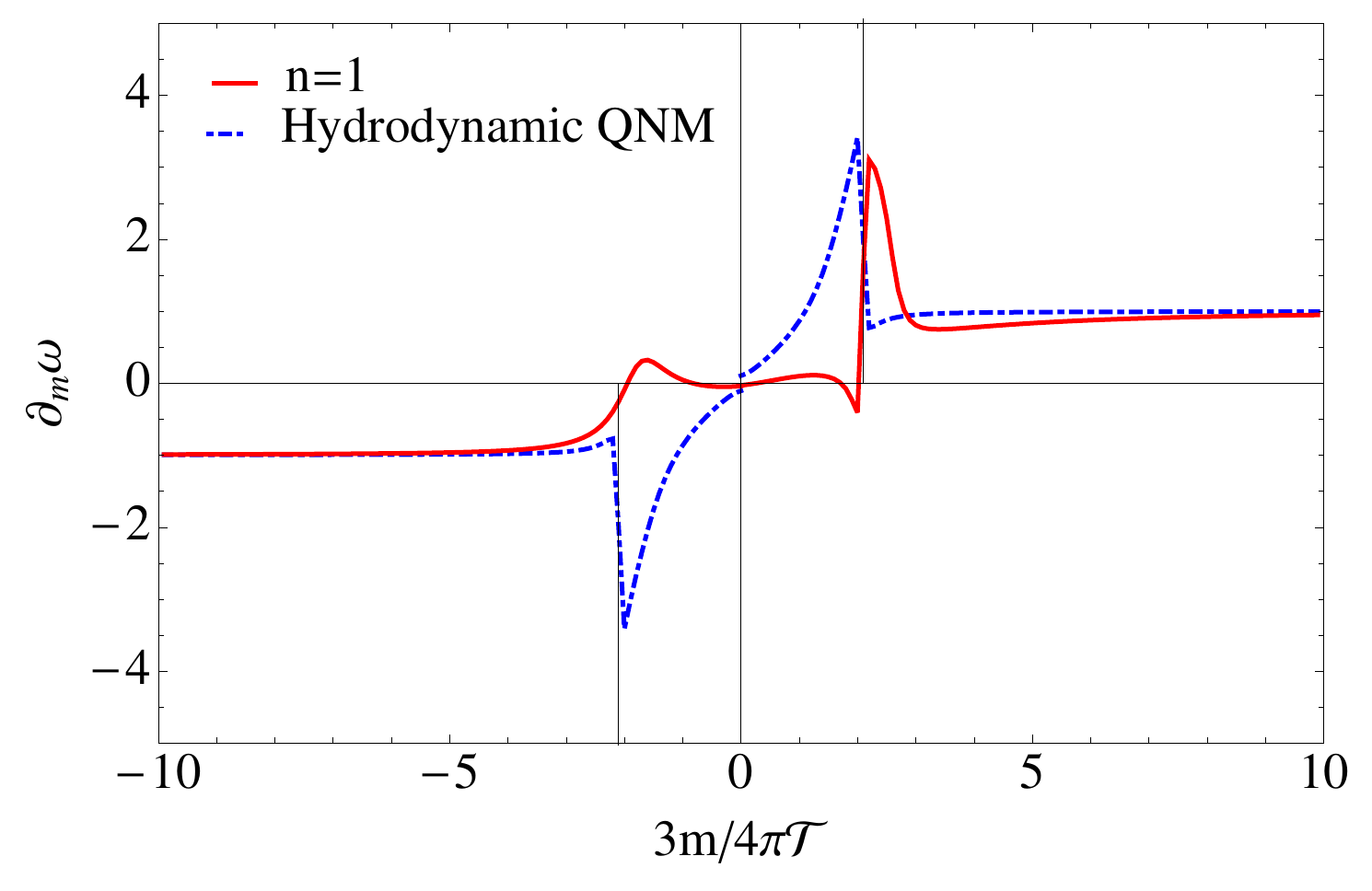}&
\includegraphics[width=7.5cm,angle=0]{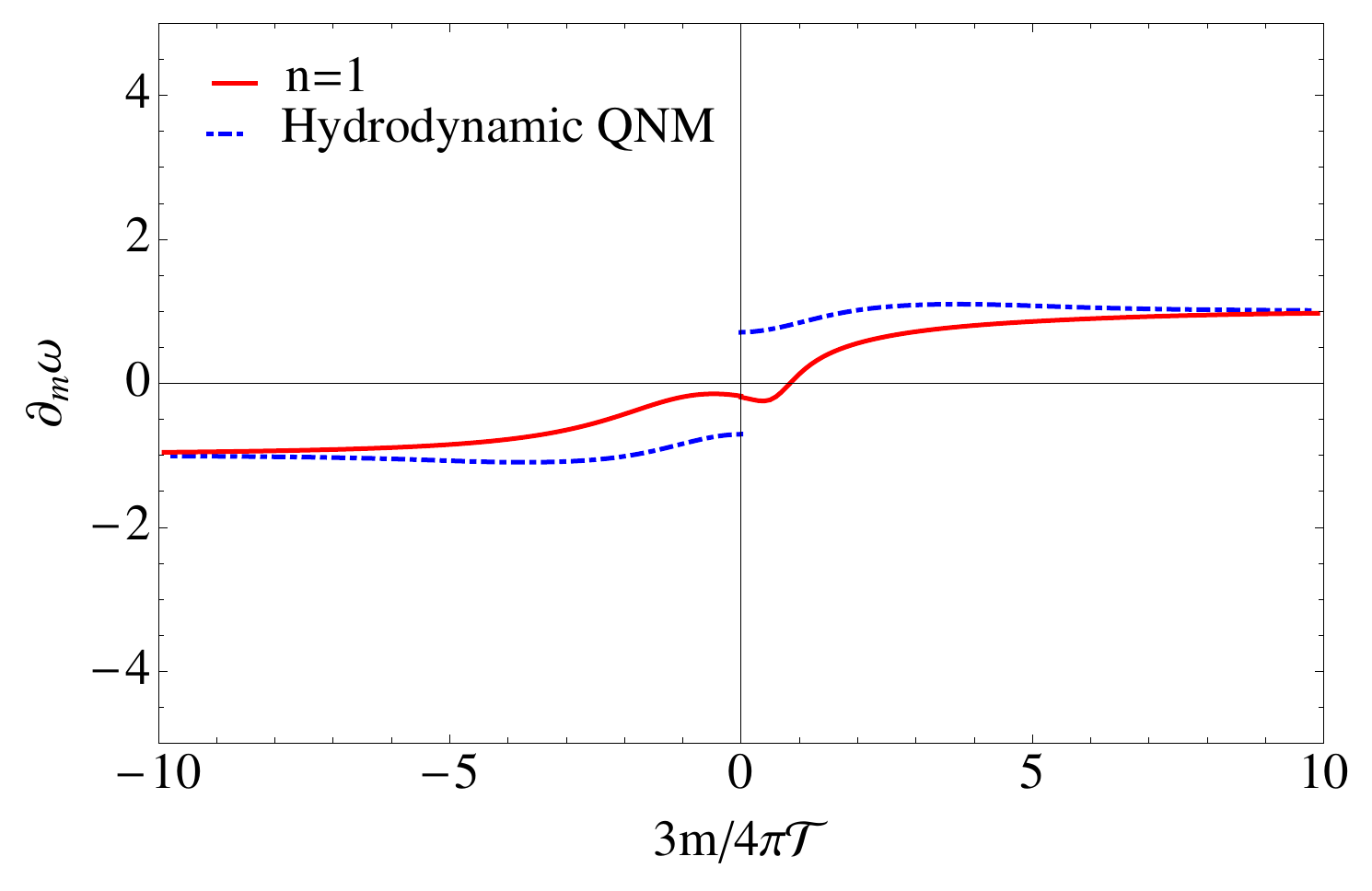}\\
\end{tabular}
\caption{The absolute values of the hydrodynamic and the first non-hydrodynamic QNMs
of the transverse sector as a function of $\mathfrak m$ for $\mathfrak{q}=0$,
with $a\alpha=0.1$ (top left panel), and $a\alpha=1/\sqrt{2}$ (top right panel).
The bottom panels show the slope of the curves  $\mathfrak{w}=\mathfrak{w}(\mathfrak m)$, given by 
$\partial_{\mathfrak{m}}{\mathfrak{w}}$, for both cases shown in the top panels.
The vertical lines in the left panels (top and bottom) are located at 
$\mathfrak{m}\approx 2.1$ and $\mathfrak{m}\approx -2.1$.}
\label{modo1q0}
\end{figure*}

Another fact worthy to be noted is the interconnection between two different 
neighbor modes, as it can be observed in Fig.~\ref{modosaxial}, where a lower 
lying mode grows and reaches the first upper lying mode. This effect occurs at
small $\mathfrak {q}$ and positive $\mathfrak m$  and is more evident for the 
imaginary parts of the frequencies. We still do not know the origin of such an 
effect that does not happen in the static black string case, and a physical 
interpretation for it is also missing. 

To get more information from Fig.~\ref{modosaxial} we take two dimensional slices
by choosing some fixed values of the wavenumber $\mathfrak q$. The slices 
for $\mathfrak{q}=0,\,1,\,2$ and $3$ of the first non-hydrodynamic mode are 
displayed in Fig.~\ref{modo1q0a}. As already mentioned, the asymmetry of the real
(left panel) and imaginary (right panel) parts of the frequencies in relation to
$\mathfrak m$ is a consequence of the rotation. The curves of $\mathfrak{w}_{I}$ 
against $\mathfrak m$ show the peculiar behavior observed in Fig.~\ref{modosaxial}
for $\mathfrak{m}\gg 1$: the asymptotic behavior of the modes for small $\mathfrak q$
is very different from that for large $\mathfrak q$. Clearly there is a sharp jump at
some intermediate value of $\mathfrak q$. However, in the $-\mathfrak{m}$
direction, the frequencies converge to the same value for all $\mathfrak q$. 

On basis of Fig.~\ref{modo1q0} an additional analysis of the slice
$\mathfrak q=0$ is developed in the sequence. As mentioned above, we take the 
hydrodynamic QNM together with the first non-hydrodynamic mode because these 
are the dominant modes throughout the time evolution of the gravitational 
perturbations (see, for instance, Ref.~\cite{Janik:2014zla}). In the top 
panels of Fig.~\ref{modo1q0} we display the absolute value of the QNM
frequencies, $\mathfrak{w} =\sqrt{\mathfrak{w}_R^2 +\mathfrak{w}_I^2}$, as 
a function of the wavenumber component $\mathfrak{m}$ for $a\alpha=0.1$ (top 
left panel) and $a\alpha=1/\sqrt{2}$ (top right panel). In these 
graphics, it is possible to observe the different behaviors of the quasinormal
frequencies in the low- and high-wavenumber regimes. To quantitatively measure 
these differences we calculate the slope of each curve through the relation
\begin{equation}
\partial_{\mathfrak{m}}\mathfrak{w}\equiv \frac{\partial\mathfrak{w}}
{\partial\mathfrak{m}} \approx\frac{\Delta \mathfrak{w}}{\Delta \mathfrak{m}}
= \frac{\mathfrak{w}_{i+1}-\mathfrak{w}_{i}}
{\mathfrak{m}_{i+1}-\mathfrak{m}_{i}}.
\end{equation}
Here $\mathfrak{w}_i$ and $\mathfrak{w}_{i+1}$ are numerical solutions 
calculated at neighbor wavenumber values $\mathfrak{m}_i$ and 
$\mathfrak{m}_{i+1}$, respectively. The bottom left panel in Fig.~\ref{modo1q0} 
shows the corresponding results for $a\alpha=0.1$. In this figure we see a sharp 
jump close to the point $\mathfrak{m}\approx 2.1$ for both the hydrodynamic and 
the first non-hydrodynamic modes. A vertical line was plotted at this point. In the 
opposite direction, the hydrodynamic QNM still has this sharp jump at 
$\mathfrak{m}\approx -2.1$, while the non-hydrodynamic mode has a smooth 
transition. An important feature is that the dispersion relation of the 
non-hydrodynamic mode is approximately constant --the slope being approximately 
zero-- between the two vertical lines. Asymptotically, for large wavenumbers, 
the dispersion relations tend to be linear, which is  characteristic behavior of the
relativistic regime (see also Ref.~\cite{WitczakKrempa:2013ht}).
Repeating the same analysis for $a\alpha=1/\sqrt{2}$, we observe a smooth transition
in the results displayed in the bottom right panel of Fig.~\ref{modo1q0}.

\begin{figure*}[ht!]
\begin{tabular}{*{2}{>{\centering\arraybackslash}p{.45\textwidth}}}
\includegraphics[width=7.5cm,angle=0]{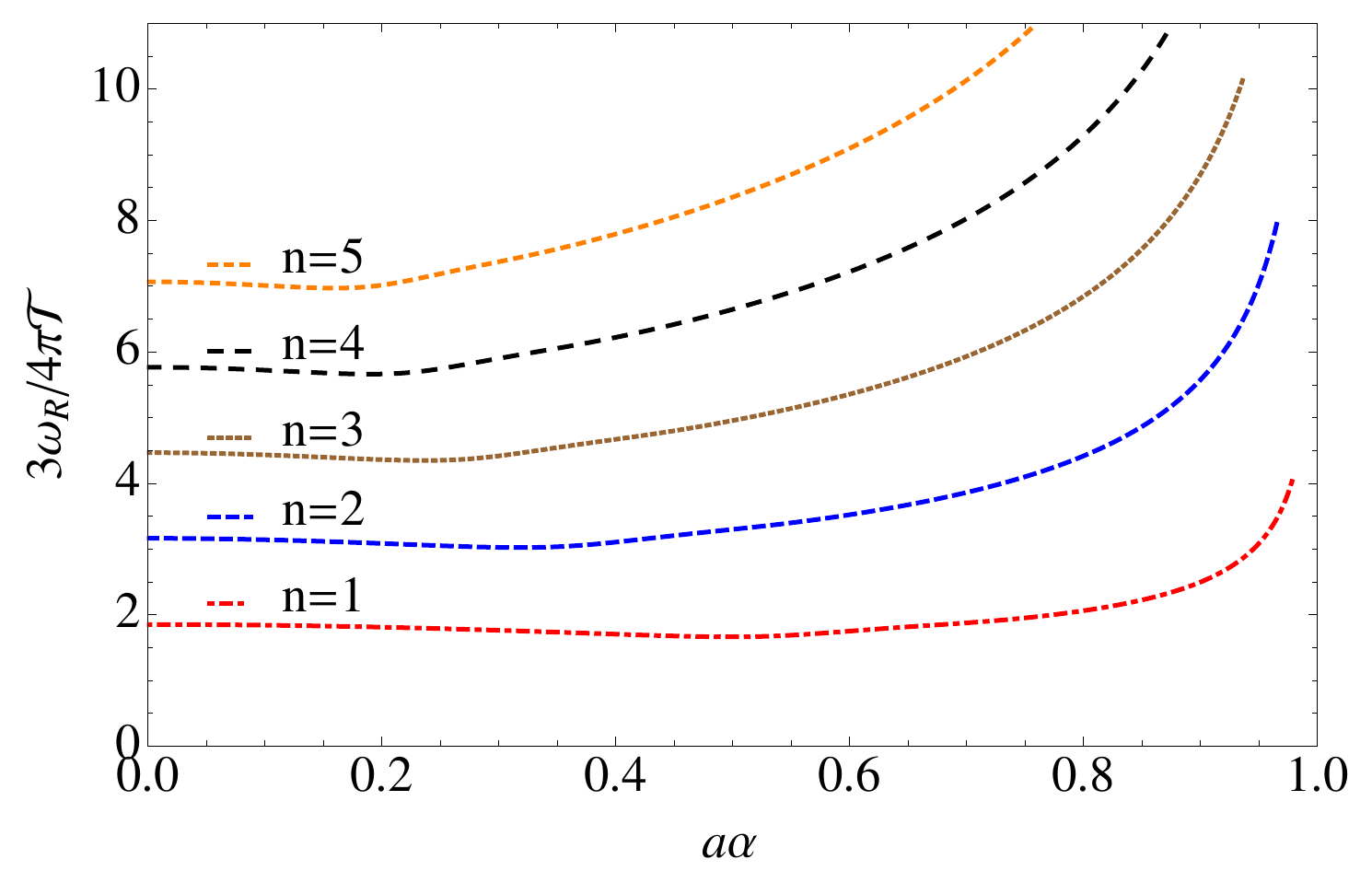}& 
\includegraphics[width=7.5cm,angle=0]{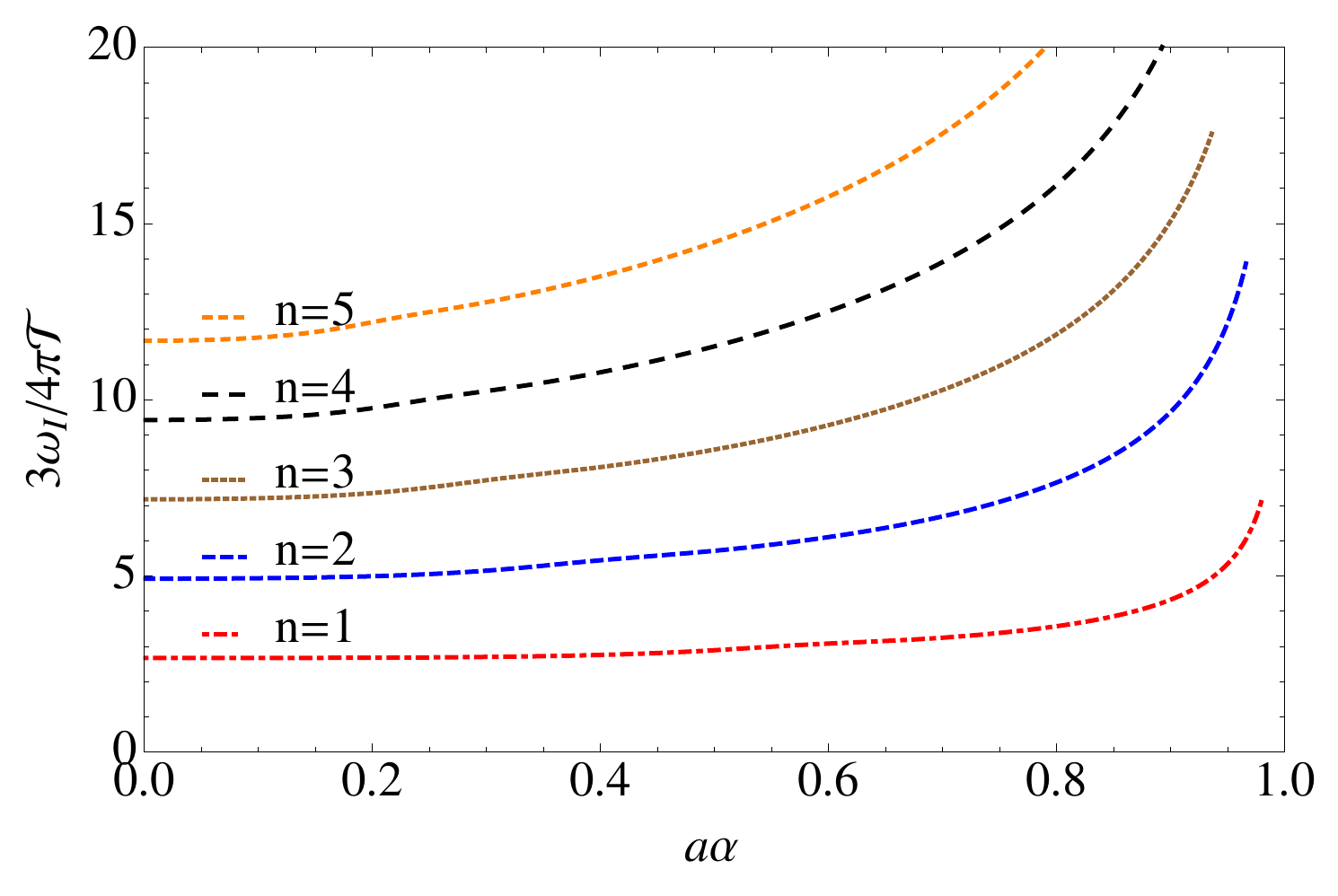}\\
\end{tabular}
\caption{The first five non-hydrodynamic QNMs of the transverse sector as a 
function of the rotation parameter for $\mathfrak{m}=\mathfrak{q}=0$.}
\label{modo1vara}
\end{figure*}
\begin{table*}[ht!]
\begin{ruledtabular}
\begin{tabular}{ccccccc}
&\multicolumn{2}{c}{$a\alpha=0.2$} & 
\multicolumn{2}{c}{$a\alpha=1/\sqrt{2}$} & 
\multicolumn{2}{c}{$a\alpha=0.8$} 
\\ \hline
 $n$&$\quad\;\;\mathfrak{w}_{{R}}\quad\;\;$ &
$\quad\;\;\mathfrak{w}_{{I}}\quad\;\;$ &
$\quad\;\;\mathfrak{w}_{{R}}\quad\;\;$ &
$\quad\;\;\mathfrak{w}_{{I}}\quad\;\;$ &
$\quad\;\;\mathfrak{w}_{{R}}\quad\;\;$ &
$\quad\;\;\mathfrak{w}_{{I}}\quad\;\;$\\
\hline 
1 & 1.80874 & 2.67375 & 1.88332 & 3.25760 & 2.05936 & 3.56685 \\
2 & 3.08284 & 4.98585 & 3.88903 & 6.73601 & 4.41405 & 7.64535 \\
3 & 4.35865 & 7.34553 & 5.98003 & 10.35772 & 6.84171 & 11.85020 \\
4 & 5.66173 & 9.75472 & 8.09551& 14.02184 & 9.28790 & 16.08711 \\
\end{tabular}
\end{ruledtabular}
\caption{The first four non-hydrodynamic QNMs of the transverse sector for 
$a\alpha=0.2$, $a\alpha=1/\sqrt{2}$, and $a\alpha=0.8$, by setting 
$\mathfrak{q}=\mathfrak{m}=0$.}
\label{taba02}
\end{table*}

The absolute values of the QNM frequencies, plotted in Fig.~\ref{modo1q0}, exhibit
two different regimes. This behavior is confirmed numerically by plotting the slope of
the dispersion relations, given by $\partial_{\mathfrak{m}}\mathfrak{w}$, as shown in the bottom
panels of Fig.~\ref{modo1q0}. A sharp jump for $a\alpha=0.1$ and a 
smooth jump for $a\alpha=0.71$ are observed. Such a result might mean that there exists 
some kind of ``phase transition" in the dual field theory: one phase dominated by a 
typical QNM behavior and another phase by a linear relativistic 
dispersion relation, as claimed in Ref.~\cite{WitczakKrempa:2013ht}.

Figure~\ref{modo1q0} also shows another important consequence of the 
rotation. In the case of $a\alpha=0.1$ a transition from the hydrodynamic-like 
to the relativistic behavior is observed, an effect that was previously 
reported in Ref.~\cite{WitczakKrempa:2013ht} for the QNMs of electromagnetic 
perturbations. As the rotation parameter increases to $a\alpha=1/\sqrt{2}$, 
such a transition is no longer evident. 
In the extremely rotating case, $a\alpha= 1$, 
the dispersion relation becomes linear, meaning that the relativistic behavior 
is dominant and no transition is observed.

To finish the analysis of the transverse-sector, in Fig.~\ref{modo1vara} we 
display the evolution of the first five non-hydrodynamic QNMs as a function of 
the rotation parameter $a\alpha$. For simplicity we choose 
$\mathfrak{m}=0=\mathfrak{q}$. These zero-wavenumber modes are interpreted as 
gravitational waves propagating in the radial direction. Additional information 
on the behavior of these modes in terms of the rotation parameter can be seen in 
table~\ref{taba02}, where we write the frequencies of the first four 
non-hydrodynamic modes for some selected values of the rotation parameter. It is 
clearly seen that the frequency grows with $a\alpha$.

\begin{figure*}[ht!]
\begin{tabular}{*{2}{>{\centering\arraybackslash}p{.45\textwidth}}}
\includegraphics[width=7.5cm,angle=0]{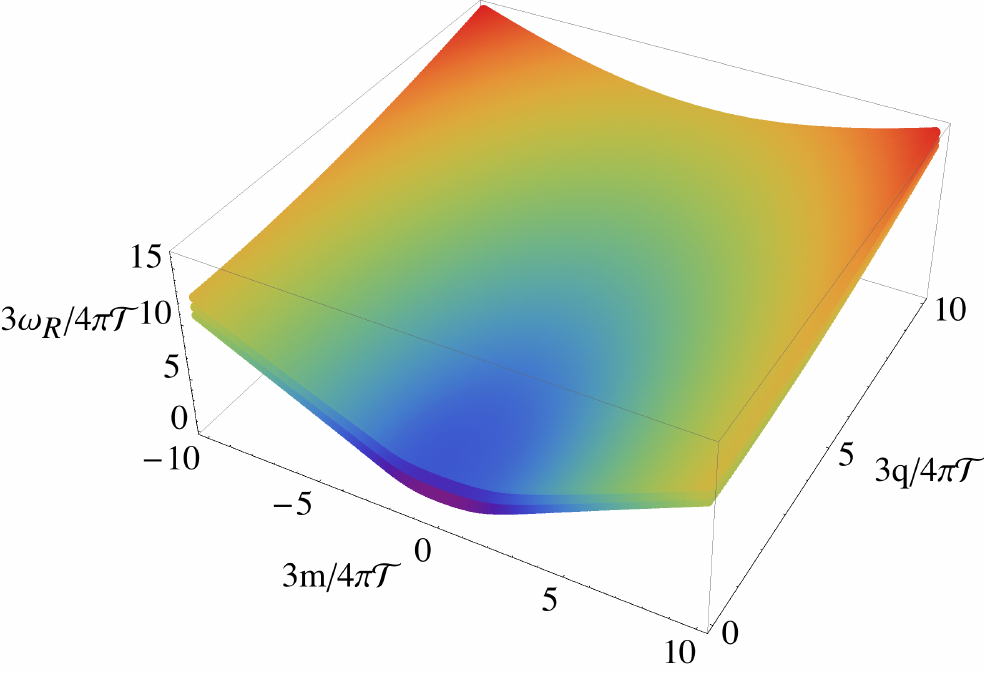}&
\includegraphics[width=7.5cm,angle=0]{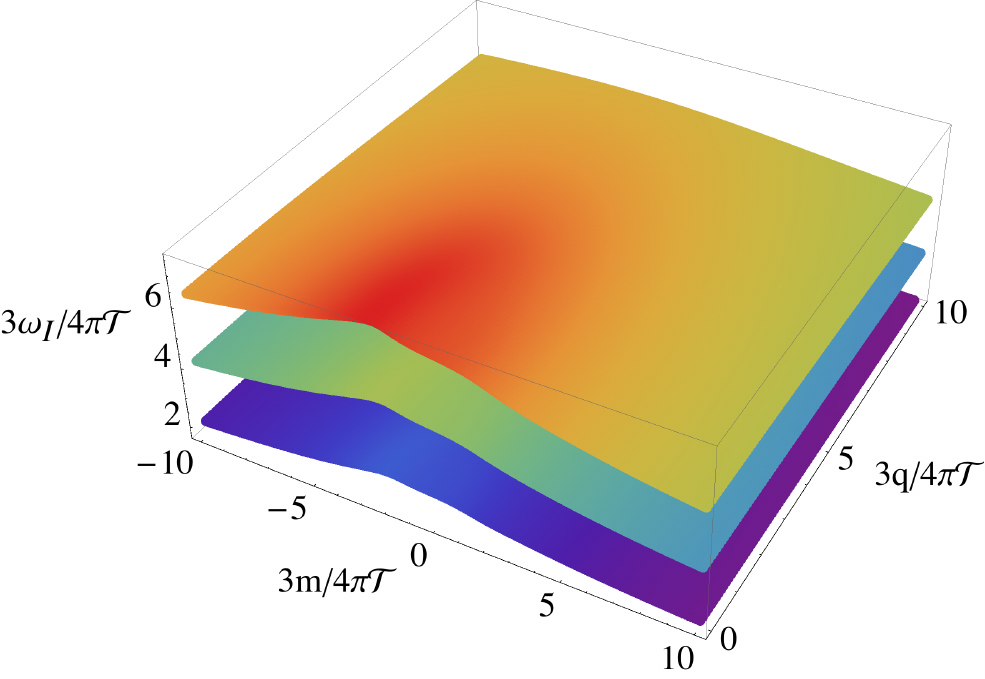}\\
\end{tabular}
\caption{\label{scalarmodo1vara} 
The first three non-hydrodynamic modes of the 
longitudinal sector of perturbations. The left (right) panel shows the real 
(imaginary) part of the frequency. These results were obtained by setting 
$a\alpha=0.1$.}
\end{figure*}

\begin{table*}[ht!]
\begin{ruledtabular}
\begin{tabular}{ccccccc}
&\multicolumn{2}{c}{$a\alpha=0.2$} & 
\multicolumn{2}{c}{$a\alpha=1/\sqrt{2}$} & 
\multicolumn{2}{c}{$a\alpha=0.8$} 
\\ \hline
 $n$&$\quad\;\;\mathfrak{w}_{{R}}\quad\;\;$ &
$\quad\;\;\mathfrak{w}_{{I}}\quad\;\;$ &
$\quad\;\;\mathfrak{w}_{{R}}\quad\;\;$ &
$\quad\;\;\mathfrak{w}_{{I}}\quad\;\;$ &
$\quad\;\;\mathfrak{w}_{{R}}\quad\;\;$ &
$\quad\;\;\mathfrak{w}_{{I}}\quad\;\;$\\
\hline 
1 & 1.80350 & 2.66783 & 1.24011 & 2.17621 & 1.16478 & 2.01591\\
2 & 3.05672 & 4.94665 & 2.89498 & 5.01419 & 3.22866 & 5.59221\\
3 & 4.28620 & 7.22177 & 4.93568 & 8.54886 & 5.62604 & 9.74459\\
4 & 5.50904 & 9.47953 & 7.03808 & 12.19031& 8.06416 & 13.96753\\
\end{tabular}
\end{ruledtabular}
\caption{The first four non-hydrodynamic QNMs of the longitudinal sector with 
zero wavenumbers ($\mathfrak{q}=\mathfrak{m}=0$) for $a\alpha=0.2$, 
$a\alpha=1/\sqrt{2}$, and $a\alpha=0.8$.}
\label{taba03}
\end{table*}
\begin{figure*}[ht!]
\begin{tabular}{*{2}{>{\centering\arraybackslash}p{.45\textwidth}}}
\includegraphics[width=7.5cm]{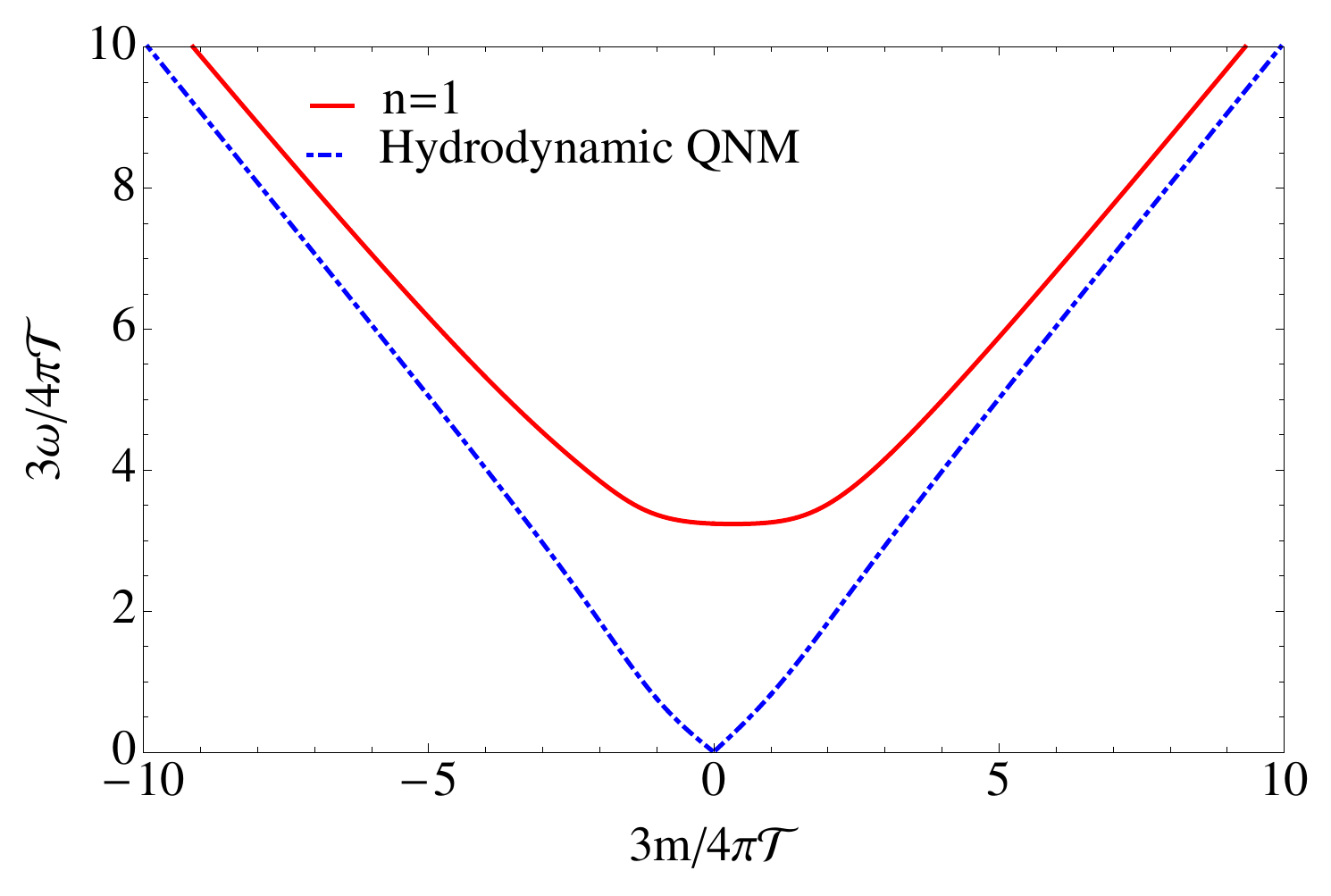}&
\includegraphics[width=7.5cm]{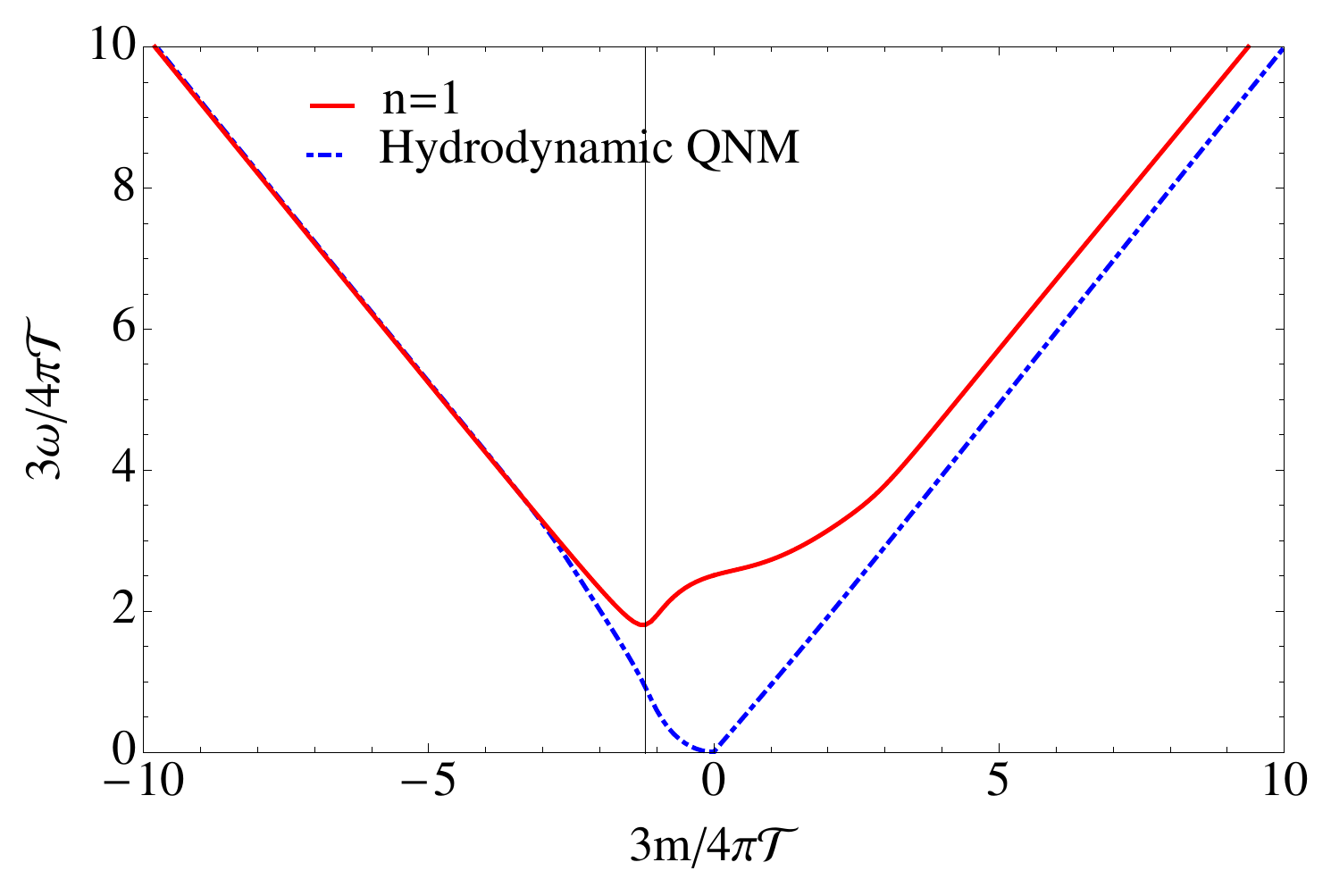}\\
\includegraphics[width=7.5cm]{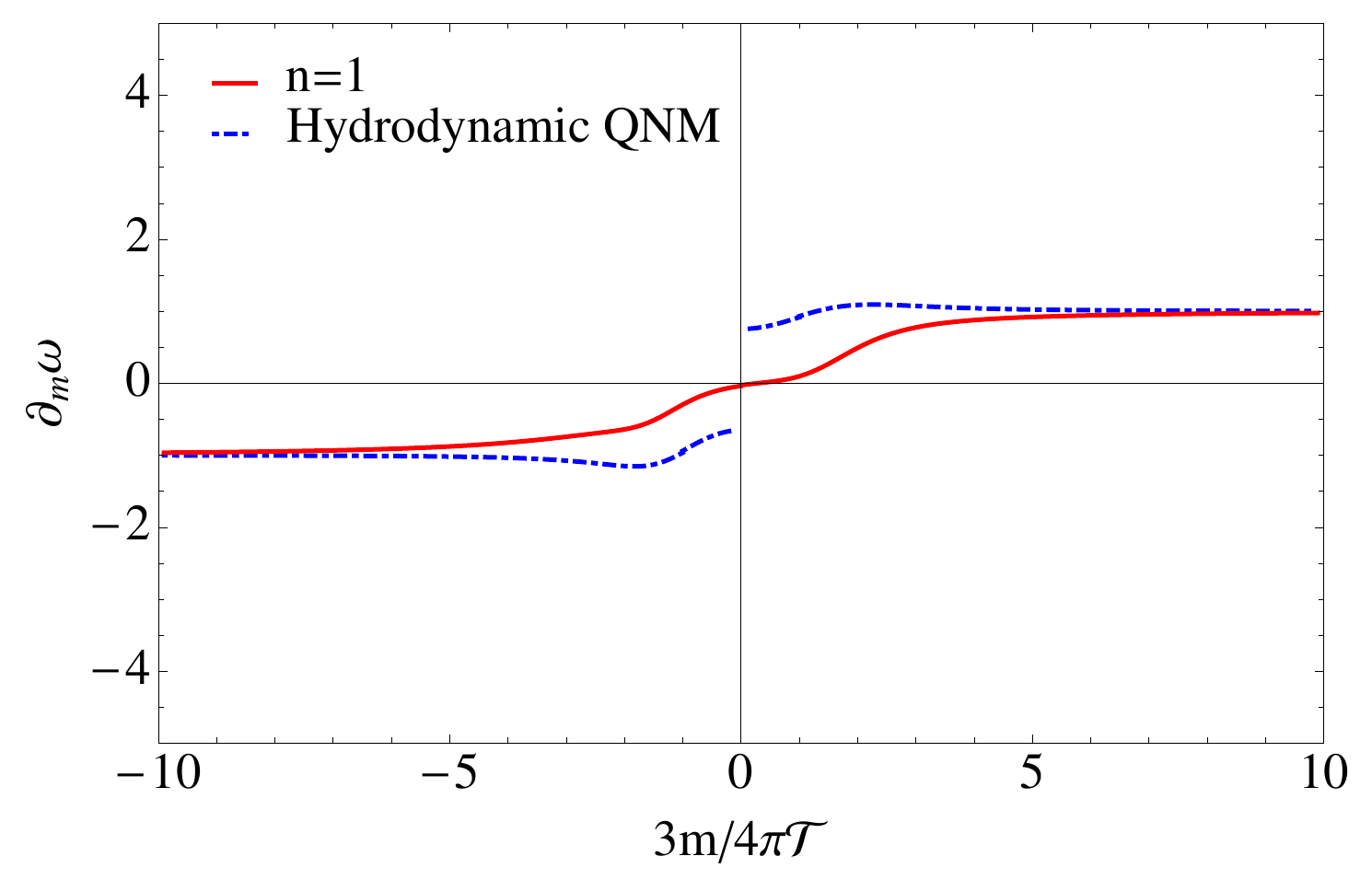} &
\includegraphics[width=7.5cm]{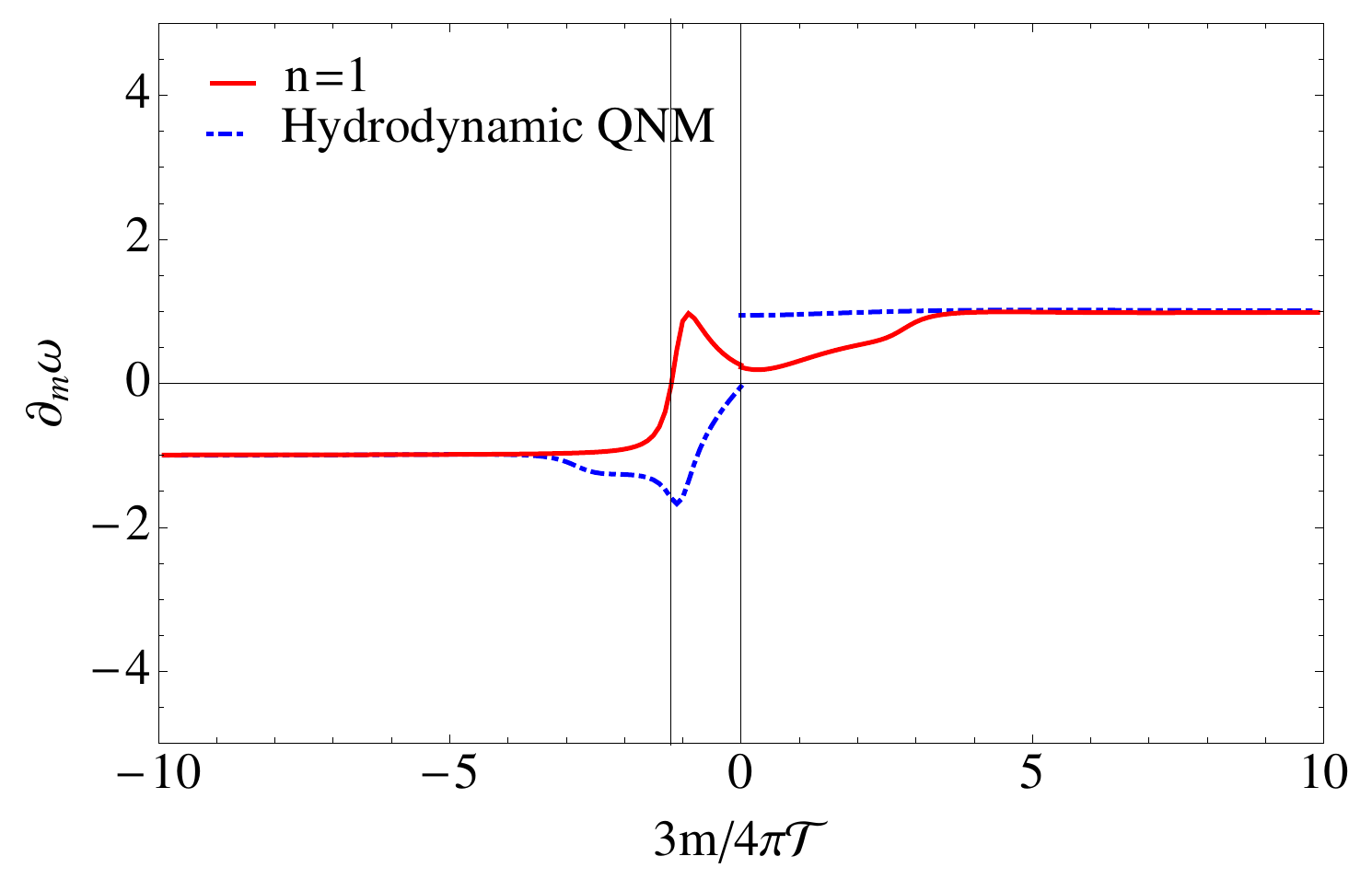}\\
\end{tabular}
\caption{The top panels show the absolute value of the frequency, 
$\mathfrak{w}=\sqrt{\mathfrak{w}_R^2 + \mathfrak{w}_I ^2}$,  for the 
hydrodynamic and the first non-hydrodynamic QNMs
of the longitudinal sector as a function of $\mathfrak m$, 
with $\mathfrak{q}=0$, and for two values of the rotation parameter:
$a\alpha=0.1$ (top left panel) and 
$a\alpha=1/\sqrt{2}$ (top right panel). The bottom panels show
the slope of the curves presented in the top panels.
The right panels show a vertical line at
$\mathfrak{m}=-1.2$.}
\label{polarmodo1q0}
\end{figure*}
\begin{figure*}[ht!]
\begin{tabular}{*{2}{>{\centering\arraybackslash}p{.45\textwidth}}}
\includegraphics[width=7.5cm]{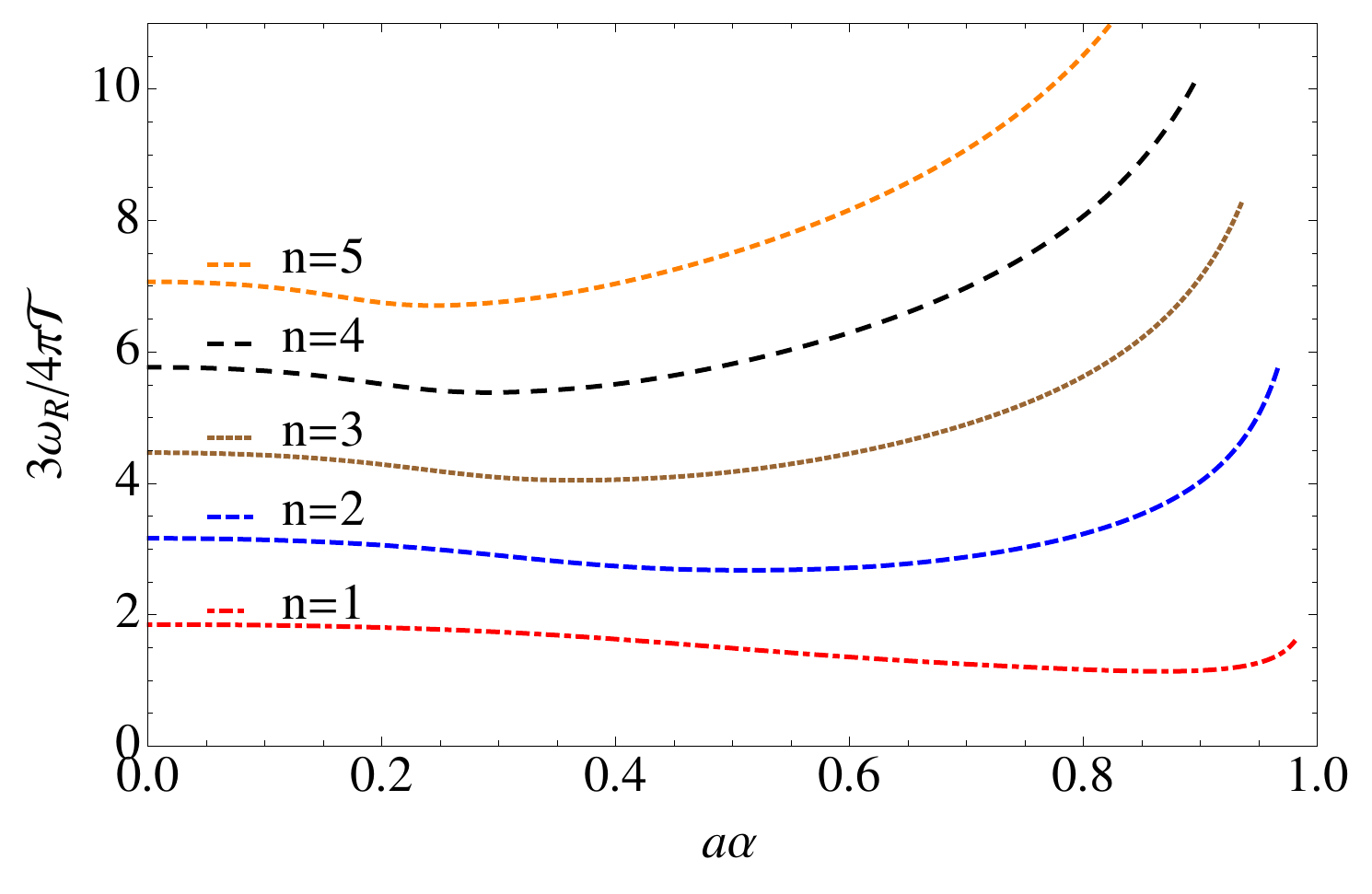}&
\includegraphics[width=7.5cm]{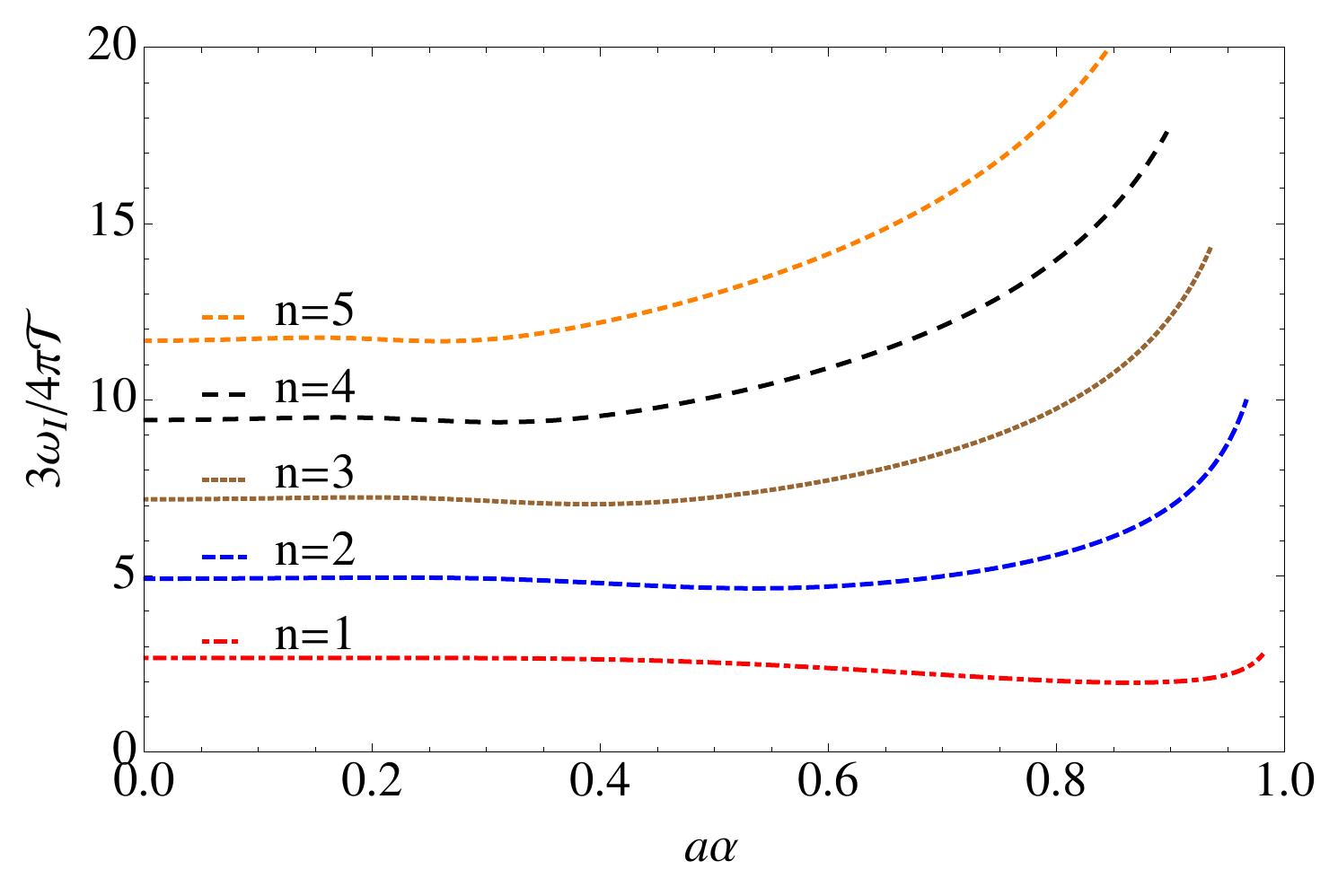} \\
\end{tabular}
\caption{The first five non-hydrodynamic QNMs of the longitudinal sector 
as a function of the rotation parameter for $\mathfrak{m}=\mathfrak{q}=0$.}
\label{scalarmodo1vara1}
\end{figure*}

\subsection{Longitudinal sector}

The non-hydrodynamic QNM frequencies of the longitudinal sector are 
obtained by solving numerically the differential 
equation \eqref{fund-eq-scalar}. Here we solve this
equation for wavenumber values in the intervals 
$\mathfrak{m}\in [-10,10]$ and $\mathfrak{q}\in [0,10]$.
We display the numerical
results of the first three non-hydrodynamic modes 
in Fig.~\ref{scalarmodo1vara} for 
$a\alpha=0.1$. As reported in Ref.~\cite{Miranda:2008vb}, 
the real and imaginary parts of the frequencies of the static
black string are symmetric under the changes $\mathfrak{m}\to -\mathfrak{m}$
and $\mathfrak{q}\to -\mathfrak{q}$. 
Here, the results show that the symmetry $\mathfrak{m}\to -\mathfrak{m}$
is broken because the rotation introduces a preferred direction. 
The asymmetry in the real part of the frequency may be interpreted as a Doppler shift.

Differently from the transverse sector, cf. Fig.~\ref{modosaxial}, no 
interconnections between adjacent modes were observed. In Table \ref{taba03} we 
write selected values of the frequency of the first four non-hydrodynamic 
modes; to obtain these results we set $\mathfrak{m}=0=\mathfrak{q}$.

As it was done for the transverse sector, to get additional information of the 
numerical results we plot slices of the absolute value of the frequency as a 
function of $\mathfrak m$ for $\mathfrak{q}=0$. In the left top panel of 
Fig.~\ref{polarmodo1q0} we plot the hydrodynamic and the first non-hydrodynamic 
quasinormal frequencies for $a\alpha=0.1$. We also calculate the slope of the curves by means of 
numerical approximations for $\partial_{\mathfrak{m}}{\mathfrak{w}}$. 
The results are displayed in the 
left bottom panel of Fig.~\ref{polarmodo1q0}. 
For a (non-)hydrodynamic QNM, it is observed a smooth transition between a
(non-)hydrodynamic like behavior for small wavenumbers to a
linear relativistic behavior in the regime of large wavenumbers.

The right top panel of Fig.~\ref{polarmodo1q0} displays the results for 
$a\alpha=1/\sqrt{2}$. This figure shows the hydrodynamic 
and the first non-hydrodynamic quasinormal modes along the $\varphi$ direction.
The slope $\partial_{\mathfrak{m}}\mathfrak{w}$ was 
also calculated and the result is displayed in the bottom right panel of
that figure. The plot shows a smooth transition from the small to the large wavenumber regimes.

To complete the analysis of the longitudinal sector we show in Fig. \ref{scalarmodo1vara1} 
the evolution of the first five non-hydrodynamic QNMs as a function of the rotation parameter for $\mathfrak{m}=0=\mathfrak{q}$. The behavior of these modes with the 
rotation parameter is similar to the case of the transverse sector.
\begin{figure*}[ht!]
\begin{tabular}{*{2}{>{\centering\arraybackslash}p{.45\textwidth}}}
\includegraphics[width=7.5cm]{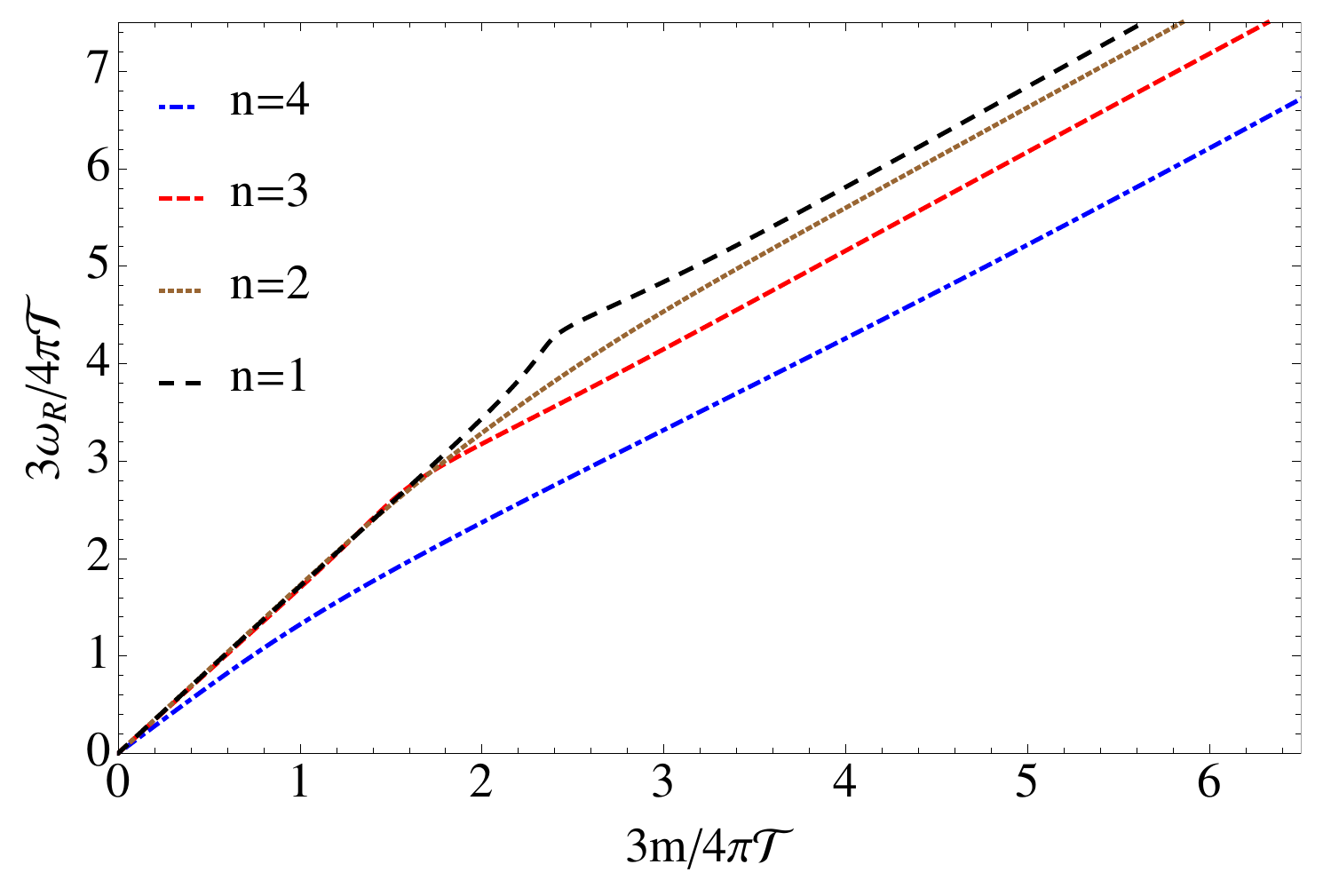}&
\includegraphics[width=7.5cm]{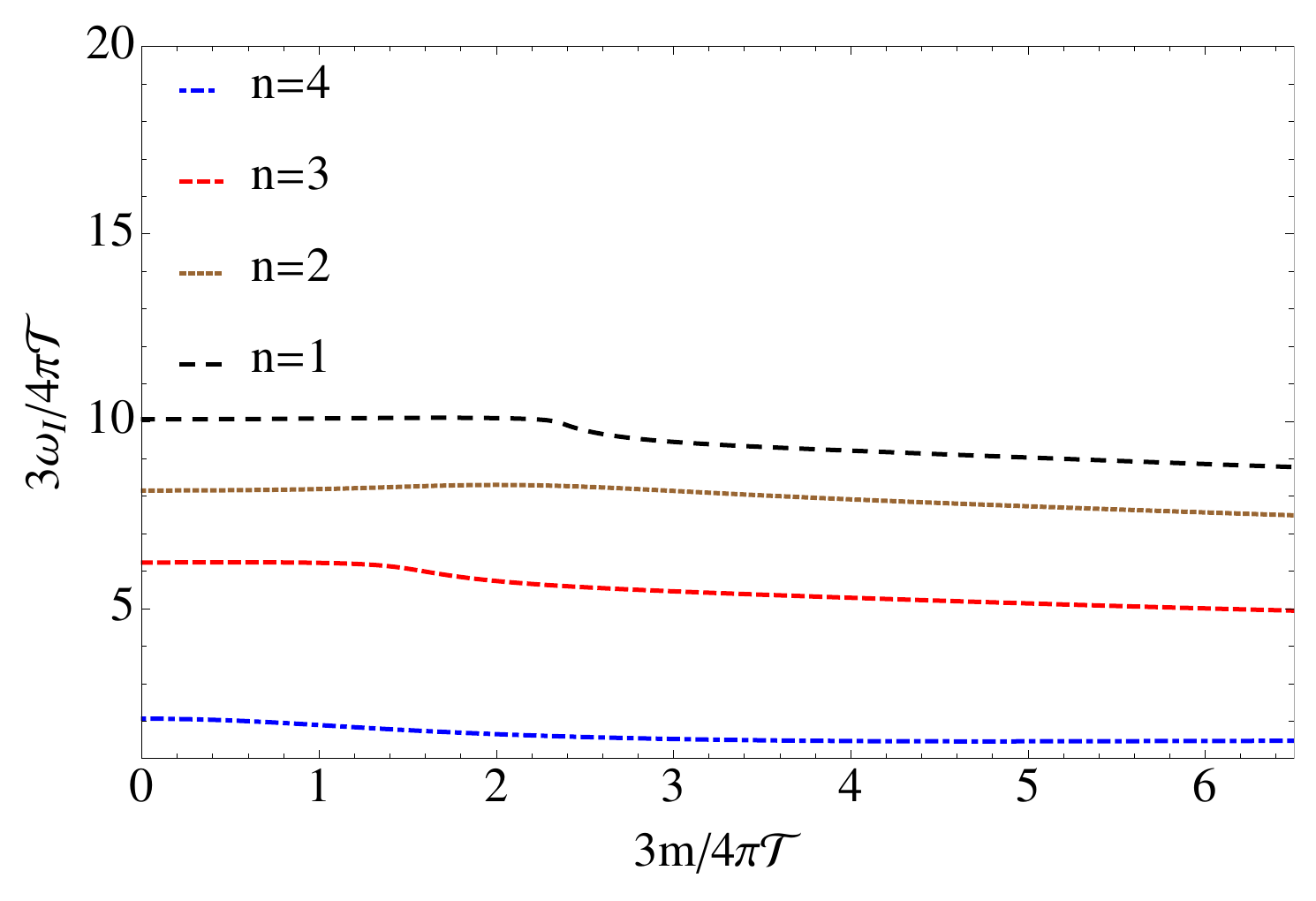}\\
\includegraphics[width=7.5cm]{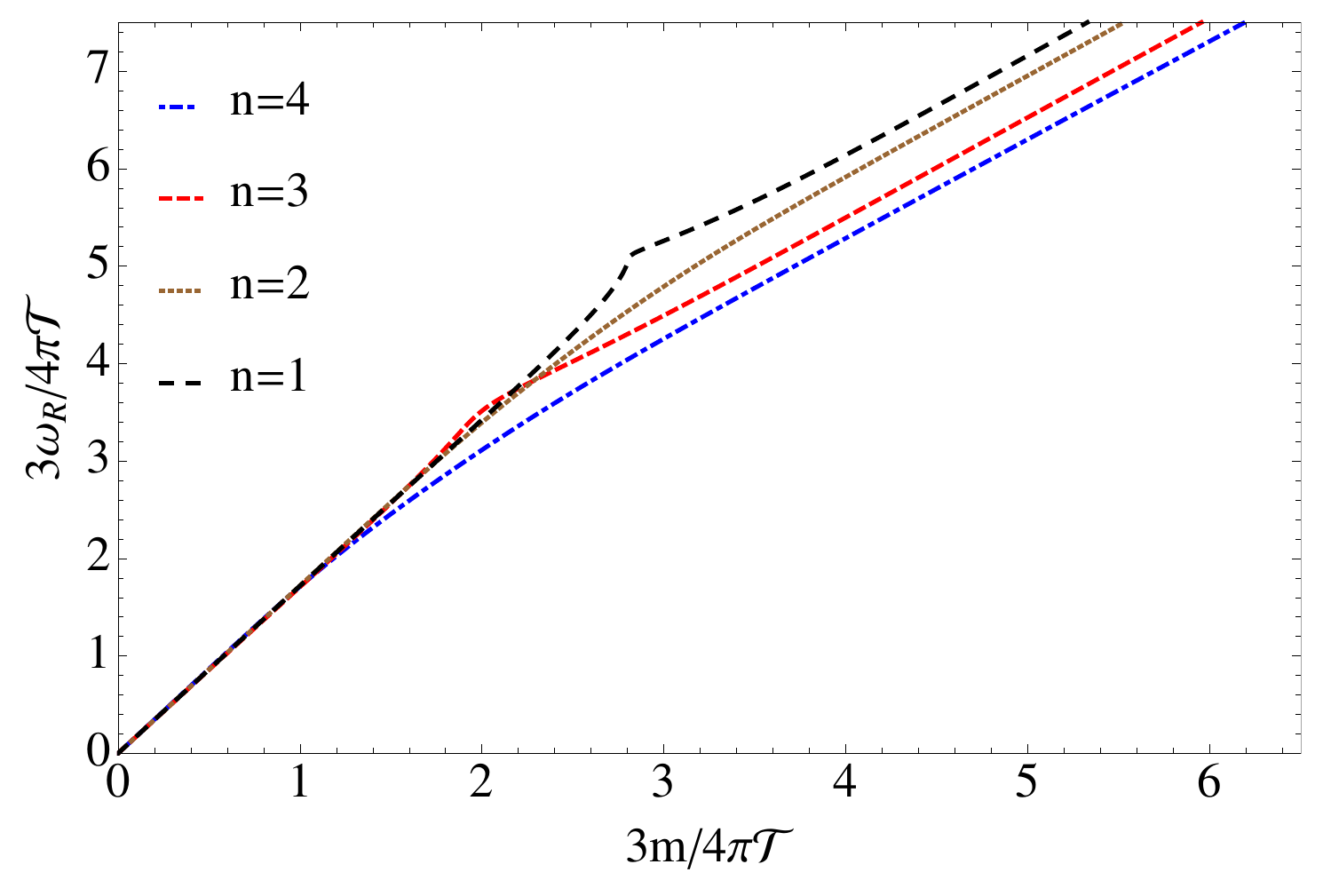}&
\includegraphics[width=7.5cm]{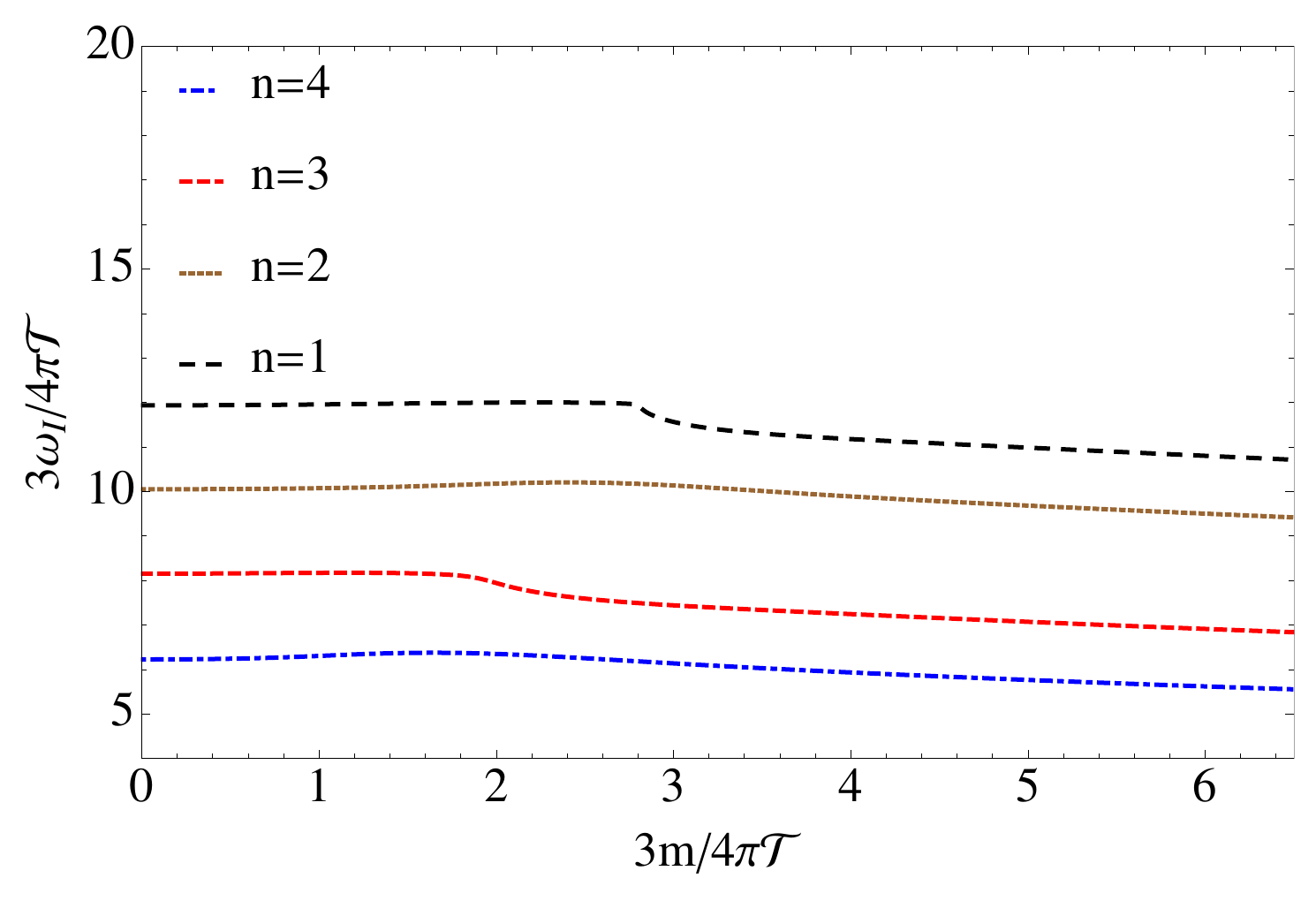}\\
\end{tabular}
\caption{First four modes that emerged in the transverse (top panel)
and longitudinal (bottom panel) sector, 
as a function of the wavenumber $\mathfrak{m}$ 
for $\mathfrak{q}=0$ and $a\alpha=1/\sqrt{2}$.}
\label{NewmdesEvolu}
\end{figure*}
\begin{figure*}[ht!]
\begin{tabular}{*{2}{>{\centering\arraybackslash}p{.45\textwidth}}}
\includegraphics[width=7.5cm]{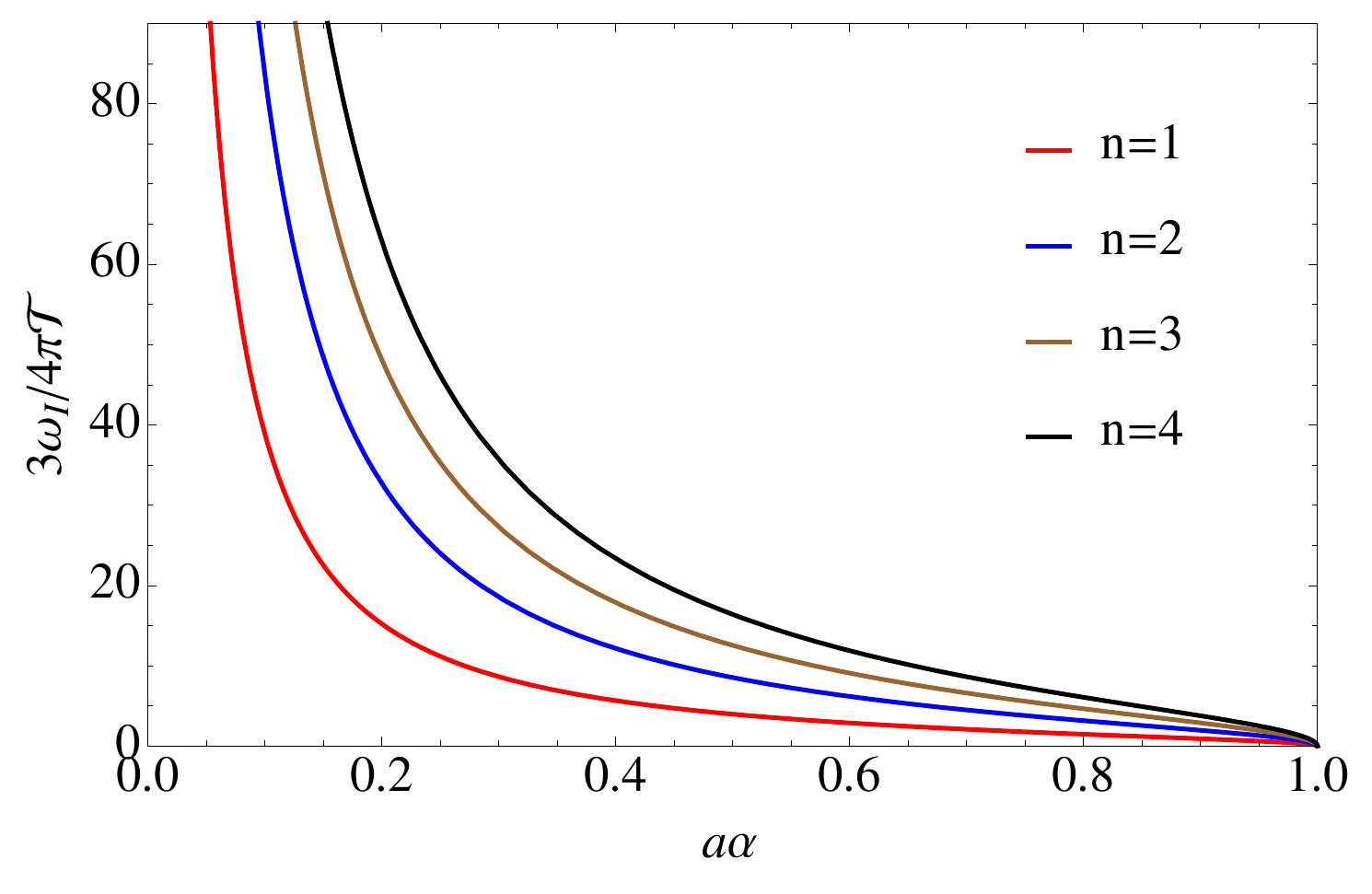}&
\includegraphics[width=7.5cm]{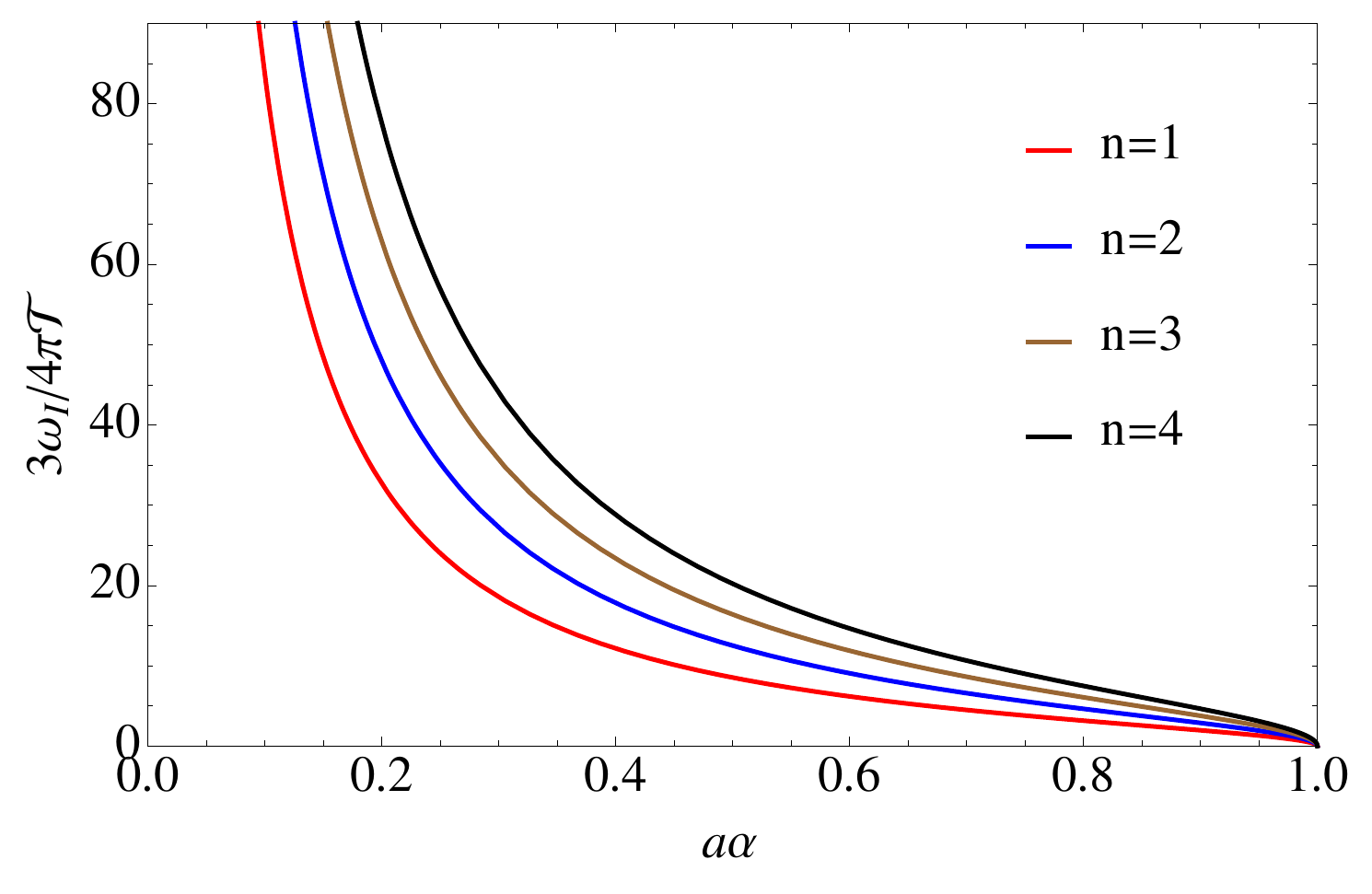}\\
\end{tabular}
\caption{The frequencies of the first fourth modes that emerged in the transverse 
(left panel) and longitudinal (right panel) sectors as a function 
of the rotation parameter for $\mathfrak m=\mathfrak q=0$. }
\label{fig-newmodes}
\end{figure*}

\section{Emergence of a new QNM class and the algebraically special modes}
\label{sec-purelyDamped}

As commented previously, the QNMs of the rotating black
strings are closely related to those of the static black strings. 
Therefore, for most of the modes, the dispersion relations of the static
black string QNMs can be recovered from the dispersion relations
of the rotating black string QNMs in the limit of $a\alpha\rightarrow 0$.
However, during the search for non-hydrodynamic QNMs, new
solutions arise in both sectors of the gravitational perturbations.
These new frequencies do not have any similar in previous
studies on static black strings (see, for instance, 
Refs.~\cite{Miranda:2005qx,Miranda:2008vb,Morgan:2009pn}), and
arise when the numerical search for QNMs is performed close the imaginary axis.
This is true, in particular, for zero wavenumbers $\mathfrak{m}=0=\mathfrak{q}$.
Figure~\ref{NewmdesEvolu} displays the evolution of the real and imaginary parts
of the frequency of the transverse (top panel)
and longitudinal (bottom panel) sectors, respectively, as a function
of the wavenumber component $\mathfrak m$.

Another characteristic property of these modes is the behavior close
to the static limit $a\to 0$. In order to see that 
we plot in Fig.~\ref{fig-newmodes} the first four 
QNMs of the transverse (left panel) and longitudinal 
(right panel) sectors as a function
of the rotation parameter. These frequencies raise to infinity when 
$a\alpha \to 0$, i.e., the value of the frequency diverges when the
rotation parameter goes to zero. In fact, this is the reason 
why these modes are not found in the static black-string analysis.
The numerical results indicate the existence of an infinite sequence of such modes.

It is worth mentioning at this point that it has appeared in the literature a
few works studying purely damped modes of the Kerr black hole in
asymptotically flat spacetimes \cite{Cook:2016fge,Cook:2016ngj}.
They report on the appearance of frequencies that have 
the same behavior as presented in Fig.~\ref{fig-newmodes}, but 
no physical interpretation was given for such perturbation modes.
On the other hand, in the present case of rotating black strings in
an asymptotically AdS background,
the solutions have a clear physical interpretation, namely, they are
gravitational quasinormal modes. 
In the holographic context, it was recently obtained a similar result in the
study of perturbations of a five-dimensional 1-R charged black hole close to 
a critical point \cite{Finazzo:2016psx}.

\begin{figure*}[ht!]
\begin{tabular}{*{2}{>{\centering\arraybackslash}p{.45\textwidth}}}
\includegraphics[width=7.5cm]{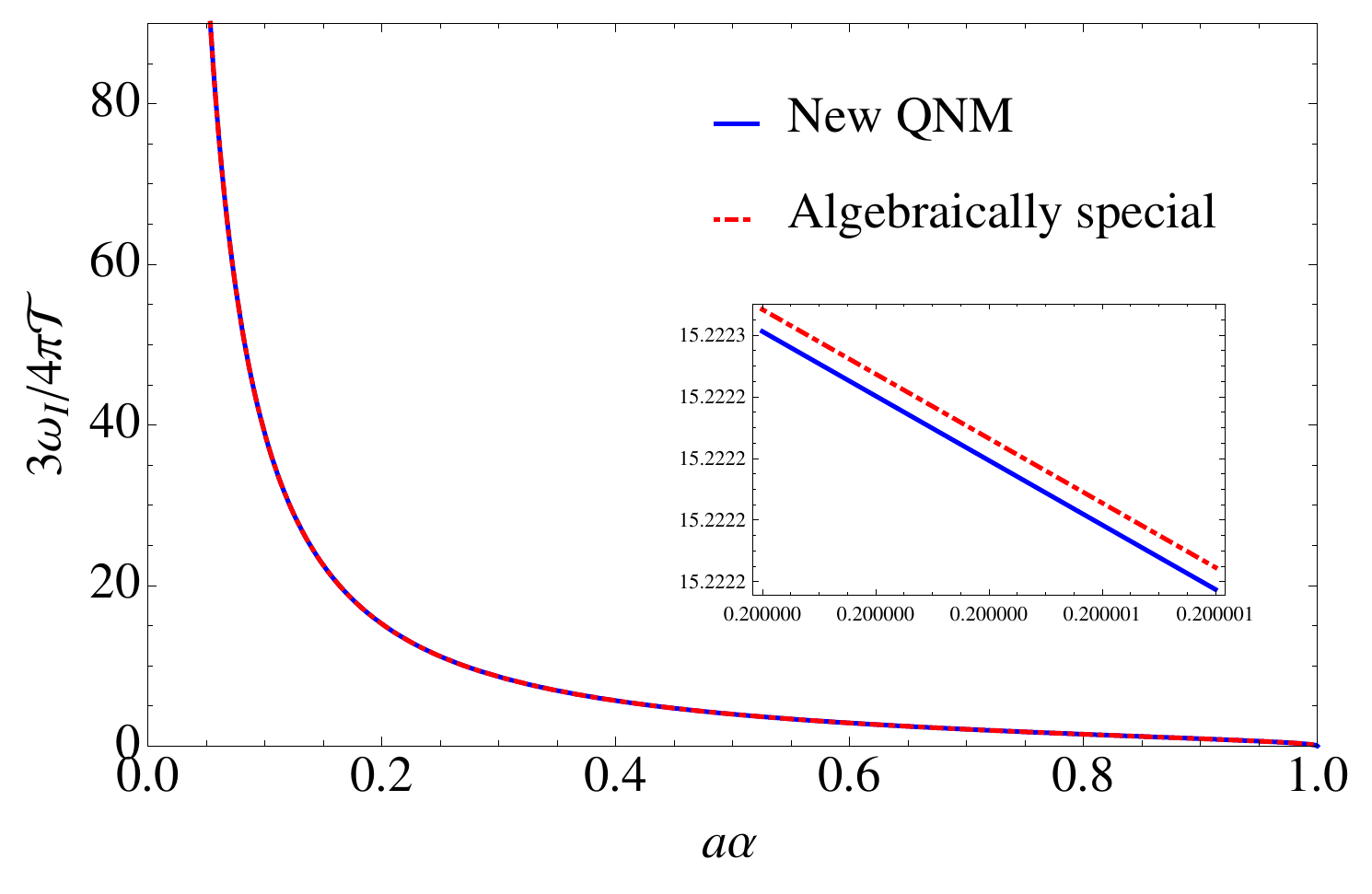}&
\includegraphics[width=7.5cm]{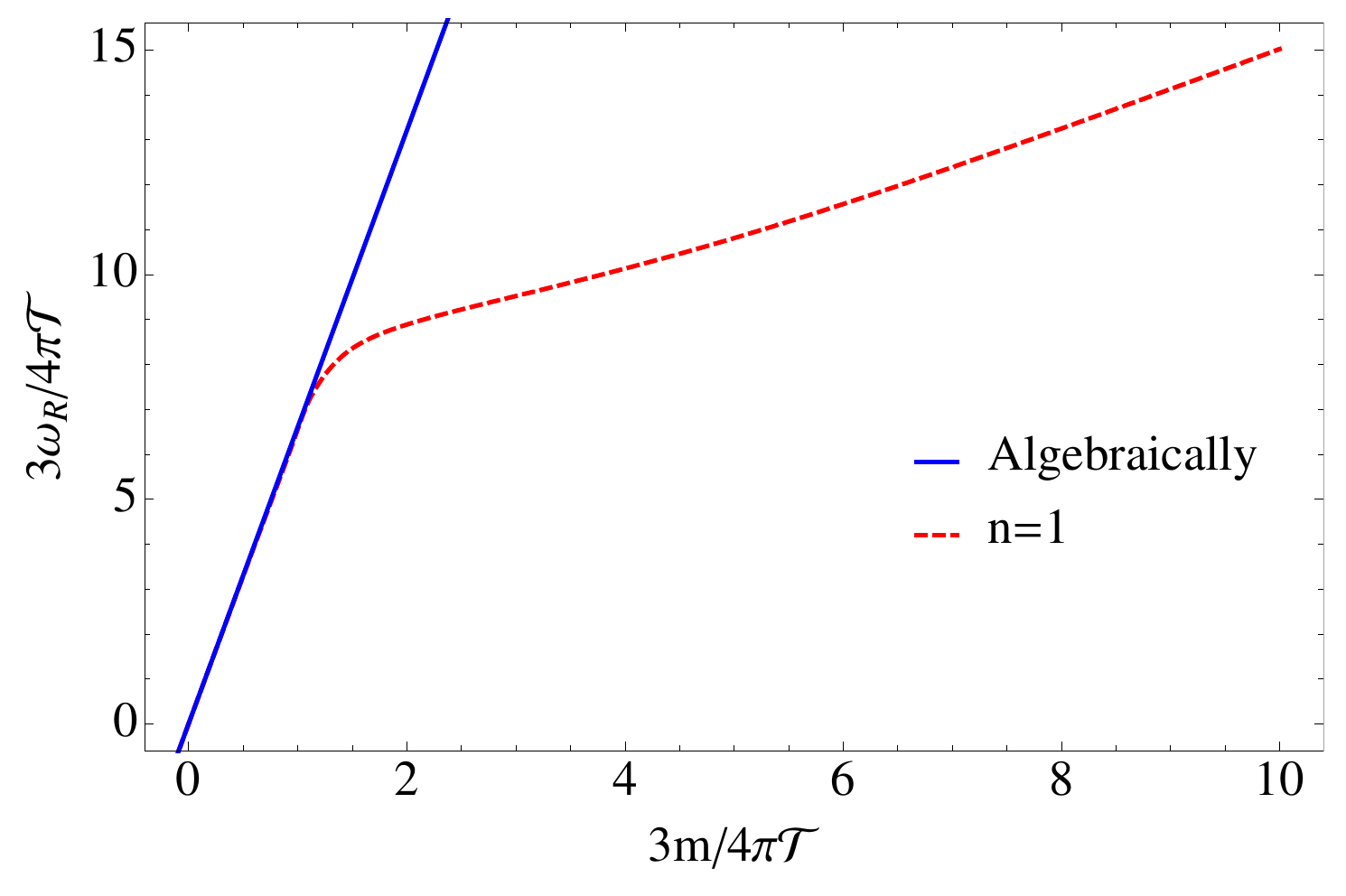} \\
\end{tabular}
\caption{Left panel: The algebraically
special frequency (dashed line) and the first new quasinormal 
mode (solid line) as a function of the rotation parameter.
Right panel: The dependence of the real parts of the frequencies of
the first new purely damped mode (dashed line), and 
of the algebraically special
mode (solid line) with the wavenumber value $\mathfrak{m}$. 
To get the results shown on the left panel
we set $\mathfrak{m}=0=\mathfrak{q}$, while to get the results on
right panel we set $\mathfrak q=0$ and $a\alpha=0.2$.}
\label{AlgebraEvol}
\end{figure*}

In complement to the previous analysis, we now explore a possible relationship
between the modes of Fig.~\ref{fig-newmodes} and the algebraically 
special modes \cite{Chandrasekhar:1984,Chee:1994,MaassenvandenBrink:2000iwh}. 
For the definition and more details on the algebraically special perturbations 
in the Kerr and Schwarzschild black holes see, 
e.g., \cite{Couch:1973,Wald:1973,Onozawa:1997,Berti:2003,Dias:2013hn}. 
It is verified numerically that the first ($n=1$) QNM frequency of the transverse sector
(left panel in Fig.~\ref{fig-newmodes}) approaches one of the algebraically
special frequencies in the limit of small values of the rotation parameter.
This is shown in the left panel of Fig.~\ref{AlgebraEvol}. 
In order to verify such a relationship, we start reviewing the dispersion 
relation of an algebraically special mode. 
As found in previous works \cite{Miranda:2008vb}, static black strings possess 
an algebraically special frequency given by
\begin{equation}\label{AlgebraEq}
\mathfrak{\bar w}=-\frac{i}{6}\left(\mathfrak{\bar q}^2+ 
\mathfrak{\bar m}^2\alpha^2\right)^2.
\end{equation}
Hence, by replacing relations Eq.~\eqref{boost2} into Eq.~\eqref{AlgebraEq} 
we get some of the algebraically special frequencies of the 
rotating black strings. Such frequencies now have nontrivial dispersion relations 
\begin{equation}\label{algebraically2}
\gamma\mathfrak{w}=
-i\frac{\gamma^4\alpha^4\,(\mathfrak{m}-a\mathfrak{w})^4}{6},
\end{equation}
where to simplify the analysis we have set $\mathfrak{q}=0$.
Solving Eq.~\eqref{algebraically2} for $\mathfrak{w}$ and
expanding the solutions around $\mathfrak{m}=0$, we get 
\begin{equation}
\label{algebraically31}
\begin{aligned}
\mathfrak{w}^{a}_{1}
=&\frac{{6}^{1/3}\,e^{i\,5\pi/6}}{{(a\alpha)^{4/3}\, 
\gamma}}+\left(\frac{4}{3a\alpha}-\frac{a\alpha}{3}\right)
\mathfrak{m}\alpha+ 
\\
&+\frac{2^{2/3}\,e^{i\pi/6}}
{3^{7/3}\,(a\alpha)^{2/3}\,\gamma^3}(\mathfrak{m}\alpha)^2+\cdots\,,
\end{aligned}
\end{equation}
\begin{equation}\label{algebraically32}
\begin{aligned}
\mathfrak{w}^{a}_{2}=
&-\frac{{6}^{1/3}\,e^{i\,\pi/2}}{{(a\alpha)^{4/3}\, \gamma}}+
\left(\frac{4}{3\,a\alpha}-\frac{a\alpha}{3}\right)\mathfrak{m}\alpha
\\
&-\frac{2^{2/3}\,e^{i\,\pi/2}}
{3^{7/3}\,(a\alpha)^{2/3}}(\mathfrak{m}\alpha)^2+\cdots\,,
\end{aligned}
\end{equation}
\begin{equation}
\begin{aligned}
\label{algebraically33}
\mathfrak{w}^{a}_{3}=&a\alpha(\mathfrak{m}\alpha)
-\frac{e^{i\pi/2}}{6\, \gamma^5}(\mathfrak{m}\alpha)^4+\cdots\,,
\end{aligned}
\end{equation}
\begin{equation}
    \begin{aligned}
\label{algebraically34}
\mathfrak{w}^{a}_{4}
=&\frac{{6}^{1/3}\,e^{i\,\pi/6}}{{(a\alpha)^{4/3}\,\gamma}}
+\left(\frac{4}{3a\alpha}-\frac{a\alpha}{3}\right)\mathfrak{m}\alpha+
\\
&+\frac{2^{2/3}\,e^{i\,5\pi/6}}
{3^{7/3}\,(a\alpha)^{2/3}\,\gamma^3}(\mathfrak{m}\alpha)^2+\cdots\,,
\end{aligned}
\end{equation}
where the ellipses stand for higher order corrections and the superscript $a$ 
refers to the algebraically special frequencies. From these results we observe 
the existence of non-vanishing real and imaginary parts of the frequencies 
$\mathfrak{w}^{a}_{i}\, (i=1,\, ..., \,4)$, contrary to the static case, where 
the algebraically special frequencies are purely imaginary numbers. We discard 
the solutions $\mathfrak{w}^{a}_{1}$ and $\mathfrak{w}^{a}_{4}$ because they 
have negative imaginary parts, representing unstable modes. We also discard 
$\mathfrak{w}^{a}_{3}$ because it does not present the same behavior of the 
modes of Fig. \ref{fig-newmodes}, whose frequencies are large for small rotation 
parameters. The remaining solution, $\mathfrak{w}^{a}_{2}$, assumes large values 
for small rotation parameters and the imaginary part is positive. 
It is worth noticing that this solution becomes 
purely damped when the wavenumber $\mathfrak{m}$ is zero. This particular 
solution is important because it coalesces into the first purely damped 
transverse QNM in the limit of zero angular momentum,  $a\alpha\to 0$. In 
fact, this singular behavior was our guide to try to connect the new-class 
frequencies to the algebraically special frequencies. We realize that the first 
new transverse-sector quasinormal mode is very close to the algebraically 
special mode for small values of the rotation parameter, deviating from it as 
the rotation parameter increases. Quantitatively the difference between these 
two frequencies can be seen in Table~\ref{tab-newmodes}. These results are also 
shown in the left panel of Fig.~\ref{AlgebraEvol}.
\begin{table}[ht!]
\begin{ruledtabular}
\begin{tabular}{ccc}
 $a\alpha$&
 $\quad\;\;\mathfrak{w}_{{I}} (\text{new mode})\quad\;\;$ 
 & $\quad\;\;\mathfrak{w}_{{2}}^{a} 
 (\text{algebraic mode})\quad\;\;$ \\
\hline 
0.1 & 38.95244 & 38.95244 \\
0.3 & 8.63127 & 8.63130 \\
0.5 & 3.97043 & 3.96541 \\
0.7& 2.10127 & 2.08788 \\
0.9& 1.03380 & 0.91153 \\
0.999& 0.09796 & 0.08135 \\
\end{tabular}
\end{ruledtabular}
\caption{The imaginary parts of the frequencies of the new-class
transverse-sector QNMs and the algebraically special
modes with $\mathfrak{q}=\mathfrak{m}=0$.}
\label{tab-newmodes}
\end{table}

Here we highlight the existence of an algebraically special mode of frequency 
\eqref{AlgebraEq} associated to the longitudinal perturbations. However, such a 
mode does not appear in the QNM spectrum of the longitudinal (polar) 
perturbations of a static black string \cite{Cardoso:2001vs,Miranda:2005qx}. 
Hence, as it could be expected, it was not found a connection between the 
frequencies \eqref{algebraically31}-\eqref{algebraically34} and the new-class 
QNMs of the longitudinal perturbations.

At this point one could ask if there exists a relation between the 
algebraically special frequencies, $\mathfrak{w}^{a}_{2}$, and the new 
family of QNMs for non-vanishing values of $\mathfrak{m}$ (or $\mathfrak{q}$).
In order to answer such a question, we plot in Fig.~\ref{AlgebraEvol} the joint
results of the real part of the frequency of the algebraically special mode and
the first new-class mode (right panel) as a
function of the wavenumber component $\mathfrak{m}$. For small wavenumber values, the 
slope of the dispersion relation curve is precisely the coefficient of the real 
part of the frequency in Eq.~\eqref{algebraically32}, and it is the same 
for both frequencies. 
This is true at least for values of the rotation parameter smaller than 
$a\alpha=0.5$. Doing the same analysis for several values of the rotation 
parameter (we do not present the whole results here), we arrive at the 
conclusion that, at least for the first new mode, the modes are numerically 
indistinguishable. Contrary to what happens in the static case, where the 
hydrodynamic QNM approaches asymptotically to the algebraically special 
frequency for large wavenumber values, here the opposite situation is 
verified. The first new QNM, $n=1$, approaches to the algebraically special 
frequency as the rotation parameter decreases, cf. Fig.~\ref{fig-newmodes}. We 
also realize that the evolution of the first new QNM frequency as a function of 
the wavenumber is approximately the same as the algebraically special frequency 
for small values of the wavenumber, as it can be seen in the right panel of 
Fig.~\ref{AlgebraEvol}.

We cannot say much about the new modes in the 
longitudinal sector because, in this case, we do not have an analytical
expression for the static black string as the agebraically special frequency
\eqref{AlgebraEq}. However, the behavior of the new QNMs of the longitudinal
perturbations seems to be similar to those of the transverse-sector QNMs.

\section{Final comments and conclusion}
\label{secfinal}

We developed a detailed analysis of the gravitational quasinormal modes
of rotating black strings. Firstly we explored the relation between the 
Regge-Wheeler-Zerilli (RWZ) and Kovtun-Starinets (KS) gauge-invariant quantities 
to get the equations of motion in the KS variables.
After that, we obtained solutions of the differential equations
for the hydrodynamic quasinormal modes through analytic and numeric methods.
The analytic solutions show the corrections in the dispersion relations due to rotation.
An example is the appearance of a nonzero term associated to convection
in the real part of the shear mode frequency.
On the other hand, the sound wave mode has an intricate dispersion relation due to the
mixture of relativistic effects like the Lorentz contraction, the time dilation, and the transformation
of the sound velocity, which become evident by considering some particular cases, cf. for instance, 
Figs.~\ref{DampingAxial2} and \ref{Sound2}. In both sectors, the Doppler effect 
appears in our results as a consequence of the finite angular momentum of the black string.

The non-hydrodynamic QNMs were investigated by means of numerical methods. We 
found that the transverse sector exhibits an interconnection between successive 
modes, cf. Figs.~\ref{modosaxial} and \ref{modo1q0a}. Additional information was 
obtained by calculating the absolute values of the frequencies, where it was 
observed a crossover from a (non-)hydrodynamic like behavior to the linear 
relativistic scaling for small values of the rotation parameter, cf. the top 
left panel in Fig.~\ref{modo1q0}. By increasing the value of the  rotation 
parameter, the crossover is no longer observed, cf. top right panel in 
Fig.~\ref{modo1q0}. This result is expected since for large values of the 
rotation parameter the system presents a typical relativistic behavior. The 
results for the longitudinal sector do not present an interconnection between 
successive modes. 

Additionally, we have found a new class of quasinormal frequencies that does not 
have a similar counterpart in the QNM spectra of static black strings. These 
modes appear in both sectors and may be identified due to their peculiar 
dependence on the rotation parameter, cf. Fig.~\ref{fig-newmodes}. In the 
transverse sector, the first mode of this new family of QNMs has a similar 
behavior as the algebraically special frequency, and approaches the last one in 
the limit of zero rotation parameter.  We also have shown the evolution of the 
first new mode with the wavenumber, cf. the right panel of
Fig.~\ref{AlgebraEvol}. It is worth mentioning also that these new frequencies 
are purely damped in the limit of zero wavenumbers.

We finish this work by mentioning one of our outlooks concerning the present work. 
The immediate goal 
is to explore how the above results change when the electric charge is 
introduced in the rotating black strings. We may follow the study on 
gravitoelectromagnetic perturbations of rotating charged black strings of Ref. 
\cite{Miranda:2014vaa} for this purpose.

\section{Acknowledgments}

We thank Alejandra Kandus, Alfonso Ballon-Bayona, Jorge Noronha and Saulo Diles for 
useful discussions and suggestions during the development of this work.
V. T. Z. thanks financial support from Conselho Nacional de Desenvolvimento Cient\'\i fico
e Tecnol\'ogico (CNPq, Brazil), Grant No.~308346/2015-7, and from Coordena\c{c}\~ao de
Aperfei\c{c}oamento do Pessoal de N\'\i vel Superior (CAPES, Brazil),
Grant No.~88881.064999/2014-01. L. A. H. M. thanks financial support from Funda\c{c}\~ao de
Amparo \`a Pesquisa do Estado de S\~ao Paulo (FAPESP, Brazil), Grant No. 
2013/17642-5, and from Programa Nacional de P\'os-Doutorado (PNPD/CAPES, Brasil).

\appendix

\section{Relativistic wave vectors}
\label{sec-RelativisticWaveVectors}
The main goal of this appendix is to find the dispersion relations 
of sound waves propagating in a moving fluid from the known dispersion
relations in the fluid rest frame. The motivation for this alternative
analysis is to get a better understanding of the dispersion 
relations obtained in the study of the gravitational perturbations of
rotating black strings,
e.g., Eq.~\eqref{rotscalar} of Sec.~\ref{sec-analiticSound}.
For that, we start reviewing the sound-wave propagation in a medium 
(which is identified as the CFT plasma in the AdS/CFT duality) and 
the wave vector transformation under a Lorentz boost.

Let $\bar K $ and $K$ denote two inertial 
reference frames. Later on we shall identify $\bar K$ as the
rest frame and $K$ as the moving frame with respect to the fluid.
The relations between the four-vector components $\bar k^{\mu}$ 
(defined in the frame $\bar K$) and $k^{\,\mu}$ (in the frame $K$) is 
$k^{\,\mu}=\Lambda^{\mu}_{\,\,\nu}\,\bar k^{\nu}$, 
where $\Lambda^{\mu}_{\,\,\nu}$ is the 
Lorentz transformation matrix. In the present case, the fluid lives 
in a $(2+1)$-dimensional Minkowski spacetime. Explicitly,
we may write (see, e.g., Ref. \cite{Jackson:1998}):
\begin{equation}\label{eq30}
\left\{
\begin{aligned}
&\omega=\gamma(\bar \omega-\bm{v}\cdot\bm{\bar k}),\\
&k_{\scriptscriptstyle{\parallel}}
=\gamma(\bar k_{\scriptscriptstyle{\parallel}}
-v\,\bar \omega),\\
&\bm{k}_{\scriptscriptstyle{\perp}}
=\bm{\bar k}_{\scriptscriptstyle{\perp}},
\end{aligned}
\right.
\end{equation}
where $\bm{v}$ is the velocity of frame $\bar K$ with respect to $K$,
$\bar k_{\scriptscriptstyle{\parallel}}$ ($k_{\scriptscriptstyle{\parallel}}$)
and $\bm{\bar k}_{\scriptscriptstyle{\perp}}$ 
($\bm{k}_{\scriptscriptstyle{\perp}}$) are, respectively, the parallel and 
perpendicular components of the wave vector with respect to $\bm{v}$.
In order to compare with the results presented in the previous 
sections, we identify $\bm{v}=\bm{a}\alpha$, where boldfaced symbols are used 
to indicate the spatial parts of vectors in the $(2+1)$-dimensional Minkowski spacetime.
The form of Eq.~\eqref{eq30} guarantees the invariance of the phase 
of the plane waves. For a discussion on this point see, for instance, 
Ref.~\cite{Houlrik:2009}. The angular frequency in the rest 
frame $\bar K$ is $\bar\omega=
\bm{\bar k}\cdot\bm{\bar u}_{\scriptscriptstyle{p}}$, where 
$\bm{\bar u}_{\scriptscriptstyle{p}}$ is the phase 
velocity, with longitudinal projection given by 
$\bm{\bar u}_{\scriptscriptstyle{p}}^{\scriptscriptstyle{L}}=
(\bm{\bar k}\cdot\bm{\bar u}_{\scriptscriptstyle{p}})\bm{\bar k}/\bar k^2$.  
The transverse components of the phase velocity are indefinite 
\cite{Houlrik:2009}. Analogously, in the moving frame $K$, the angular 
frequency is $\omega=\bm{k}\cdot\bm{u}_{\scriptscriptstyle{p}}$.
From here on, in this appendix, we also use the relations
$\bar\omega=\bm{\bar k}\cdot\bm{\bar u}_{\scriptscriptstyle{p}}
=\bar k\,\bar c$ in frame $\bar K$, and 
$\omega=\bm{k}\cdot\bm{u}_{\scriptscriptstyle{p}}
= k\,c$ in frame $K$.

Our next task is then to transform
the identity $\bar\omega = \bar c\,\bar k$ from $\bar K$ to
the new frame $K$, i.e., we are interested in obtained the transformed quantities 
$\omega$, $c$ and $k$ entering the identity
$\omega = c\, k$ in frame $K$. By applying the inverse of Lorentz 
transformation~\eqref{eq30} to such an identity it follows 
\begin{equation}\label{linearterm}
\begin{split}
\gamma\,\left(\omega-v\,k_{\scriptscriptstyle{\parallel}}\right)
=\bar c \sqrt{\gamma^2\left(k_{\scriptscriptstyle{\parallel}}
-v\,\omega\right)^2
+\bm{  
k}^{2}_{\scriptscriptstyle{\perp}}}\,.
\end{split}
\end{equation}
To complete the transformation, it is necessary to find the transformation 
from the speed $\bar c$ in frame $\bar K$ to the speed $c$ in the 
new frame $K$. 
We then solve Eq.~\eqref{linearterm} for the frequency $\omega$ and use 
the identity $\omega=c\,k$ to get
\begin{equation}\label{speedsound}
\begin{split}
\hspace*{-.2cm}
c\;=\;&\frac{(1-{\bar c\,}^2)v\cos\theta}{1-v^2\bar{ c\,}^2}\\
&
\pm\frac{\bar{c}\,\sqrt{1-v^2}\,\sqrt{1-v^2\bar{c\,}^2
-\left(1-\bar{c}^2\right)v^2\cos^2\theta}}
{1-v^2\bar{ c\,}^2},
\end{split}
\end{equation}
where we used the relations $k_{\scriptscriptstyle{\parallel}}
=k\,\cos{\theta}$ and $|\bm{k}_{\scriptscriptstyle{\perp}}|
=k\,\sin{\theta}$. 
A similar result to Eq.~\eqref{speedsound} was 
found in the context of relativistic (super)fluids in Ref.~\cite{Haber:2015exa}.

Now we are ready to deal with the dispersion relations
in both reference frames. We know the dispersion relation of a 
sound wave propagating in a dissipative medium in the 
rest frame of the fluid (see, for instance, Ref.~\cite{Kanadoff:1963}), 
as being
\begin{equation}\label{eq33}
\bar \omega
=\bar c \sqrt{\bar k^{2}_{\scriptscriptstyle{\parallel}}
+\bm{ \bar  
k}^{2}_{\scriptscriptstyle{\perp}}} 
-\frac{i\,(\zeta+\eta)}{2(\epsilon+p)} \left(\bar 
k^{2}_{\scriptscriptstyle{\parallel}}
+\bm{ \bar  k}^{2}_{\scriptscriptstyle{\perp}}\right)
+\mathcal{O}(\bar 
k^{3},\bar k^{4},\cdots).
\end{equation}
Quantities $\zeta$, $\eta$, $\epsilon$ and $p$ are, respectively, the bulk viscosity, 
shear viscosity, energy density and the pressure of the fluid. The additional terms, i.e.,
$\mathcal{O}(\bar k^{3},\bar k^{4},\cdots)$, are subleading 
terms in the regime where the frequency and wavenumber are of the same order.
Here, it is interesting to point out that in a CFT fluid $\epsilon=2p$ and
$\zeta=0$.

In order to build the dispersion relation in the frame $K$, we start with the 
expression of the invariant wave phase written in both 
frames \cite{Houlrik:2009},
\begin{equation}\label{eq31}
\phi=
\bar\omega\,\bar t-\bm{\bar k}\cdot\bm{\bar r}=
\omega\,t-\bm{k}\cdot\bm{r}.
\end{equation}
Taking the derivative of Eq.~\eqref{eq31} with respect to time $t$, considering a
constant phase and using Lorentz transformations, it follows
\begin{equation}\label{eq32}
\omega=\bm{k}\cdot\bm{u}_{\scriptscriptstyle{p}}
+\frac{1}{\gamma(1 
-\bm{v}\cdot\bm{\bar u}_{\scriptscriptstyle{p}})}
\left(\bar \omega-\bm{\bar 
k}\cdot\bm{\bar u}_{\scriptscriptstyle{p}}\right),
\end{equation}
where $\bm{u}_{\scriptscriptstyle{p}}
=d\bm{r}/dt$ and $\bm{\bar u}_{\scriptscriptstyle{p}}
=d\mathbf{\bar r}/d\bar{t}$ are the 
wave velocities, i.e., 
the phase velocities in  each reference frame. Notice 
that, at first order
approximation in the frequencies and wavenumbers, Eq.~\eqref{eq32} is 
trivially satisfied. However, we now are assuming that the dispersion
relations are no longer linear. This means that the relation 
$\bar \omega = \bar c\, \bar k$ (or $\omega = c\, k$ in frame $K$) 
are first order approximations to the full dispersion relation.

Finally, the dispersion relation in the new frame can be obtained by 
substituting the relation 
$\bm{k}\cdot \bm{u}_{\scriptscriptstyle{p}}
\equiv k\, c$ and Eq.~\eqref{eq33} 
into \eqref{eq32},
\begin{equation}\label{eq34}
\begin{split}
\omega&= c\,\sqrt{k^2_{\scriptscriptstyle{\parallel}}
+\bm{k}^2_{\scriptscriptstyle{\perp}}}
-\frac{i\,(\zeta+\eta)}
{2\gamma(\epsilon+p) (1-\bm{v}\cdot\bm{\bar u}_{\scriptscriptstyle{p}})} \\ 
& \times\left[\gamma^2k^2_{\scriptscriptstyle{\parallel}}
\left(1-\frac{v\,c}{\cos\theta}
\right)^2+\bm{k}^2_{\scriptscriptstyle{\perp}}\right]+\cdots.
\end{split}
\end{equation}
Notice that $\bm{v}\cdot\bm{\bar u}_{\scriptscriptstyle{p}}
=v\,\bar{c}\,\cos\bar\theta$, where $\cos\bar{\theta}
=\bar k_{\scriptscriptstyle{\parallel}}/\bar{k}$.
Eq.~\eqref{eq34} is equivalent to 
Eq.~\eqref{rotscalar} obtained in 
Sec.~\ref{sec-analiticSound}. Here we prove this equivalence
at least for the linear term in the 
wavenumbers. We start by replacing  
Eq.~\eqref{speedsound} into Eq.~\eqref{eq34} to get
\begin{equation}
\begin{split}
\omega\;=\; &\frac{(1-{\bar c\,}^2)v\,k_{\scriptscriptstyle{\parallel}}}
{1-v^2\bar{ c\,}^2} \pm\frac{\bar{c}\,\sqrt{1-v^2}}{1-v^2\bar{ c\,}^2} \\
\; & 
\times 
\sqrt{(1-v^2\bar{c\,}^2)k^2
-\left(1-\bar{c}^{\,2}\right)v^2(k_{\scriptscriptstyle{\parallel}})^2}\,,
\end{split}
\end{equation}
where it was taken into account the $\pm$ signs of Eq.~\eqref{speedsound}.
Then, by setting $k_{\scriptscriptstyle{\parallel}}=k\cos\,\theta=\alpha\,m$, 
$k_{\scriptscriptstyle{\perp}}=k\sin\,\theta=q$, $\bar c=1/\sqrt{2}$, $v=a\alpha$, and 
$(a\alpha)^2=1-1/\gamma^2$, after simplications we obtain
\begin{equation}\label{DRLinearWave}
\begin{split}
\hspace*{-.2cm}
\omega\;=\;\frac{a\alpha\,\gamma^2}{1+\gamma^2}\alpha m
\pm\frac{1}{1+\gamma^2}\sqrt{2\alpha^2m^2+(1+\gamma^2)q^2}.
\end{split}
\end{equation}
Notice that the right-hand side of \eqref{DRLinearWave} is precisely the 
linear term of Eq.~\eqref{rotscalar}. The proof of the 
equivalence for the higher order terms requires additional 
algebraic manipulations and we do not present here.

By comparing the results \eqref{eq33} and \eqref{eq34}, we obtain a little more
clarification about the imaginary part of the frequency. Setting $k_\perp =0$
(to simplify the analysis) in Eq.~\eqref{eq34} and considering the $\pm $ signs
in Eq.~\eqref{speedsound} we obtain the relation
\begin{equation}\label{diffusioncoef}
\omega_{I}=
\frac{(1-v\,\bar c)}{\gamma\,(1+v\,\bar{c})^2}\bar \omega_{I} 
\end{equation}
for the plus sign, and
\begin{equation}\label{diffusioncoef2}
\omega_{I}=
\frac{1}{\gamma\,(1-v\,\bar{c})}\bar \omega_{I}
\end{equation}
for the minus sign. In the last two relations, $\bar \omega_{I}$ 
is the imaginary part of the frequency in the
fluid rest frame $\bar K$, namely,
\begin{equation} \label{barD}
\bar \omega_{I}=\frac{1}{2}\frac{\eta+\zeta}{\epsilon+p}
\,{\bar k}^{2}_{\parallel}\,.
\end{equation}
The imaginary part of the frequency in the moving frame $K$, $\omega_{I}$,
is modified (contracted) by the Lorentz factor $\gamma$, and it is also modified
due to the relative motion by the factors $(1-v\,{\bar c})$ and $(1+v\,{\bar c})$.

Taking into account that the damping 
time $\tau$ goes as the 
inverse of the imaginary part of the wave 
frequency, i.e. $\tau = 1/\omega_{I}$, for \eqref{diffusioncoef}
it follows  
\begin{equation}\label{eq35}
\tau=\gamma^3\frac{(1+{v}\,{\bar c})^2}{(1-{v}\,{\bar c})}\bar \tau,
\end{equation}
while for \eqref{diffusioncoef2} it gives
\begin{equation}\label{eq35b}
\tau=\gamma(1-{v}\,{\bar c})\bar \tau,
\end{equation}
where $\bar\tau$ and $\tau$ are the damping times 
in the frames $\bar K$ and $K$, respectively. 
Equations \eqref{eq35} and \eqref{eq35b} are valid for 
arbitrary values of the velocity $v$. By taking 
$v=\bar c$ into Eq.~\eqref{eq35b}, it becomes 
$\tau=\bar\tau/\gamma$ which is the time dilation 
effect.

Furthermore, yet by comparing equations \eqref{eq33} 
and \eqref{eq34} we see that 
the parallel wavenumber component changes 
as follows
\begin{equation}\label{eq36}
\bar k_{\scriptscriptstyle{\parallel}}=\left(1-\frac{v\,c}
{\cos\theta}\right) 
\gamma\,k_{\scriptscriptstyle{\parallel}}.
\end{equation}

Considering the particular situation where 
$k_{\scriptscriptstyle{\perp}}=0$
or, equivalently $\theta=0$, 
the plus sign solution of Eq.~\eqref{speedsound} can be written
as $c=(v+\bar c)/(1+v\,\bar c)$. Replacing into 
Eq.~\eqref{eq36} it reduces to
\begin{equation}\label{eq36a}
k_{\scriptscriptstyle{\parallel}}
=(1+v\,\bar c)\gamma\,\bar k_{\scriptscriptstyle{\parallel}}. 
\end{equation}
On the other hand,
by considering the minus sign solution and taking
$\theta =0$ into  Eq.~\eqref{speedsound} it
can be written as $c=(v-\bar c)/(1-v\,\bar c)$, so that 
Eq.~\eqref{eq36} reduces to
\begin{equation}\label{eq36b}
k_{\scriptscriptstyle{\parallel}}
=(1-v\,\bar c)\gamma\,\bar k_{\scriptscriptstyle{\parallel}}.
\end{equation}
Replacing $v=\bar c$ in Eq.~\eqref{eq36b} it reduces to 
$k_{\scriptscriptstyle{\parallel}}
=\bar k_{\scriptscriptstyle{\parallel}}/\gamma$ which 
is the Lorentz contraction. It is worth to point out 
that by doing $v=\bar c$ the observer $K$ is comoving with
the frame of the wave front.

\end{document}